# The UV Legacy Library of Young Stars as Essential Standards (ULLYSES) Large Director's Discretionary Program with Hubble. I. Goals, Design, and Initial Results

Julia Roman-Duval,[1] William J. Fischer,[1] Alexander W. Fullerton,[1] Jo Taylor,[1] Rachel Plesha,[1] Charles Proffitt,[1] TalaWanda Monroe,[1] Travis C. Fischer,[2] Alessandra Aloisi,[1] Jean-Claude Bouret,[3] Christopher Britt,[1] Nuria Calvet,[4] Joleen K. Carlberg,[1] Paul A. Crowther,[5] Gisella De Rosa,[1] William V. Dixon,[1] Catherine C. Espaillat,[6] Christopher J. Evans,[7] Andrew J. Fox,[2,8] Kevin France,[9] Miriam Garcia,[10] Scott W. Fleming,[1] Elaine M. Frazer,[1] Ana I. Gómez De Castro,[11] Gregory J. Herczeg,[12,13] Svea Hernandez,[2] Alec S. Hirschauer,[1,14] Bethan L. James,[15] Christopher M. Johns-Krull,[16] Claus Leitherer,[1] Sean Lockwood,[1] Joan Najita,[17] M.S. Oey,[18] Cristina Oliveira,[1] Tyler Pauly,[1] I. Neill Reid,[1] Adric Riedel,[1] David R. Rodriguez,[1] David Sahnow,[1] Ravi Sankrit,[1] Kenneth R. Sembach,[1] Richard Shaw,[1] Linda J. Smith,[1] S. Tony Sohn,[1] Debopam Som,[1] Leonardo Úbeda,[1] and Daniel E. Welty[1]

[1]Space Telescope Science Institute
3700 San Martin Drive
Baltimore, MD 21218, USA
[2]AURA for ESA, Space Telescope Science Institute
3700 San Martin Drive
Baltimore, MD 21218, USA
[3]Aix-Marseille University, CNRS, CNES, LAM
13388 Marseille, France
[4]Department of Astronomy, University of Michigan, 1085 South University Avenue, Ann Arbor, MI 48109, USA
[5]Dept Physics & Astronomy
Hounsfield Road, University of Sheffield
Sheffield S3 7RH, UK
[6]Institute for Astrophysical Research, Department of Astronomy, Boston University, 725 Commonwealth Avenue, Boston, MA 02215, USA
[7]European Space Agency (ESA), ESA Office, Space Telescope Science Institute
3700 San Martin Drive
Baltimore, MD 21218, USA
[8]Department of Physics & Astronomy, Johns Hopkins University, 3400 N. Charles St., Baltimore, MD 21218
[9]Laboratory for Atmospheric and Space Physics, University of Colorado Boulder
Boulder, CO 80303, USA
[10]Centro de Astrobiologia, CSIC-INTA, Dpto. de Astrofisica, Instituto Nacional de Tecnica Aeroespacial, Ctra. de Torrejon a Ajalvir, km 4, 28850 Torrejon de Ardoz (Madrid), Spain
[11]Space Astronomy Research Group-AEGORA, Fac CC Matematicas
Universidad Complutense de Madrid
Plaza de Ciencias 3
28040 Madrid, Spain
[12]Kavli Institute for Astronomy and Astrophysics, Peking University
Beijing 100871, China
[13]Department of Astronomy, Peking University
Beijing 100871, China
[14]Department of Physics & Engineering Physics, Morgan State University, 1700 East Cold Spring Lane, Baltimore, MD 21251, USA
[15]AURA for ESA, Space Telescope Science Institute
3700 San Martin Drive
Baltimore, MD 21218, USA
[16]Department of Physics and Astronomy, Rice University
6100 Main Street
Houston, TX 77005, USA
[17]NSF's NOIRLab
950 N. Cherry Avenue
Tucson, AZ 85719, USA
[18]University of Michigan, 1085 S. University Ave
Ann Arbor, MI 48109

## ABSTRACT

Specifically selected to leverage the unique ultraviolet capabilities of the *Hubble Space Telescope*, the Hubble Ultraviolet Legacy Library of Young Stars as Essential Standards (ULLYSES) is a Director's Discretionary program of approximately 1000 orbits — the largest ever executed — that produced a UV spectroscopic library of O and B stars in nearby low metallicity galaxies and accreting low mass stars in the Milky Way. Observations from ULLYSES combined with archival spectra uniformly sample the fundamental astrophysical parameter space for each mass regime, including spectral type, luminosity class, and metallicity for massive stars, and the mass, age, and disk accretion rate for low-mass stars. The ULLYSES spectral library of massive stars will be critical to characterize how massive stars evolve at different metallicities; to advance our understanding of the production of ionizing photons, and thus of galaxy evolution and the re-ionization of the Universe; and to provide the templates necessary for the synthesis of integrated stellar populations. The massive star spectra are also transforming our understanding of the interstellar and circumgalactic media of low metallicity galaxies. On the low-mass end, UV spectra of T Tauri stars contain a plethora of diagnostics of accretion, winds, and the warm disk surface. These diagnostics are crucial for evaluating disk evolution and provide important input to assess atmospheric escape of planets and to interpret powerful probes of disk chemistry, as observed with ALMA and JWST. In this paper we motivate the design of the program, describe the observing strategy and target selection, and present initial results.

## 1. INTRODUCTION

Stars are important components of galaxy evolution and ingredients for life. Stars enrich their parent galaxy with newly minted heavy elements while injecting energy and momentum into their surroundings through winds, jets, supernova (SN) explosions, and ionizing radiation. This stellar feedback contributes to galaxy evolution by regulating star formation, enriching gas in heavy elements, expelling chemically enriched gas from galaxies into the circum- and inter-galactic media and redistributing gas and dust within galaxies, inducing metallicity gradients, etc. Heavy elements formed in the cores of massive stars eventually coalesce into protoplanetary disks and planets forming around low-mass stars. The process of mass accretion onto low mass stars results in energetic radiation in the UV and X-ray, which influences the evolution and chemistry of the proto-planetary disks, and is a determining factor in the habitability of planets within them.

The overarching scientific goals and design of the Ultraviolet Legacy Library of Young Stars as Essential Standards (ULLYSES) program were determined in partnership with the astronomical community through an ultraviolet (UV) Legacy Working Group (UVLWG) and report[1] and a Science Advisory Committee (SAC)[2] composed of leading experts in the fields of massive stars, population synthesis, and accreting low mass stars. The goal of ULLYSES was to form a complete reference sample that can be used to create spectral libraries capturing the diversity of stars, ensuring a legacy dataset for a wide range of astrophysical fields.

The immediate objective of the massive star observations is to create a spectral library necessary to characterize their winds (velocity law, density structure, and mass loss rate, see Hawcroft et al. 2023; Krtička et al. 2024; Parsons et al. 2024a) and photospheres (temperature, gravity, chemical abundances, rotational velocity, see Martins et al. 2024), as a function of metallicity, spectral type, and luminosity class (the parameter space). While these stellar properties strongly influence the evolution of massive stars, detailed comparison with model predictions have been hampered by the haphazard coverage of parameter space for stars with different metallicities (Leitherer et al. 2010; Wofford et al. 2016; Crowther 2019).

ULLYSES addresses this shortcoming by obtaining UV spectra of 94 targets in the LMC (50% solar metallicity Russell & Dopita 1992; Tchernyshyov et al. 2015, and references therein) and 60 stars in the SMC (20% solar metallicity Russell & Dopita 1992; Tchernyshyov et al. 2015, and references therein) with the Cosmic Origins Spectrograph (COS) and Space Telescope Imaging Spectrograph (STIS) onboard the *Hubble Space Telescope* (HST). The ULLYSES observations complement and leverage existing archival HST (88 stars in the LMC, 73 stars in the SMC) and *Far Ultraviolet Space Explorer* (FUSE) data (for 155 stars in the LMC and SMC). In addition, three early-type stars in each of NGC 3109 (∼15% solar metallicity Evans et al. 2007) and Sextans A (∼10% solar metallicity Kaufer et al. 2004) were observed to explore the effects of even lower metallicities on stellar and mass loss properties. The ULLY-SES spectral library provides the templates necessary for the synthesis of integrated stellar populations at all redshifts (Chisholm et al. 2019; Crowther & Castro 2024) that are or will be accessible to HST, the *James Webb Space Telescope* (JWST), and the next generation of Extremely Large Telescopes (ELTs), and advance our understanding of Lyman-continuum escape and the reionization of the Universe.

The low mass star component of the program aims to constrain accretion physics in TTS. The far-UV (FUV) spectra reveal the dynamics and morphology of accretion shocks and winds (see review by Schneider et al. 2020), including emission in highly excited lines such as C IV (e.g. Lamzin 1998; Ardila et al. 2013) and wind absorption in neutral and low-ionization lines (e.g. Xu et al. 2021). The accretion rate can be derived from modeling a star's near-UV (NUV)/optical/near-infrared (NIR) continuum (Calvet & Gullbring 1998; Gullbring et al. 1998; Ingleby et al. 2013; Robinson & Espaillat 2019). FUV and NUV spectra of TTS also encode information about the energetics and kinematics of outflows and jets through the C II multiplet (at 1335 Å) and semi-forbidden and forbidden UV lines such as O III] 1663, N III] 1750, C III] 1908, Si III] 1892, C II] 2326, and [O II] 2471 (Gómez de Castro & Verdugo 2003; Gómez de Castro & Ferro-Fontán 2005; Lopez-Martinez & Gómez de Castro 2015). Additionally, FUV radiation determines the thermal structure of the disk and ultimately its photo-evaporation and dispersal (Pascucci et al. 2022; Arulanantham et al. 2021, 2023; Gangi et al. 2023; France et al. 2023). In turn, the timescale for disk dispersal influences the formation, evolution, and atmospheric escape of planets (Owen & Wu 2016), while the UV irradiation on planets plays a major role in their atmospheric photo-chemistry and escape (France et al. 2018; Feinstein et al. 2022). UV spectra of TTS therefore contain a plethora of diagnostics for disk evolution and planet habitability that will be needed to interpret the powerful probes of disk chemistry observed with the Atacama Large Millimeter Array (ALMA) and JWST.

Prior to ULLYSES, most HST spectroscopy of T Tauri stars targeted objects above 0.5 $M_\odot$ in Taurus (e.g. France et al. 2012; Ingleby et al. 2013). High quality IUE (Valenti et al. 2000; Johns-Krull et al. 2000) and HST ACS/SBC (Yang et al. 2012) spectra established UV radiation fields and correlations with accretion rate, but lacked the spectral resolution necessary to interpret lines and the near-simultaneous optical coverage to place lines in context of the accretion rate. ULLYSES complements the archival samples by uniformly sampling accretion rates and stellar masses (the parameter space for this component) and expanding the spectroscopic library of

T Tauri stars to masses 0.05 $M_\odot < M_* < 0.5$ $M_\odot$ with COS and STIS UV-optical-NIR spectra of 58 Tauri stars in nearby star-forming regions at different evolutionary stages, enabling a complete picture of the evolution of accretion. Additionally, ULLYSES performed a comprehensive UV study of accretion variability on various time scales by devoting 100 orbits to monitoring four prototypical objects (TW Hya, BP Tau, RU Lup, and GM Aur). Those four objects were monitored with four observations per rotational period over three consecutive rotation periods. This pattern is repeated over two epochs separated by a year. These monitoring observations inform accretion variability over timescales ranging from minutes to years (Wendeborn et al. 2024a,b).

The legacy of the program goes well beyond its immediate goals, as reflected by the multitude and diversity of accepted *general observer* (GO) and *archival* (AR) HST programs utilizing or complementing ULLYSES data (16 programs in HST Cycles 27-31). In addition to programs directly addressing stellar physics, several programs are tackling key questions about the ISM and CGM thanks to their absorption imprint on the massive star spectra. Specifically, the massive star spectra are being used to characterize interstellar chemical abundances, dust depletions (e.g. Roman-Duval et al. 2022), and extinction at low metallicity (e.g. Gordon et al. 2024), as well as the structure and kinematics of the Milky Way, LMC, and SMC halos, and ram-pressure and stellar feedback in the LMC (e.g., Zheng et al. 2024).

In this paper, we describe the design and initial results of this unprecedented effort. The execution, data reduction and calibration, and the design and generation of high-level science data products will be described in Roman-Duval et al. (in prep, hereafter Paper II). In Section 2 of this paper, we present the observing strategy adopted for the observations of the high and low mass stars. In Section 3, we provide the general strategy used for the target selection, while Sections 4 and 6 provide the details of the target selection for the massive stars and low-mass stars, respectively. In Section 7, we present ancillary and coordinated programs led by the STScI team or the community that greatly enhance the scientific value of the program. In Section 8, we present some initial results from the ULLYSES program.

## 2. OBSERVING STRATEGY

### 2.1. *Massive stars in the LMC and SMC*

#### 2.1.1. *Instrumental modes*

The instrumental requirements in common which enable stellar, ISM, and CGM science for the massive star component of ULLYSES are full UV coverage at medium-resolution (R > 15,000) and high signal-to-noise ratio (S/N of 20-30 per resolution element in the continuum).

Table 1. Instrumental configurations for the stars in the ULLYSES sample

| ULLYSES Component | Spectral Type | Wavelength Range of Interest | Instrumental Mode |
|---|---|---|---|
| LMC, SMC | O | 950-1150 Å | COS G130M/1096 |
| | O-B | 1150-1750 Å | STIS E140M |
| | | | COS G130M/1291 + G160M/1611 |
| | O9 I - B9 I | 1850-2100 Å | STIS E230M/1978 |
| | | | COS G185M/1953 + G185M/1986 |
| | B5 I - B9 I | 2800 Å | STIS E230M/2707 |
| Sextans A, NGC 3109 | O-early B | 950-1800 Å | COS G140L/800 |
| T Tauri single epoch | K0 - M6 | 1150 - 1750 Å | COS G130M/1291 + G160M/1589 + G160M/1623[a] |
| | | 1750 - 10200 Å | STIS G230L/2376 + G430L/4300 + G750L/7751 |
| T Tauri monitoring | K3 - K7 | 1400 - 3150 Å | COS G160M/1589,1623 + G230L/2635, 2950 |

[a] The ULLYSES observations of T Tauri stars in Orion OB1 in fall 2020 utilized the G160M/1611 cenwave only. Later observations were modified to dither over the gap at 1600 Å where the molecular bump is located using the G160M 1589 and 1623 cenwaves.

On the stellar side, the 940-1750 Å range contains a plethora of diagnostics for the winds and abundances of O and B stars (e.g., C III $\lambda$1175, Fe III $\lambda$1207, $\lambda$1240, N V $\lambda\lambda$1238, 1242, Si IV $\lambda\lambda$1393, 1402, C IV $\lambda$1548, 1550, He II $\lambda$1640, Fe IV forest between 1550 and 1700 Å; see Leitherer et al. 2011), with the shorter wavelength range (940-1150 Å) being specifically interesting for O stars, for example, O VI $\lambda\lambda$1032, 1038 (see, e.g., Evans et al. 2004a) and P V $\lambda\lambda$ 1118, 1128 (Crowther et al. 2002; Fullerton et al. 2006). Measurements of abundances in massive stars require resolving powers greater than R = 10,000-15,000. The 1150-1750 Å wavelength range can be covered at medium resolution by either COS G130M + G160M or STIS E140M, with the former geared toward fainter stars, while the latter offers higher spectral resolution and a smaller aperture (0.2×0.2") which can be advantageous in crowded fields. In order to maximize both the lifetime of COS and the spectral resolution of the observations, STIS E140M was preferentially used for stars that could be observed in three orbits or fewer; COS was used otherwise. The short wavelength range (940-1150 Å) is covered by FUSE and COS G130M/1096. Targets with archival FUSE data were preferentially selected (see Section 4.1). However, given the paucity of catalogued early O stars in the LMC and SMC, some O stars without FUSE data were selected for observations in order to cover the parameter space. In those cases, bright O stars that could be observed with COS G130M/1096 in (< 8) orbits were selected for observations with this mode.

Supergiants of spectral type O9 and later were also observed in the NUV to cover the Al III $\lambda\lambda$1855, 1863 P Cygni-like profiles and Fe III forest ($\lambda$1850-2100 Å). Indeed, the ionization fraction in the atmosphere of supergiants is such that the Fe III forest and Al III lines in the 1800-2200 Å wavelength range appear in stellar spectra for effective temperatures lower than about 30,000 K, which corresponds to the O8-O9 spectral types (Smith et al. 2002). Supergiants of spectral type O9 and later were preferentially observed with the STIS E230M/1978 (1600-2380 Å) mode. A few targets were too faint to be observed with STIS in a reasonable amount of time, and were instead observed with COS G185M/1953 and 1986 central wavelength settings.

Finally, the Mg II $\lambda\lambda$2796, 2804 doublet is an important wind diagnostic for B supergiants of spectral type

B5 and later (Bates & Gilheany 1990), while this diagnostic fades for earlier spectral types (Gurzadian 1975). Therefore, B5-B9 supergiants were observed with the STIS E230M/2707 mode.

From an ISM perspective, the observing strategy outlined above covers the wavelength range 940-2400 Å in spectra toward a large sample of O and early B stars, which are ideal for interstellar line measurements owing to their large rotational velocities (typically $> 80$ km s$^{-1}$) and relatively smooth continua. This wavelength coverage includes all relevant UV diagnostics for interstellar absorption line measurements in the UV ($H_2$ Lyman-Werner bands, Fe II $\lambda\lambda 1143$, 1144, $\lambda\lambda$ 2249, 2260, H I $\lambda 1216$, Mg II $\lambda\lambda 1239,1240$, S II $\lambda\lambda 1250$, 1253, C II $\lambda 1334$, Si II $\lambda\lambda\lambda 1260$, 1526, 1808, Zn II $\lambda\lambda 2026$, 2062 etc; see Table 2 in Roman-Duval et al. 2021).

From a CGM perspective, diagnostics of the disk-halo interface and galaxy-scale winds, such as O VI $\lambda\lambda 1032$, 1038, Si III $\lambda 1206$, Si IV $\lambda\lambda 1393$, 1402, and C IV $\lambda\lambda 1548$, 1550 (Tumlinson et al. 2017) are covered by the above observing strategy.

We note that, at lifetime position 4 of COS FUV where most of the ULLYSES spectra are taken, there is a "gain-sag" detector gap caused by Ly-$\alpha$ airglow imprinted on the detector by the general use of the G130M/1291 central wavelength (hereafter, cenwave). This gap can be spectrally dithered over by the use of all four FP-POS and/or different cenwaves for G160M. However, only FP-POS 3 and 4 are allowed for G130M/1291 in the framework of the COS2025 usage policies[3]. Therefore, a detector gap exists around the position of redshift zero Ly-$\alpha$ (1216 Å) in G130M/1291 data. Ly-$\alpha$ $\lambda 1216$ Å is used to measure the column density of hydrogen in the ISM, which in turn serves as the normalization for interstellar abundance measurements. Fortunately, the Ly-$\alpha$ interstellar absorption feature in the LMC and SMC is much broader than the detector gap induced by gain-sag (e.g., Roman-Duval et al. 2019), such that only the very bottom trough of the absorption profile, which is not used in column density measurements, is masked out from the spectra. The gain-sag detector gap is therefore not a concern for ISM science, nor for CGM or stellar astrophysics.

#### 2.1.2. Signal-to-noise

Both the SAC and UVLWG recommended that a S/N of 30 per resolution element (resel) with COS (R = 15,000) and 20 per resel with STIS (R = 30,000) was necessary to achieve the scientific goals of the ULLY-SES program. Those numbers are in line with many observational programs focused on massive stars, the ISM, or CGM. In practice, S/N and exposure times are computed for a given wavelength, and with throughputs and source spectra varying with wavelengths, some scientific compromises have to be made on the fraction of the spectral range reaching the goal S/N versus the orbit cost of an observation. For massive stars in the LMC and SMC, we found that targeting the S/N requirements at the wavelengths outlined in Table 2 constitute a pragmatic and effective approach.

### 2.2. Massive stars in low metallicity galaxies

The stellar, ISM, and CGM UV spectroscopic diagnostics for O and early B stars in the low-metallicity galaxies targeted by ULLYSES (NGC 3109 and Sextans A) are similar to those for the LMC/SMC stars. However, at distances of 1.27 Mpc and 1.32 Mpc respectively, the flux of stars in NGC 3109 and Sextans A are reduced by almost three orders of magnitude compared to the LMC and SMC. Such faint stars cannot be observed in a few orbits with medium-resolution gratings at a S/N of 20-30 per resel, even with COS. For the very low metallicity stars in Sextans A and NGC 3109, the requirement was therefore S/N $> 15$ over the 1130-1600 Å wavelength range (which includes the N V and C IV wind lines) at R $\sim$ 3000. This can be accomplished by either the use of the low resolution COS G140L/800 setting, which has the broadest wavelength coverage, or the use of the G130M + G160M settings, with reduced wavelength coverage but the added ability to bin the spectra to any resolution between R $\sim$ 3,000 and 15,000. The widest wavelength coverage offered by COS G140L/800 was ultimately chosen, since it includes the important O VI wind lines.

Given the expensive orbit cost of the spectroscopic observations for stars in Sextans A and NGC 3109, and the relatively large uncertainty on their UV flux as estimated from V-band magnitudes and spectral type, pre-imaging of fields around the six target stars was obtained using the UVIS channel of the Wide Field Camera 3 (WFC3) onboard HST. The pre-imaging observations used the widest possible combinations of wide-band filters (F225W, F275W, F336W, F475W, F814W) to sample the flux at short wavelengths while also providing sufficient coverage to enable accurate determinations of extinction. In general, the combination of short and long wavelengths photometry in star + dust SED modeling allows us to disentangle the effects of stellar mass, age, and reddening (Gordon et al. 2016).

**Table 2.** S/N Requirements for massive stars in the LMC and SMC

| Instrumental mode | Wavelength | S/N per resel | Resel[a] |
|---|---|---|---|
| | Å | | pixel |
| COS G130M/1096 | 1080 | 16-20 | 9 |
| COS G130M/1291 | 1150 | 30 | 6 |
| COS G160M/1611 | 1590 | 30 | 6 |
| COS G185M/1953 | 1860 | 30 | 3 |
| COS G185M/1986 | 1980 | 30 | 3 |
| STIS E140M/1425 | 1200 | 20 | 2 |
| STIS E230M/1978 | 1800 | 20 | 2 |
| STIS E230M/2707 | 2800 | 20 | 2 |

[a]Size of a resolution element in pixels

### 2.3. Single epoch ("survey") T Tauri stars

#### 2.3.1. Instrumental modes

Three key diagnostics for accretion flows and shocks in T Tauri stars are the N V ($\lambda\lambda1238, 1242$), C IV ($\lambda\lambda1548, 1550$), and Mg II ($\lambda\lambda2796, 2803$) lines. These lines are broad and complex, and in principle require medium resolution. Medium resolution observations of T Tauri stars can be efficiently obtained with COS in the FUV (N V, C IV). However, medium-resolution NUV observations (Mg II) with sufficient wavelength coverage require STIS, which has lower sensitivity than COS. Medium resolution observations of a large sample of TTS with STIS would therefore be prohibitively expensive. The NUV spectra of the ULLYSES TTS were therefore obtained with the low resolution modes of STIS. The NUV-optical-NIR continuum is typically modeled to measure the accretion rate in T Tauri stars, along with the stellar mass and extinction.

Thus, the TTS in the survey sample were observed with the COS G130M, COS G160M, STIS G230L, STIS G430L, and STIS G750L instrumental modes. As for the massive stars, the G130M/1291 setting is affected by gain-sag holes in the Ly$\alpha$ area. Given the ability to dither over those gaps and the width of the Ly$\alpha$ profile of TTS, these small gaps in wavelength coverage do not impact the science goals. For the COS G160M observations, we selected the 1589 (FP-POS 3 and 4) and 1623 (FP-POS 1 and 2) cenwaves in order to maximize coverage and exposure time over the 1600 Å $H_2$ bump (France et al. 2017; Espaillat et al. 2019). For the STIS observations, the 52"×2" aperture was used for its superior throughput and photometric accuracy, unless prohibited by the presence of a close neighbor or other bright object protection issue.

We note a few exceptions to the observing strategy adopted for the vast majority of the sample. In some instances, a few stars did not pass the bright object protection checks for COS G130M/1291 and/or G160M due to bright emission lines (Ly-$\alpha$, Si IV, C IV). Those were instead observed with STIS G140L (which has much less stringent count rate limits) and COS G130M/1222 (which places Ly-$\alpha$ in the detector gap). The STIS G140L setting provides similar wavelength coverage to COS G130M+G160M, but at coarser spectral resolution (R ∼ 1,000). The COS G130M/1222 mode complements the low-resolution spectra by providing medium resolution (R ∼ 15,000) over the N V doublet. Furthermore, early observations obtained in the fall of 2020 ($\sigma$ Ori and Ori OB1 regions) initially used the 1611 cenwave of COS G160M only. It was realized that this setting placed the interesting molecular bump around 1600 Å in the detector gap. Subsequently, the observing strategy was modified to use the combination of G160M/1589 and G160M/1623 (see paragraph above), which offers full wavelength coverage without any gaps. The specific targets affected by these changes are listed in Section 6.1.

In all cases, the full wavelength range (FUV-NUV-optical-NIR) is obtained nearly simultaneously (within 24h to 48h) to ensure that the spectra can be used to constrain the stellar and accretion properties, given the variable nature of accretion in T Tauri stars.

#### 2.3.2. Signal-to-noise

The S/N was agreed upon with community experts and is listed in Table 3. It is based on trade-offs between exposure time and scientific usability.

**Table 3.** S/N Requirements for survey T Tauri Stars

| Instrumental mode | wavelength Å | S/N per resel | Resel pixel |
|---|---|---|---|
| COS G130M/1291 | peak of N V ($\lambda\lambda1238$, 1240) | 15 | 6 |
| COS G160M/1589+1623 | peak of C IV ($\lambda\lambda1548$, 1550) | 20 | 6 |
| STIS G230L | peak of Mg II ($\lambda\lambda$ 2796, 2803) | 20 | 2 |
| STIS G430L | continuum at 4000 Å | 20 | 2 |
| STIS G750L | continuum at 5700 Å | 20 | 2 |

### 2.4. T Tauri stars monitored with HST

For the four stars monitored spectroscopically with HST, each observation is one orbit long and covers both FUV and NUV. The key diagnostics in this case are the C IV ($\lambda\lambda1548$, 1550) and Mg II ($\lambda\lambda2796$, 2803) emission lines, the NUV continuum, from which the mass accretion rate can be estimated and compared to emission line fluxes. To achieve this wavelength coverage in one orbit, these stars were observed with COS G160M (1589 FP-POS 3 and 4 and 1623 FP-POS 1 and 2 to maximize coverage of the $H_2$ bump at 1600 Å) and COS G230L. For the NUV, we selected the 2635 (2436 — 2833 Å) and 2950 (2750 — 3150 Å) cenwaves in order to cover Mg II ($\lambda\lambda2796$, 2803), Si II ($\lambda1808$), and the understudied forbidden and semi-forbidden Si III] ($\lambda1892$), [C III] ($\lambda1907$), C III] ($\lambda1909$) lines (Gómez de Castro & Ferro-Fontán 2005; López-Martínez & Gómez de Castro 2014). The use of two cenwaves allows for correction of vignetting effects on the left edge of stripe B for the 2950 cenwave. Even with one orbit per observation including G160M and G230L, the S/N per exposure in the C IV and Mg II line is well above 30, allowing for multiple time samples per exposure to be extracted as part of the time-series ULLYSES products, which are described in a future paper.

In order to provide sufficient time coverage of accretion variability, while ensuring that variations due to stellar rotation can be disentangled from accretion variability, the optimal cadence was determined to be four observations per rotation period over three rotation periods, with this pattern repeated over two epochs separated by about one year.

### 2.5. LCO photometric monitoring of T Tauri stars

The T Tauri stars observed with HST as part of the ULLYSES program exhibit substantial variability (fac-tors of a few), driven in large part by the stochastic nature of the accretion process. Understanding this variability is not only a key goal of the ULLYSES program, but also a functional requirement to preserve the safety of the HST UV detectors. To fulfill this requirement, we designed an LCO (Las Cumbres Observatory) photometric monitoring program in the u' (monitoring stars only), V and i' bands (all stars, survey and monitoring), coordinated with HST. This program ensures that all low-mass targets benefit from photometric monitoring in the months leading to the HST observations, and that they are not in an accretion burst state that would compromise the safety of the COS and STIS UV detectors. In addition, the photometric monitoring also comes with the key scientific benefit of a uniform cadence in the same photometric bands for all targets, thus maximizing the scientific return of both the HST and LCO observations.

The program primarily utilizes the SBIG imager on the 0.4m telescope network in both time critical and queue modes distributed in equal proportions. A small fraction of the images were obtained with the Sinistro imagers on the 1m telescope network at times when outages affected multiple 0.4m telescopes in the network. We monitored the targets in the V and i' bands, with the V band tracking accretion and the V - i' color providing an estimate of the varying extinction, which is required to derive accretion rates. Additionally, V and i' photometry constrain the contributions from photospheric emission and tie the monitoring to measurements previously reported in the literature. The photometric LCO observations of the four stars monitored with HST also includes the u' band for a more accurate measurement of the varying accretion rate. Monitoring all survey stars in the u' band would have been prohibitive in terms of exposure time.

In order to capture both the long and short term accre-tion variability of T Tauri stars, the LCO observations executed with the following cadence:

- 10 observations per rotational period over the period(s) corresponding to the HST observations (three rotational periods for monitoring stars, one rotational period for survey stars). Rotation periods for many stars were derived from TESS data and provided by Javier Serna (priv. comm.[4]). For the survey stars, a rotational period of 10 days is assumed if the period is not known

- 15 minute cadence during the HST observations

- One observation per day for 10 days before and after the rotational period(s) corresponding to the HST observations

- One observation per day for 10 days approximately three months before and three months after the HST observations

Here, an observation refers to the group of exposures taken in two (survey stars) or three (monitoring stars) bands ($u'$, $V$, $i'$) at a given time. Frequent bursts are commonly associated with accretion onto classical T Tauri stars (Alencar et al. 2010; Cody et al. 2014, e.g., ). The 15 minute cadence during the HST observations is driven by findings from previous short-cadence (1-2 minutes) optical observations of TW Hya, one of the targets in the ULLYSES sample, which revealed variability down to 11-15 min timescales associated with accretion Äúbursts Äù from variable mass loading onto the star (Siwak et al. 2018). The longitudinal coverage of the 0.4m telescopes allows for flexible scheduling and subsequently for the high cadence monitoring we aimed to achieve.

Exposure times were estimated in order to detect changes of 50% in the optical at the $5\sigma$ level, which requires S/N = 10 in the observations. Changes this large imply changes in the UV line flux by a factor of two. Typical exposure times range from 20 to 90 seconds in the $V$ band and 10 to 30 seconds in the $i'$ band. For the monitoring stars, exposure times in the $u'$ band are set to 500s. Our observations in most cases are dominated by overheads (see LCO documentation[5] for a description of overheads), and we therefore set the minimum exposure times in $V$ and $i'$ for the 0.4m network to 20s and 10s, respectively. For exposures with the 1m network, the exposure times used for the 0.4m were typically divided by four.

Fortunately, most ULLYSES T Tauri stars have coverage in SDSS, SkyMapper, or AAVSO/APASS, allowing the LCO frames to be flux-calibrated using non-variable field stars with SDSS, SkyMapper, or AAVSO/APASS $u'$, $V$, and $i'$ magnitudes. In cases where no such coverage exists, or sufficiently bright stars are not present in the field, extra flux calibration fields were observed at the same cadence as the science observations. Separate calibration fields were obtained only for the $u'$ band images of the monitoring stars.

### 2.6. Exposure time estimates

The HST ETC (https://etc.stsci.edu) is not scriptable and therefore not practical to estimate exposure times for thousands of objects — this is the size of the initial catalog of target candidates for massive stars, as we will see in the following sections. In order to overcome this limitation, the ULLYSES core implementation team (CIT) developed a custom, scriptable ETC, the UBETT (ULLYSES batch-mode exposure time tool). The UBETT essentially consists of a comprehensive wrapper around PySynphot. The UBETT can be run in batch mode and includes a wider variety of stellar models than the STScI ETC, for example PoWR (Hainich et al. 2019), WM-Basic (Pauldrach et al. 2001), and CM-FGEN (Hillier & Miller 1998).

Models of the UV spectrum of all low- and high- mass star candidates were input into the UBETT in order to estimate the required exposure time during the target selection phase. During the technical implementation phase, the standard ETC was used on the models SEDs to determine exposure times and perform bright object checks. The methodology used to model and predict the spectra is described in Appendix A.

## 3. TARGET SELECTION: GENERAL STRATEGY AND COMMUNITY INPUT

The target selection for the low- and high-mass components of ULLYSES followed similar strategies, which consisted of:

1. Assembling a comprehensive list of target candidates from catalogs of relevant objects

2. Estimating preliminary exposure times to estimate the cost of the observation in orbits

3. Refining the final sample based on scientific merit as quantified by the relevance of the parameter space covered, the availability of ancillary data (in particular HST and FUSE for the massive stars), and the efficiency of observation (i.e., the total exposure time including all instrumental modes needed to satisfy the observing strategy)

---

[4] https://www.tessextractor.app/

[5] https://lco.global/observatory/instruments/

While the implementation team at STScI performed the target selection, members of the scientific community were invited to provide input through a request for target proposals placed in September 2019. The STScI team received four responses to the call for low-mass stars, and seven responses for high-mass stars. Responses ranged in scope from specific target samples to general recommendation about the types of objects to include. Proposals were taken into account during the target selection process to the extent that they were compatible with the recommendations from the UVLWG report.

In the next few sections, we describe in more detail the target selection procedure for each component of the program. Excerpts from the target tables are listed in Section B.

## 4. TARGET SELECTION FOR THE MASSIVE STARS IN THE LMC AND SMC

The UVLWG recommended populating a grid of spectral types and luminosity classes with medium-resolution UV observations of OB stars in the LMC and SMC, probing metallicities ∼50% and ∼20% solar (see details on the metallicity measurements in Table 4). Specifically, the sample was to target early O to B1 dwarfs and giants, as well as early O to B9 supergiants with at least four stars per spectral type and luminosity class bin to sample the individual variations within each bin (e.g., mass-loss rate, rotational velocity).

Additionally, the UVLWG advocated including a few Wolf-Rayet (WR) stars in each galaxy. The rationale for including WR stars is that 1) these stars can dominate wind feedback and hard ionizing radiation; and 2) their strong emission-line signatures provide key starburst mass and age diagnostics.

The UVLWG report did not emphasize the need to include binarity explicitly as a criterion for selection. Since the incidence of binarity is thought to be as high as 75% among massive stars (Sana et al. 2009), a sample selected solely on spectral type and luminosity class will implicitly contain a significant number of undetected binaries. In this sense, the initial sample provides appropriate templates for applications. Nevertheless, in response to inquiries from the community and after consultation with the SAC, single "snapshot" observations [6] of several interacting binaries were added to the sample to provide a glimpse of the properties of "stripped star" candidates that may have undergone significant mass-transfer (Götberg et al. 2019). In the final ULLYSES sample, approximately 12% of the LMC and SMC stars have established spectroscopic orbits.

In the next three sub-sections, we describe the target selection process for the OB stars, WR stars, and interacting binaries included in the ULLYSES sample of stars in the LMC (182 stars: 94 observed under the auspices of ULLYSES, 88 archival observations) and SMC (133 stars: 60 from ULLYSES, 73 archival). Tables B1 and B2 list astrophysical parameters for a subset of the samples for the LMC and SMC, respectively. Similar information for the full sample is available online in machine-readable form. Hertzsprung-Russell (H-R) diagrams for the LMC and SMC targets are presented in Figure 1, while distributions of targets per (spectral type, luminosity class) bin are shown in Figure 2.

### 4.1. Dwarfs, giants, and supergiants

In order to assemble a comprehensive list of OB target candidates in the LMC and SMC, we searched the literature for massive stars and their parameters — spectral type (SpT), luminosity class (LC), selective extinction E(B-V), optical photometry — using the list of catalogs provided by the SAC and STScI experts. The catalogs used for the target selection are listed in Table 5. Sources of supplementary information used to complete individual catalogs are noted in the comments of Table 5. Objects with very uncertain and/or incomplete spectral types and luminosity class and/or incomplete photometry in the B and V bands were removed from the list. Extinction was computed by taking stellar intrinsic colors from the calibrations of Martins & Plez (2006) for O-type stars and Fitzgerald (1970) for B-type stars. The different catalogs were merged after eliminating duplicate entries, which were identified by matching coordinates. We note that the stellar properties used in the target selection rely on a the assumption of single stars (no binaries), unless otherwise reported in the literature. Since the incidence of binarity can be high (up to 75%) among massive stars (Sana et al. 2009), we expect the sample to contain a significant number of undetected binaries, which could modify the true stellar properties of the sample.

Archival data from the *Mikulski Archive for Space Telescopes* (MAST, which includes, e.g., HST, FUSE, and the International Ultraviolet Explorer, IUE) and the European Space Observatory (ESO, which includes, e.g., the Very Large Telescope, VLT) were compiled and downloaded for each target candidate. A search radius of 2″ was used, corresponding to the approximate coordinate accuracy of the overall list of candidates. For each archival data set, fluxes and S/N were estimated from the data at wavelengths of 1160, 1360, 1700, or 2200 Å to cover the different instrumental modes used by ULLYSES.

---

[6] Time series observations of any massive binary system would be expensive in terms of HST orbits, and would detract from the primary goal of populating (spectral type, luminosity class) bins with representative spectra.

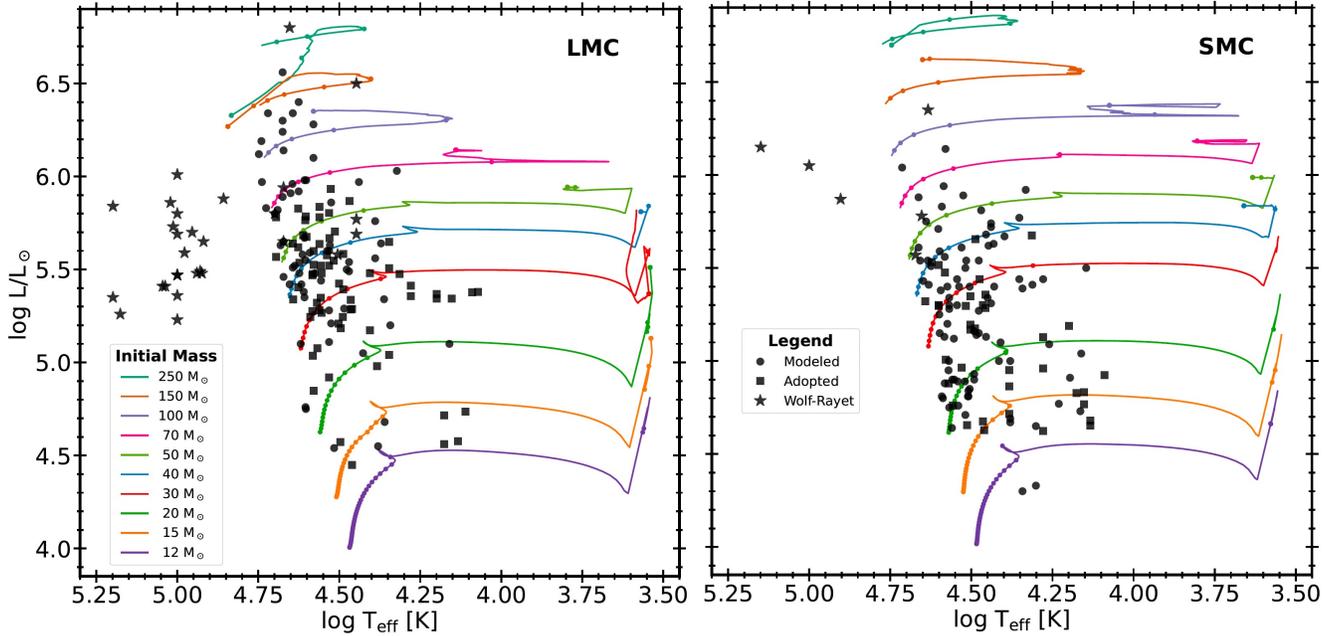

**Figure 1.** Distribution of ULLYSES targets in H-R diagrams for the LMC (left) and SMC (right). The locations of individual stars were determined from model parameters available in the literature at the outset of ULLYSES ("Modeled") or by using the spectral-type calibrations of Doran et al. (2013, "Adopted"). Single-star evolutionary tracks from the BoOST grid (Szécsi et al. 2022) computed for an initial rotational velocity of 100 km s$^{-1}$ are also shown. Small dots along the tracks indicate intervals of 0.5 Myr.

With a catalog of objects with spectral types, luminosity classes, selective extinction, photometry, and fluxes measured in archival data, preliminary exposure times were estimated for each target candidate using the UBETT, as described in Section A.1.

For the LMC, the LMCAverage extinction curve in PySynphot was applied with the E(B−V) value appropriate for each object. For stars located within the 30Dor region as traced by the hot dust seen in the Spitzer 24 μm image (within a 0.13° radius of RA = 5:38:32.086, DEC = −69:07:33.40), the LMC2 Supershell/30DorShell extinction law was applied (Gordon et al. 2003).

The models were normalized such that the maximum value of the models in a 60 Å wide spectral window centered on $\lambda_0$ is equal to the UV flux measured in archival UV spectra, when available. The maximum value of the model was used within the wide normalization window to circumvent cases where the models were depressed by a deep stellar line at $\lambda_0$. The normalization of the model spectra is performed at the wavelength $\lambda_0$ clos-est to the wavelength for which exposure times and S/N are computed (e.g., 1160 Å for COS G130M and STIS E140M, 1700 Å for G160M, 2200 Å for E230M, etc.). If no archival UV data is available, the model spectra are normalized to the available $B$ or $V$ photometry.

The full catalog of stars, their archival data, and exposure times were split into different tables, with one table per bin of SpT (binned by one temperature class) and LC (binned as "I", "II+III", "IV+V"). Each of these tables was examined in order to select a shortlist from the full pool of objects. The criteria used to create the shortlists were the following, in order of priority:

1. The total exposure time for COS G130M + G160M or STIS E140M must be <15,000s (roughly six orbits).

2. Targets in the N 11 (LMC) and NGC 346 (SMC) regions, particularly those included in the community proposal submitted by Bestenlehner et al. (see Section 4.1) are prioritized to ensure a co-eval sub-sample of targets.

3. Targets with archival HST COS (G130M, G160M) and/or STIS (E140M, E230M) data (or targets planned to be observed in Cycle 26 and 27 with modes similar to ULLYSES) are prioritized.

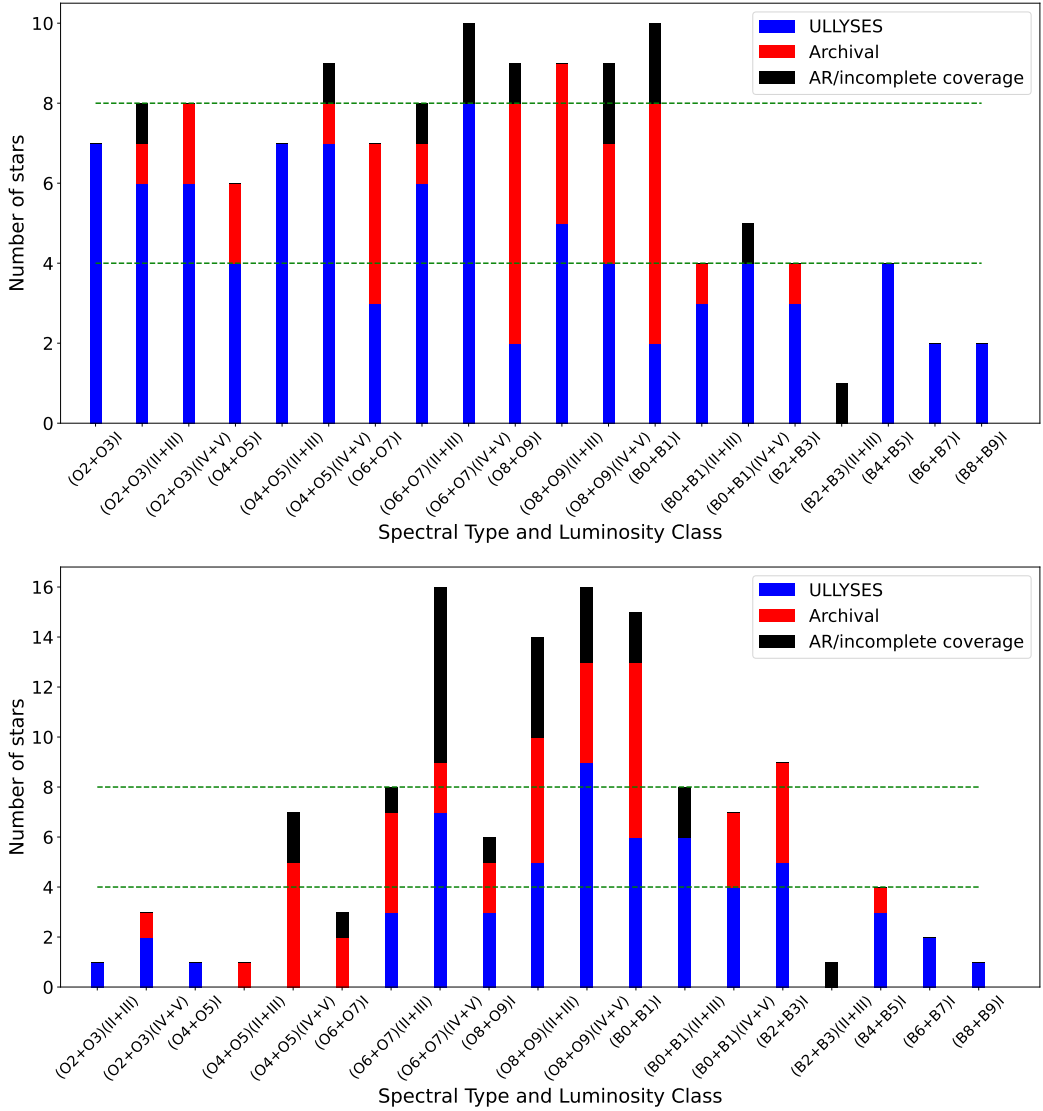

**Figure 2.** Distribution of the targets in the ULLYSES sample in (spectral type, luminosity class) bins for the LMC (top) and SMC (bottom). "AR/Incomplete coverage" denotes archival targets with data that do not necessarily conform to the S/N or wavelength requirements for new ULLYSES data, but are still relevant to the goals of the program.

4. Targets with FUSE data are preferred over otherwise equivalent targets without FUSE data.

5. If there are no other UV archival data, targets with IUE archival data are preferred.

6. Targets with VLT data (e.g., VFTS catalog) are preferred over otherwise equivalent targets.

7. Preserve stars with unusual spectral classifications (e.g., Of?p Ofnp, Be)

8. Upon examination of the fields in Aladin, using DSS, 2MASS and GALEX images, the COS or STIS aperture does not contain any bright star other than the target.

Within each (SpT, LC) bin, each target candidate was evaluated against the criteria above, which resulted in the selection of a "shortlist" of target candidates.

Once shortlists of target candidates for each bin of SpT/LC were selected, rotational velocities were populated when available, using the literature catalogs of rotational velocities listed in Table 6.

Lastly, the final selection of one to five stars per bin was performed. Following the recommendations of the UVLWG, O stars were prioritized with three to five stars in each bin, while early B stars (B0-4) were allocated two to three stars per bin and B5-9 supergiants were populated with one object per bin. First, all stars with full archival coverage in the UV were identified in each bin of SpT/LC. If a bin had the required number of stars from

**Table 4.** Parameters of the low metallicity galaxies included in the ULLYSES sample

| Galaxy | Distance[a] | [O/H][b] | [Fe/H][c] |
|--------|------------|----------|-----------|
| | Mpc | | |
| LMC | 0.05 | -0.26±0.11 | -0.22±0.08 |
| SMC | 0.062 | -0.62±0.08 | -0.69±0.08 |
| IC 1613 | 0.725 | -0.86±0.08 | -0.67±0.09 |
| NGC 3109 | 1.27 | -1.00±0.07 | −0.67±0.13 |
| WLM | 0.932 | -0.93±0.12 | -0.93±0.12 |
| Sextans A | 1.32 | -1.09±0.02 | -0.99±0.04 |
| Leo A | 0.824 | -1.35±0.08 | -1.35±0.08 |
| Leo P | 1.72 | -1.59±0.04 | ⋯ |

Note—All references below derived abundances in blue supergiants, except when noted otherwise

[a] References — LMC: Schaefer (2008); SMC: Hilditch et al. (2005); IC1613: Hatt et al. (2017); NGC 3109: Hosek et al. (2014) ; WLM: McConnachie et al. (2005) ; Sextans A: Dolphin et al. (2003) ; Leo A: Lešcinskaitė et al. (2021) ; Leo P: McQuinn et al. (2013)

[b] Assumes $12 + \log O/H = 8.76$ for the Sun (Lodders 2003). References — LMC: Tchernyshyov et al. (2015); SMC: Tchernyshyov et al. (2015); IC1613: Bresolin et al. (2007); NGC 3109: Evans et al. (2007); WLM: Bresolin et al. (2006) ; Sextans A: Kaufer et al. (2004) ; Leo A: Urbaneja et al. (2023); Leo P: Skillman et al. (2013) (H II region, direct method)

[c] Assumes $12 + \log Fe/H = 7.54$ for the Sun (Lodders 2003).References — LMC: Tchernyshyov et al. (2015); SMC: Tchernyshyov et al. (2015); IC1613: Tautvaišienè et al. (2007) ; NGC 3109: Hosek et al. (2014) ; WLM: Bresolin et al. (2006) ; Sextans A: Kaufer et al. (2004) ; Leo A: Urbaneja et al. (2023)

archival data alone, no additional stars were selected. Otherwise, we examined the distribution of temperatures, luminosity classes, and rotational velocities in the bin and attempted to sample those parameters diversely, prioritizing stars with partial UV coverage (FUSE or HST medium-high resolution in particular). Rapid rotators and special stars without UV heritage were selected, so long as their exposure times were reasonable. In all cases, the trade-offs between scientific value (particularly SpT, rotational velocity, etc.) and exposure times were evaluated, and a judgement on which target would make the most efficient and valuable use of HST time was made. We note that the sample of massive stars heavily leverages archival HST data to achieve the required coverage in wavelength and astrophysical parameter space. As a result, a significant fraction of the massive star observations cover multiple epochs combining archival and new data.

Targets in N11 and NGC 346 recommended in the community proposal submitted by Bestenlehner et al. were also prioritized during the down-selection process, so long as they were competitive with other, scientifically equivalent targets. In some cases, this was not the case and alternative targets had shorter exposure times and/or more UV archival coverage. In such cases, the stars identified by Bestenlehner et al. were rejected and replaced by more optimal, similar targets in those star-forming regions. In total, ten stars in N11 and 19 objects in NGC 346 (including archival objects) sampling a range of spectral types and luminosity classes were selected in the LMC and SMC samples, respectively. We note that five targets initially selected for observation were dropped from the sample approximately half-way through the implementation. Those targets are Sk -67 51 (LMC), BI 128 (LMC), VFTS 96 (LMC), ST92 5-52 (LMC), and AV 255 (SMC). Indeed, exposure times were recomputed during Cycle 28 with updated time dependent sensitivity calibration files and improved stellar models, resulting in an increase in the orbit cost of the program. This sample update was therefore undertaken to ensure that the observations of high priority targets slated for Cycle 29 did not exceed the orbit allocation of the program. Those specific targets were selected to be removed from the sample either because sufficient diversity of targets in their "bin" of temperature and luminosity class was already achieved, or because refined predictions of the exposure time showed that they would be quite expensive. The first "deselection" criterion was helped by our closer look at archival data sets that almost met the ULLYSES criteria in terms of wavelength coverage and S/N (see Section 4.4).

### 4.2. Wolf-Rayet stars

Compilations of WR stars were assembled from catalogs available in the literature, in particular Hainich et al. (2014) for the LMC and Hainich et al. (2015) and Shenar et al. (2016) for the SMC. As for the OB stars, lists of archival data associated with the target candidates were compiled using our software. WR stars with archival FUSE observations and bright UV fluxes were prioritized. The WR target selection optimized archival coverage with FUSE while sampling the temperature range along the WN and WC sequences. Seven LMC and four SMC WR stars were selected for observations, though additional archival WR targets and spectra were added to the sample (see Section 4.4). The final sample of WR stars (including archival data) is shown in Figure 1.

### 4.3. *Close binaries*

We selected two close binaries in both the LMC and SMC from literature studies of known or suspected interacting binary systems, LMCe055-1 and VFTS 66 in the LMC, and SMC AB6 and NGC346 ELS 013 in the SMC. Those four stars had the best observational evidence of suspected interaction.

In the LMC, LMCe055-1 (Massey et al. 2017; Smith et al. 2018) is a prime candidate for a massive star + stripped star binary with a short period of two days (Graczyk et al. 2011). This is the only known binary in the Smith et al. (2018) sample of WN3/O3 and WN4/O4 stars. VFTS066 is a rapid rotator (330 km s$^{-1}$) and short period contact binary from Mahy et al. (2020), and thus a prime candidate for an interacting binary system.

In the SMC, SMC AB6 (also known as AV 332, Sk 108) is the shortest period WR binary and part of a quintuple system (Shenar et al. 2018). According to the same study, SMC AB6 likely experienced non-conservative mass transfer in the past. NGC346 ELS 013 is a close short period binary from (Ritchie et al. 2012) experiencing mass transfer.

### 4.4. *Archival sample of massive stars in the LMC and SMC*

Additional massive stars with archival UV spectra were later identified in the LMC and SMC from systematic MAST queries. These additional datasets do not always match the ULLYSES instrumental set-ups or S/N requirements, but nevertheless provide significant scientific value to the astrophysical fields addressed by the ULLYSES program. An example would be a super-giant in the LMC and SMC with only COS G130M and/or G160M coverage, but no NUV coverage. Such stars and associated datasets were included in the ULLYSES sample (after the archival datasets were vetted for quality issues). In total, 53 additional targets were added to the LMC sample and 28 to the SMC sample. Astrophysical parameters (e.g., SpT, LC, binarity) were identified from the literature when available for all archival targets, and are listed in Tables B1 (LMC) and B2 (SMC). Those archival targets are shown as black in the histograms presented in Figure 2.

## 5. MASSIVE STARS IN VERY LOW METALLICITY GALAXIES

In this Section, we describe the process for selecting massive stars for observations in nearby galaxies with metallicities lower than the SMC. Several galaxies were considered, and stars in Sextans A and NGC 3109 were ultimately selected for observation. Additionally, purely archival targets in other lower metallicity galaxies or systems (the MC Bridge connecting the LMC and SMC, IC 1613, WLM, Leo A, and Leo P) are included in the sample. Tables for targets (both archival and observed by ULLYSES) are listed in Appendix B, in particular Tables B3 (MC Bridge), B6 (IC 1613), B7 (WLM), B8 (Leo A), and B9 (Leo P).

### 5.1. *Massive stars observed by the ULLYSES program in low metallicity galaxies NGC 3109 and Sextans-A*

Massive stars in low-metallicity dwarf galaxies beyond the LMC and SMC remain difficult to observe, and there are consequently few catalogs of OB stars with spectral types determined from optical-IR ground-based spectroscopy (though their number has grown since the ULLYSES target selection phase). Such catalogs include galaxies Sextans A (Kaufer et al. 2004; Camacho et al. 2016; Garcia et al. 2019; Lorenzo et al. 2022), NGC 3109 (Evans et al. 2007; Hosek et al. 2014), WLM (Bresolin et al. 2006; Urbaneja et al. 2008), and IC 1613 (Bresolin et al. 2007; Tautvaišienė et al. 2007; Garcia et al. 2009; Garcia & Herrero 2013; Garcia et al. 2014). Key information about nearby very low metallicity galaxies that were considered is listed in Table 4. We note that, while the UVLWG report advocated observing massive stars in Sextans B for its low metallicity, there are currently no catalogs of spectral types in this galaxy.

Based on the optical photometry and extinction reported in these catalogs, preliminary exposure times were computed with COS G140L, G130M, and G160M for all stars accessible to HST UV spectroscopy (roughly $V$-mag < 20 and spectral type earlier than B1). Stars with exposure times greater than 50 ks (about 20 orbits) were rejected. The UV archival coverage of each galaxy was then examined. IC 1613 has received significant attention in the last decade, with 15 stars observed with a combination of the G140L, G130M, and G160M gratings of COS. It was therefore decided to focus the observing effort on other galaxies. WLM has been sparsely observed (two stars observed with COS). However, other stars in this galaxy from the Bresolin et al. (2006) catalog turn out to be unreasonably expensive to observe.

This leaves Sextans A and NGC 3109, the two galaxies targeted by ULLYSES. Only six stars in Sextans A had been observed with COS prior to ULLYSES, despite additional stars in the relevant catalogs being accessible to COS. The Sextans A ULLYSES sample includes the three additional OB stars observable with the G140L/800 setting within the orbit allocation: [VPW98] 1670, [VPW98] 1805, and [VPW98] 410 (aka, s2, s4, and s8 in García et al. (2019) and s003, s004, and s071 in Lorenzo et al. (2022)).

**Table 5.** Catalogs for massive stars in the LMC and SMC used to assemble the target sample

| Galaxy | Catalog | Designation | Tables | Comments |
|---|---|---|---|---|
| LMC | Massey (2002) | M2002 | 4, 5 | |
| LMC | Evans et al. (2015a) | AAO | 2 | |
| LMC | Parker et al. (1992) | PGMW | 10, 11 | |
| LMC | Massey et al. (2000) | various | 2 | |
| LMC | | CTIO85 | 1 | |
| LMC | Evans et al. (2006a) | N11, NGC2004 | 6, 7 | |
| LMC | Blair et al. (2009) | various | 1 | Photometry partially revised from Massey (2002), Zaritsky et al. (2002, 2004) |
| LMC | Evans et al. (2011) | VFTS | 5 | Spectral types from Walborn et al. (2014, O stars) and Evans et al. (2015b, B star) |
| LMC | Crowther & Walborn (2011) | various | 3 | |
| LMC | Ramachandran et al. (2018a) | N206-FS | 3 | Photometry from Zaritsky et al. (2004) |
| LMC | Ramachandran et al. (2018b) | N206-FS | A1 | Photometry from Zaritsky et al. (2004) |
| LMC | Bonanos et al. (2009) | various | 3 | |
| LMC | Fariña et al. (2009) | N159, N160 | 1 | Photometry from Zaritsky et al. (2004) |
| SMC | Massey (2002) | M200 | 6 | |
| SMC | Evans et al. (2004b) | 2dFS | 2 | Photometry from Massey (2002), Zaritsky et al. (2002), Udalski et al. (1998) |
| SMC | Massey et al. (1989) | MPG | | Photometry partially revised from Massey (2002), Zaritsky et al. (2002) |
| SMC | Massey et al. (2000) | various | 2 | |
| SMC | Evans et al. (2006b) | NGC330, NGC346 | 4, 5 | |
| SMC | Blair et al. (2009) | various | 2 | Photometry partially revised from Massey (2002), Zaritsky et al. (2002, 2004) |
| SMC | Lamb et al. (2016) | M2002 | 1, 2 | Photometry from Massey (2002) |
| SMC | Ramachandran et al. (2019) | SMCSGS-FS | B1 | Photometry from Massey (2002), Zaritsky et al. (2002) |
| SMC | Bonanos et al. (2010) | various | 3 | Photometry from various sources |

NGC 3109 was unexplored with UV spectroscopy prior to ULLYSES. From the catalog of Evans et al. (2007), six objects in this galaxy were accessible to COS with exposure time of 15 orbits each: #7 (B0-1 Ia), #9 (B0-

1Ia), #11 B0 I, #20 (O8 I), #33 (O9 If) and #34 (O8 I(f)). Given the redundancy in spectral type, we selected one star for each bin of temperature and log g that minimized exposure time, leading to a final sample that includes [EBU2007] 7, [EBU2007] 20, and [EBU2007] 34.

WFC3/UVIS imaging in the F225W, F275W, F336W, F475W, and F814W filters was obtained in fields around those targets in order to estimate accurate UV spectroscopic exposure times (see Section A.2). The photometric measurements for the six ULLYSES low-metallicity targets, obtained using aperture photometry applied to the drizzled WFC3 images, are listed in Table 7, along with the SpT and E(B-V) derived from the WFC3 photometry. The target parameters from the original catalogs, such as spectral type, extinction, mass, for the very low metallicity galaxies are listed in Table B4 and B5 for NGC 3109 and Sextans A, respectively.

**Table 6.** Literature catalogs of rotational velocities for massive stars in the LMC and SMC

| Catalog | Galaxy:Region | O stars | B stars |
|---------|---------------|---------|---------|
| FUSE | LMC + SMC | Penny & Gies (2009) | Penny & Gies (2009) |
| VFTS | LMC: 30 Dor | Ramírez-Agudelo et al. (2013) | Dufton et al. (2013) |
| VLT-FLAMES | LMC: NGC 2004 | | Hunter et al. (2008) |
| VLT-FLAMES | LMC: N11 | Mokiem et al. (2007) | Hunter et al. (2007, 2008) |
| VLT-FLAMES | SMC: NGC 330 | Mokiem et al. (2006) | Hunter et al. (2008) |
| VLT-FLAMES | SMC: NGC 346 | Mokiem et al. (2006), Dufton et al. (2019) | Hunter et al. (2007, 2008); Dufton et al. (2019) |
| N206-FS | LMC: N206 | Ramachandran et al. (2018a,b) | Ramachandran et al. (2018b) |
| SMC SFS-FS | SMC: wing | Ramachandran et al. (2019) | Ramachandran et al. (2019) |
| MPG | SMC: NGC 346 | Bouret et al. (2003) | |
| AV | SMC | Hillier et al. (2003) | |
| Various | SMC | Bouret et al. (2013) | Bouret et al. (2013) |
| | LMC + SMC | Massey et al. (2005) | Massey et al. (2005) |
| Various | LMC + SMC | Crowther et al. (2002) | Crowther et al. (2002) |
| | LMC + SMC | Rivero González et al. (2012) | |

## 5.2. *Archival sample of massive stars in Sextans A, IC 1613, WLM, Leo P, and Leo A*

Additional massive stars with available archival UV spectra were later identified in IC 1613, WLM, Sextans A, Leo A (Z = 0.04 Z$_\odot$, Urbaneja et al. 2023) and Leo P (Z = 0.03 Z$_\odot$, Skillman et al. 2013), from systematic MAST queries in those galaxies. The additional datasets did not strictly match the ULLYSES instrumental set-ups, but do provide significant scientific value to the astrophysical fields addressed by the ULLYSES program. An example would be a massive star in Sextans A with only G130M archival spectra. Such stars and associated datasets were included in the ULLYSES sample (after the archival datasets were vetted for quality issues). In total, nine archival massive stars were added to the Sextans A sample. In addition, archival targets were added to the ULLYSES sample in galaxies not covered by ULLYSES observations. These low-metallicity galaxies increase the metallicity coverage of the program. Thus, two massive stars in WLM, 11 stars in IC 1613, three stars in Leo A, and one star in Leo P were added to the ULLYSES sample. These targets are covered by either COS or STIS, with low- or medium-resolution gratings. The distance and metallicity (in O and Fe) of the full sample of the ULLYSES very low metallicity galaxies are listed in Table 4. Astrophysical parameters (e.g., SpT, LC, binarity) for archival very low metallicity stars were identified from the literature when available, and are listed in Tables B6 (IC 1613), B7 (WLM), B8 (Leo A), and B9 (Leo P).

## 6. TARGET SELECTION FOR T TAURI STARS

The UVLWG recommended populating a grid of stellar mass and accretion rates with medium-resolution UV and low-resolution NUV-optical-NIR spectroscopy of young low-mass stars in nearby star-forming regions of the Milky Way, specifically targeting accreting stars with masses below 0.5 M$_\odot$.

In the next two sub-sections, we detail the target selection process for single-epoch "survey" T Tauri stars observed by the ULLYSES program (56 stars), as well as the 80 archival T Tauri stars included in the final sample, for a total of 156 survey T Tauri stars.

A small sub-set of the ULLYSES sample of T Tauri stars is listed in Tables B10, while the full sample is available online in machine-readable form. The corresponding sampling of the stellar mass and accretion rate is shown in Figure 3.

## 6.1. *Single-epoch T Tauri stars observed by ULLYSES*

In order to compile a list of TTS candidate targets, we queried a list of possible targets and their parameters (SpT, stellar mass, accretion rate, photometry) from the literature catalogs listed in Table 8. Objects

**Table 7.** WFC3 photometry and SpT and E(B-V) derived from the WFC3 photometry for the ULLYSES targets in Sextans A and NGC 3109

| Star | RA(J2000) | DEC(J2000) | F225W | F275W | F336W | F475W | F814W | SpT$_{\mathrm{WFC3}}$ | E(B-V)$_{\mathrm{WFC3}}$ |
|------|-----------|------------|-------|-------|-------|-------|-------|------|---------|
|      |           |            | mag   | mag   | mag   | mag   | mag   |      | mag     |
| NGC3109 EBU 07 | 10h02m54.69s | -26d08m59.64s | 16.25 | 16.62 | 17.23 | 18.27 | 20.12 | B0 I | 0.065 |
| NGC3109 EBU 20 | 10h03m03.22s | -26d09m21.41s | 17.04 | 17.35 | 17.90 | 18.91 | 20.67 | O8 I | 0.13 |
| NGC3109 EBU 34 | 10h03m14.24s | -26d09m16.96s | 16.99 | 17.40 | 18.03 | 19.16 | 21.07 | O8 I | 0.065 |
| Sextans A LGN s004 | 10h10m57.89s | -04d43m10.2s | 18.05 | 18.45 | 19.17 | 20.37 | 22.38 | O6 V | 0.045 |
| Sextans A LGN s003 | 10h10m58.59s | -04d43m28.9s | 18.49 | 18.84 | 19.41 | 20.47 | 20.08 | O5 V | 0.13 |
| Sextans A LGN s071 | 10h11m05.69s | -04d42m13.6s | 17.38 | 17.71 | 18.30 | 19.28 | 21.16 | B0 I | 0.077 |

NOTE—Magnitudes reported for WFC3 filters are STmag magnitudes.

without accretion rates or stellar masses were removed from the compilation. We queried the MAST archive for each target candidate using a search radius of 5", which captures coordinate uncertainties and proper motions. Target candidates with the appropriate UV-optical-NIR coverage from HST were identified.

Exposure times were then computed for each target candidate with the UBETT, using the approach described in Appendix A, Section A.3. To account for a possible underestimation of interstellar extinction $A_V$ and conservatively ensure that the desired S/N is reached, the $A_V$ value reported in the literature for each candidate target was padded by an extra 0.5 mag. This value corresponds to the level of underestimation we empirically observed by comparing archival UV spectra to models derived from the reported extinction and accretion rate estimated from VLT *X-Shooter* spectra and/or optical photometry (see Section A.3).

Based on exposure time estimates for each target candidate, the number of orbits required to observe each target was estimated, accounting for guide star and target acquisition, buffer dumps, and simulating orbit packing by allowing a 20% cut in exposure time (10% in S/N).

Targets that were unreasonably expensive to observe (more than 15 orbits) were removed from the list of potential candidates. The list of candidates was then sorted by ascending stellar mass. Within each 0.1 $M_\odot$ interval of stellar mass, we selected the T Tauri stars with a range of measured accretion rates that could be observed in the shortest exposure time. For example, we identified stars with similar masses and accretion rates, and selected the most efficient one to observe. We also kept track of the $V$-band photometry to ensure that the selected stars could be monitored from the ground at a reasonable cost. Stars with full UV coverage in the HST archive were flagged to be included in the ULLYSES database, but not re-observed.

We received a compelling proposal from the community by Thanathibodee et al. Their proposal was the most comprehensive, with 45 stars sampling different star-forming regions, accretion rates, and stellar masses, in line with the recommendations from the UVLWG (see references for source properties in Table 8). To the extent that their proposed targets had reasonable exposure times and published accretion rates and stellar masses, those objects were prioritized in the sample.

Lastly, the SAC recommended that non-accreting stars be included as a reference sample, which broadly covers the HR diagram. Several weak-line T Tauri stars (WTTS) already had HST UV-optical coverage in the archive (see Table 9), satisfying this requirement for stars of most spectral types. However, a gap in the HR diagram coverage of WTTS templates was identified, specifically for spectral types K2, M3, and M4 and later. We identified star RX J0438.6+1546 (Manara et al. 2017a) as a good K2 WTTS template, being very efficient to observe in the UV. The only M3 star that was observable in a reasonable exposure time while not violating the bright object protection policy for flaring M stars is RECX 6 (Rugel et al. 2018). No WTTS templates of spectral type M4 and later satisfying those requirements could be identified.

The last step in the TTS target selection process was to ensure that none of the candidates violate the bright object rules for COS (G130M, G160M) and STIS (G230L). The BOP clearing of TTS is described in Appendix A, Section A.4 and applied to each target. All targets cleared with STIS G230L. However, a few target candidates did not pass the BOP check for flaring M dwarfs with COS G130M/1291 (Si IV) and COS G160M (C IV): HD-104237E (M dwarf in the field), RECX 5, RECX 6, and RECX 9. Additionally, the target acquisition strategy that cleared the BOP for flaring M dwarfs for Sz115 led to unreasonable acquisition times, and no offset star was available nearby. These stars were therefore observed with STIS G140L and COS G130M/1222 (except for RECX 6) instead of COS G130M/1291 and COS G160M/1589+1623.

**Table 8.** Literature catalogs of T Tauri stars

| SF Region | Catalog |
| --- | --- |
| Lupus | Alcalá et al. (2014) |
| | Alcalá et al. (2017) |
| Chamaeleon I | Manara et al. (2017b) |
| | Manara et al. (2017a) |
| Ori OB1a, b | Calvet et al. (2005) |
| | Briceño et al. (2005) |
| | Briceño et al. (2019) |
| $\eta$ Cha | Rugel et al. (2018) |
| $\epsilon$ Cha | Fang et al. (2013) |
| $\sigma$ Ori | Hernández et al. (2014) |
| | Maucó et al. (2016) |
| CrA | Manara et al. (2014) |

**Table 9.** WTTS templates in the ULLYSES sample

| Star | SpT | Reference |
| --- | --- | --- |
| LkCa19 | K0 | Ingleby et al. (2013) |
| RX J0438.6+1546 | K2 | Manara et al. (2017a) |
| RECX 1 | K5 | Ingleby et al. (2013) |
| HBC 427 | K7 | Ingleby et al. (2013) |
| TWA 7 | M1 | Ingleby et al. (2013) |
| RECX 6 | M3 | Rugel et al. (2018) |

The resulting TTS sample of 56 targets, which includes star-forming regions Ori OB1, $\sigma$ Ori, Lupus, Chamaeleon I, $\eta$ Cha, $\epsilon$ Cha, Taurus, and CrA, is listed in Table B10. The corresponding sampling of the stellar mass/accretion rate parameter space is shown in Figure 3.

### 6.2. Archival sample of T Tauri stars

Additional T Tauri stars with available archival UV-visible-IR spectra were later identified in several star-forming regions from systematic MAST queries in Milky Way star-forming regions. Those include star-forming regions not observed by the ULLYSES program, such as Taurus, $\lambda$ Orionis, or Upper Scorpius. The additional datasets did not necessarily strictly match the ULLYSES instrumental set-ups (i.e., COS G130M + G160M, STIS G230L, STIS G430L, STIS G750L), but do provide significant science value to the astrophysical fields addressed by the ULLYSES program. An example would be a T Tauri star observed by STIS G140L and G230L only (with no optical or IR coverage). Such stars and associated datasets were included in the ULLYSES sample (after the archival datasets were vetted for quality issues). In total, 80 additional targets were added to the TTS sample. All targets, both archival and observed by ULLYSES, are listed in Table B10.

### 6.3. T Tauri stars monitored over time with HST

The single epoch observations of TTS provide insight into the dependence of the accretion process and rate on stellar parameters and age, with each observation capturing a snapshot of the state of a given star at a given time. Because TTS are variable owing to the stochasticity of the accretion process and stellar rotation, the UV Legacy Working group also recommended a monitoring campaign of four well-studied prototypical TTS to improve our understanding of accretion variability over timescales from minutes to months or years.

The criteria established to select the four prototypical T Tauri stars to be monitored over time with HST were the following.

1. Candidate objects had to have been previously observed with HST UV spectroscopy. This would ensure that the UV flux was well known, ensuring that the stars were bright enough to be observed with COS G160M and COS G230L in one orbit.

2. The magnetic field of the candidate objects (Johns-Krull 2007) had to be previously mapped (e.g., Donati et al. 2008, 2011a,b,c) or have planned mapping (Sousa et al. 2023; Bouvier et al. 2023) using spectro-polarimetry . A community proposal by C. Argiroffi recommended to include BP Tau in the monitoring TTS sample as one such objects.

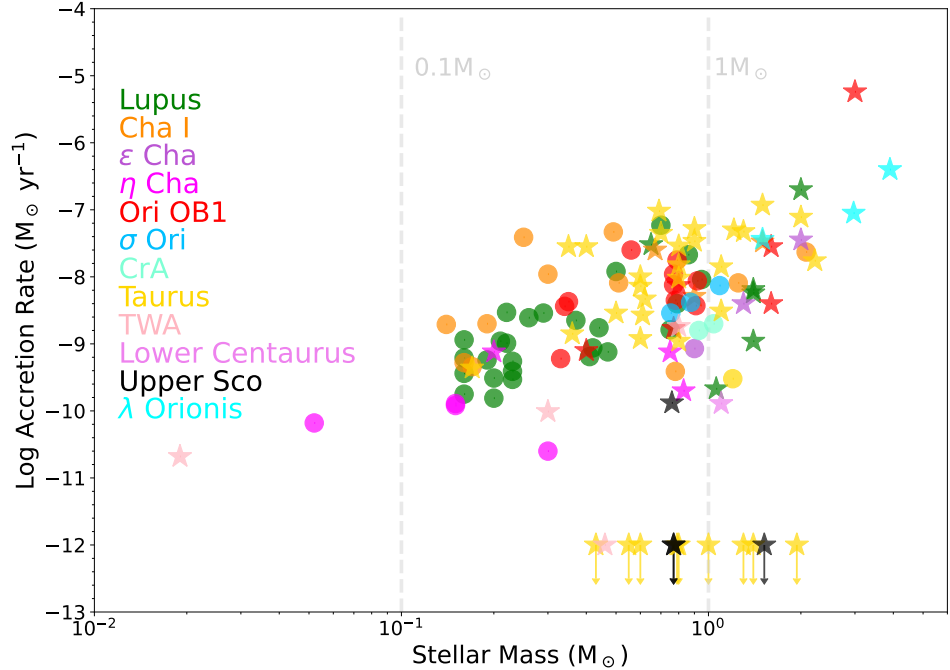

**Figure 3.** Accretion rate as a function of stellar mass in the ULLYSES TTS sample, including stars newly observed by ULLYSES (circles), which cover the first eight star-forming regions listed in the legend, and purely archival targets (stars), which probe four more star-forming regions.

3. The stars' rotational period and disk inclination (e.g., Nguyen et al. 2012; Appenzeller & Bertout 2013) should be known.

4. The objects should be accessible to telescopes in the Southern hemisphere (in particular ALMA and the VLT).

5. Objects in the TESS footprint during the epochs of monitoring (2021 and 2022) should be prioritized, as TESS provides highly complementary photometric monitoring at 10 min cadence in the *i* band.

6. Targets should be good candidates for X-ray monitoring, and objects from the C. Schneider et al. community proposal to coordinate XMM-Newton and HST monitoring should be prioritized as a result.

Based on these criteria, TW Hya, BP Tau, GM Aur, and RU Lup were selected (see Table 10). They have bright C IV fluxes, making them observable in the FUV and NUV in one orbit, and are observable from telescopes in the Southern hemisphere. All four stars were included in the C. Schneider et al. proposal.

## 7. COORDINATED PROGRAMS

The scientific value of ULLYSES is enhanced by numerous coordinated and ancillary programs with ground- and space-based observatories.

The full ULLYSES LMC/SMC sample (including the additional archival targets added later in the implementation of the program) was observed with ESO's VLT/X-Shooter in the optical-IR throughout 2020-2024 as part of the X-ShootU program (Vink et al. 2023). The combination of the UV, optical, and IR enables the comprehensive modeling of the photospheres and winds of the massive stars (Vink & Sander 2021). The X-ShootU dataset (Sana et al. 2024) is already resulting in updates in the spectral types of ULLYSES stars plotted in Figures 1 and 2(Bestenlehner et al. 2025). In addition, SNAP program 16230 (PI Massa) followed up a significant fraction of LMC and SMC massive stars with the STIS CCD in the NUV and visible, thus providing the photometric accuracy required to derive atmospheric parameters and interstellar dust extinction curves toward these sightlines. High-level science data products from SNAP-16230 and the X-ShootU programs are delivered through the ULLYSES collection at MAST.

For the low-mass stars, numerous community-led programs were coordinated with the HST observations to execute simultaneously (or nearly so). VLT/X-Shooter, ESPRESSO, and UVES observations simultaneous with HST obtained as part of the PENELLOPE program[7] enable the derivation of accurate accretion rates, masses, and extinction (Manara et al. 2021). X-Shooter HLSPs from PENELLOPE will be hosted by the ULLYSES database[8] as they become publicly available so as to ensure a maximum impact for the program. Observations with NASA's Infra-Red Telescope Facility (IRTF) were carried out in order to calibrate mid-infrared (MIR) diagnostics of accretion with contemporaneous UV observations with HST, which will enable MIR diagnostics (e.g., Br-$\alpha$) of accretion for the deeply embedded protostars that are being observed with JWST (Fischer et al., in prep).

**Table 10.** Sample of four prototypical T Tauri stars monitored with HST

| Star | RA(J2000) | DEC(J2000) | SpT [a] | V | $A_V$ [b] | $M_*$ [c] | log(dm/dt) [d] | Rotation period [e] |
|------|-----------|------------|---------|------|-----------|-----------|----------------|---------------------|
| | | | | mag | mag | $M_\odot$ | $M_\odot$ yr$^{-1}$ | days |
| BP Tau | 04h19m15.86s | +29d06m27.2s | K7 | 12.12 | 0.51 | 0.7 | -7.54 | 8.19 |
| GM Aur | 04h55m10.98s | +30d21m59.1s | K3 | 13.1 | 0.6 | 1.36 | -8.3 | 6.1 |
| TW Hya | 11h01m51.95s | -34d42m17.7s | K7 | 10.5 | 0.0 | 0.7 | -8.7 | 3.57 |
| RU Lup | 15h56m42.31s | -37d49m15.47s | K7 | 9.6 | 0.07 | 0.7 | -7.3 | 3.71 |

[a] References for SpT — BP Tau: Furlan et al. (2011) ; GM Aur: Furlan et al. (2011) ; TW Hya: Webb et al. (1999) ; RU Lup: Stock et al. (2022)

[b] References for $A_V$ — BP Tau: Furlan et al. (2011) ; GM Aur:Furlan et al. (2011) ; TW Hya: Webb et al. (1999) ; RU Lup: France et al. (2017)

[c] References for $M_*$ — BP Tau: Ingleby et al. (2013) ; GM Aur: Ingleby et al. (2013) ; TW Hya: Ingleby et al. (2013) ; RU Lup: Stock et al. (2022)

[d] References for accretion rate — BP Tau: Ingleby et al. (2013) ; GM Aur: Ingleby et al. (2013) ; TW Hya: Ingleby et al. (2013) ; RU Lup: France et al. (2017)

[e] References for rotation period — BP Tau: Percy et al. (2006) ; GM Aur: Percy et al. (2010) ; TW Hya: Huélamo et al. (2008); RU Lup: Stempels et al. (2007)

In addition, a photometric monitoring campaign with the LCO global telescope network led by STScI provides the long-term variability context for each star (see Section 2.5). This photometric monitoring is complemented by *TESS* (Transiting Exoplanet Survey Satellite) observations for 23 survey T Tauri stars and three monitoring T Tauri stars for which TESS+HST coordination was possible. The TESS observations provide month-long high cadence (15 min) monitoring of those objects in what is essentially a very wide *i*-band filter. The HOYS project also coordinated a program with amateur astronomers to monitor the survey targets (Froebrich et al. 2022).

In addition to this suite of coordinated observations, for the four stars monitored over time with HST, simultaneous observations with XMM-Newton or Chandra provide the full suite of X-ray and UV accretion shock diagnostics (Schneider et al., in prep, Guenther et al., in prep). The magnetic field structure of those four T Tauri stars was mapped using spectro-polarimetry with CFHT/SPIRou (Sousa et al. 2023; Bouvier et al. 2023; Donati et al. 2024). The ULLYSES program has helped to motivate temporal analysis that combines these datasets with archival spectra and photometry (e.g. Claes et al. 2022; Herczeg et al. 2023)

## 8. INITIAL RESULTS

The observations, data reduction, calibration, and creation of high-level science data products (HLSPs) for ULLYSES will be presented in an upcoming paper (Paper II). All data products are available from the ULLYSES search form [9] (doi:10.17909/t9-jzeh-xy14) and detailed information about the high-level science data products is available at the ULLYSES website[10].

An example of the array of analysis of accretion and disk diagnostics from ULLYSES data is presented by

**Table 11.** Cycle 27-31 programs coordinated with or complementary to ULLYSES

| Observatory | Cycle | Program ID | PI | Program Title |
|---|---|---|---|---|
| HST | 27 | GO-15967 | J. Chisholm | Constraining the Stellar Astrophysics Powering Cosmic Re-ionization: Spectral Templates of Extremely Low-metallicity Main-sequence O-stars |
| HST | 27 | GO-15891, GO-16235, GO-16786 | C. Murray | Scylla: A pure-parallel, multi-headed attack on dust evolution and star formation in ULLYSES galaxies |
| HST | 28 | GO-16233 | C. Schneider | Jets and disk scattering Äì Spatially resolved optical and FUV observations of AA Tau |
| HST | 28 | SNAP-16230 | D. Massa | A NUV SNAP program to supplement and enhance the value of the ULLYSES OB star legacy data |
| HST | 28 | AR-16148 | P. Senchyna | Painting the first empirical picture of massive stars below the metallicity of the SMC with ULLYSES |
| HST | 28 | AR-16129 | G. Herczeg | Outflows and Disks around Young Stars: Synergies for the Exploration of ULLYSES Spectra (ODYSSEUS) |
| HST | 28 | AR-16131 | D. Hillier | CMFGEN: A key spectroscopic tool for astrophysics |
| HST | 28 | AR-16133 | E. Jenkins | A comprehensive investigation of Gas-phase element abundances and extinction by dust in the LMC and SMC |
| HST | 29 | AR-16616 | J. Howk | Interstellar tomography of highly ionized gas in the MW thick disk with ULLYSES |
| HST | 29 | AR-16623 | C. Leitherer | Feasting on the Riches of Odysseus' voyage |
| HST | 29 | AR-16640 | Y. Zheng | Braving the storm, quantifying the effects of Ram Pressure and Stellar Feedback in the LMC |
| HST | 29 | AR-16602 | K. Barger | The LMC's Galactic Wind through the eye of ULLYSES |
| HST | 29 | AR-16635 | K. Tchernyshyov | The first direct measurement of CO/H2 in subsolar environments using ULLYSES data |
| HST | 30 | AR-17051 | N. Lehner | A ULLYSES Survey of the Magellanic Clouds: a Laboratory for the Physics of Interfaces between Hot and Cold Gas |
| HST | 30 | GO-17111 | M. Garcia | The winds of massive stars at the peak of the star formation history of the Universe |
| HST | 31 | GO-17491 | G. Telford | A Legacy Far-Ultraviolet Spectral Atlas of Extremely Metal-Poor O Stars |
| Chandra | 22 | 22200086 | H. Gunther | The power of space: Simultaneous X-ray and UV monitoring if an accretion low-mass star |
| XMM-Newton | AO-20 | 088206 | C. Schneider | HERA: High-Energy Radiation from Accretion in young stars |
| ESO | Period 106 | 106.20Z8 | C. Manara | PENELLOPE: the ESO data legacy program to complete the Hubble UV Legacy Library of Young Stars (ULLYSES) |
| ESO | Period 106 | 106.211Z | J. S. Vink | X-Shooting ULLYSES: The Physics of Massive Stars at Low Metallicity |
| Las Campanas | 2021b | | N. I. Morrell | Magellan/MIKE spectroscopy of ULLYSES slow rotators |
| IRTF | 2020B–2022A | 2020B098, 2021A073, 2021B080, 2022A078, 2022B059 | W. Fischer | LAERTES: L-Band Accretion Estimator Reconnaissance of TTS Emission Spectra |
| LCO | 2020B–2023A | DDT2020B-001, DDT2021A-001 | W. Fischer | Photometric monitoring of ULLYSES T Tauri stars |

Espaillat et al. (2022). For early results related to disks and winds, we refer readers to Pittman et al. (2022), France et al. (2023), Gangi et al. (2023), Arulanantham et al. (2023), Wendeborn et al. (2024a), and Wendeborn et al. (2024b).

Early results on the structure and terminal velocity of massive star winds are presented in Parsons et al. (2024b) and Hawcroft et al. (2023), while surface abundances of O stars are covered in Martins et al. (2024). A comparison between population synthesis models and template built on ULLYSES massive star spectra is presented in Crowther & Castro (2024).

In the following section, we report here on some initial results from the ULLYSES observations of low- and high-mass stars. The goal of this section is not to provide a review of community-led results from ULLYSES, but rather to report on immediate results that arose directly from the data.

### 8.1. A comprehensive library of UV-visible-NIR spectra of T Tauri stars

The sample of TTS observed by ULLYSES and combined with archival spectra covers the range of stellar mass and accretion rate recommended by the UVLWG. In Figure 4, we show a subset of UV-visible-NIR spectra of a few TTS covering the full range of stellar masses and accretion rates, highlighting the extent of the wavelength coverage and spectral quality of the ULLYSES data products, as well as the relative intensities of emission lines for different stellar masses and accretion rates. The combined UV-optical spectra allow for accurate measurements of accretion rate (Pittman et al. 2022) and provide accretion diagnostics across a wide range of temperatures (Armeni et al. 2023). Updated stellar and accretion parameters are being measured from fits to the ULLYESES and PENELLOPE spectra (Manara et al. 2021; Frasca et al. 2021; Pittman et al. 2022) and will be provided to the community.

Anecdotally, the STIS observations revealed sub-arcsecond companions for two of the Ori OB1 T Tauri stars, CVSO 109 and CVSO 165. The discovery was reported in Proffitt et al. (2021). The companions do not appear to have been documented in the literature in advance of the ULLYSES observations, although the CVSO 109 companion was subsequently listed in the Gaia DR3 release (Gaia Collaboration et al. 2016, 2021). The STIS spectra of the CVSO 109 companion suggest that the companion is much less active than the T Tauri primary, exhibiting a lack of Hα emission, and weak Mg II emission. The companion likely does not contribute any significant flux to the COS FUV spec-tra. In contrast, both of the close components of CVSO 165 show multiple emission lines. The secondary shows strong emission in multiple Balmer lines and has a significant UV upturn. Conversely, only the Hα Balmer emission line appears in the spectrum of the primary, and the NUV continuum is actually fainter than in the secondary.

### 8.2. Accretion variability and UV diagnostics in T Tauri stars

The monitoring observations of the four prototypical T Tauri stars (RU Lup, BP Tau, TW Hya, and GM Aur) provides insight into the timescales and magnitudes of accretion variability, from minutes to years. A detailed analysis of accretion variability in the four TTS monitored by ULLYSES is presented in Wendeborn et al. (2024a) and Wendeborn et al. (2024b). Here, we present initial findings resulting directly from the monitoring observations.

Figure 5 shows the flux variations of BP Tau in three UV accretion diagnostics (C IV λλ1548, 1550, He II λ1641, and Mg II λλ2796, 2803) as a function of time on timescales of days for each of the two epochs taken about one year apart. The time variations in Figure 5 are produced from the "tss" time-series products, in which one time sample corresponds to one exposure. Twelve observations sampling three rotational periods were taken during each epoch, and each observation includes four exposures (one per FP-POS). Figure 6 shows the flux variations of BP Tau in the same three UV diagnostics, but as a function of time on timescales of minutes. The time variations in Figure 6 are produced from the "split-tss" time-series products generated at the sub-exposure level, in which one time sample corresponds to about 30s, and only include the fifth observation of each epoch.

The flux variations in all three UV diagnostics on minute-timescales appear to be minimal, while variations over days- or even year-timescales can reach a factor of several. On days-timescales, the line profiles also exhibit significant variations, in particular for C IV. These changes may indicate variations in the kinematics of accretion and/or transient extinction of certain regions near the accreting star.

Figure 7 shows the peak line or continuum flux in various tracers (C IV λλ1548, 1550, He II λ 1641, Mg II λλ 2796, 2803, NUV continuum at 2000 and 2800 Å, u', V, i) as a function of time for BP Tau and RU Lup. In both stars, the flux variations in different diagnostics track each other well, possibly indicating a common physical origin. The relative strengths of the different tracers are strikingly different between BP Tau and RU Lup. In RU Lup, Mg II is by far the brightest diagnostic, with C IV over an order of magnitude fainter. Conversely in BP Tau (as well as GM Aur and TW Hya), C IV, He II, Mg II are the brightest diagnostics (and have comparable fluxes), followed by $V$, $i'$, the NUV continuum, and $u'$. These differences are potentially related to inclination and magnetic field geometry.

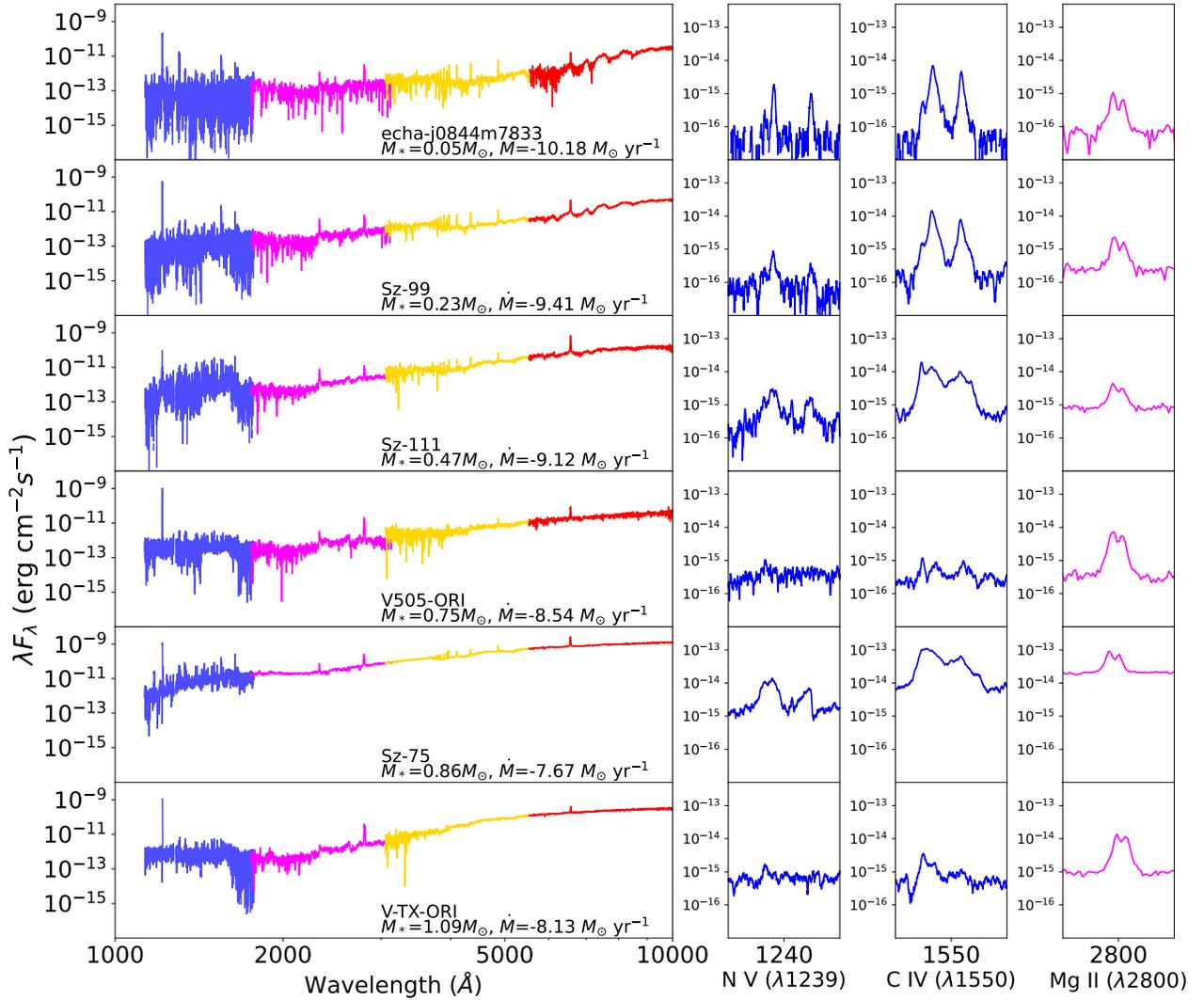

**Figure 4.** COS (G130M + G160M, blue), STIS G230L (magenta), G430L (yellow) and G750L (red) spectra of a few select TTS of increasing stellar mass, from top to bottom. The three right columns highlight some key diagnostics of accretion, namely N V $\lambda\lambda1238,1242$, C IV $\lambda\lambda1548, 1550$, and Mg II $\lambda\lambda2796, 2803$

In addition to the four stars monitored by the ULL-YSES program, accretion variability over timescales of many years can be probed by the combination of UV and optical spectroscopy acquired over different epochs.

In order to build the TTS target sample and estimate HST COS and STIS exposure times, we relied on previous accretion rate estimates obtained from optical spectroscopy, most of which was obtained with VLT/X-Shooter (Alcalá et al. 2014; Manara et al. 2014; Alcalá et al. 2017; Manara et al. 2017b,a), and photometry (for Orion, see Calvet et al. 2005). Indeed, as explained in Section A.3, the correlation between accretion rates and peak UV line fluxes observed in three templates (DN Tau, DR Tau, and V836 Tau) for which simultaneous COS FUV and STIS NUV-optical spectra were available (Ingleby et al. 2013) was used to predict the peak UV line fluxes (and exposure times) of the ULLYSES TTS sample based on the X-Shooter spectroscopy or optical photometry.

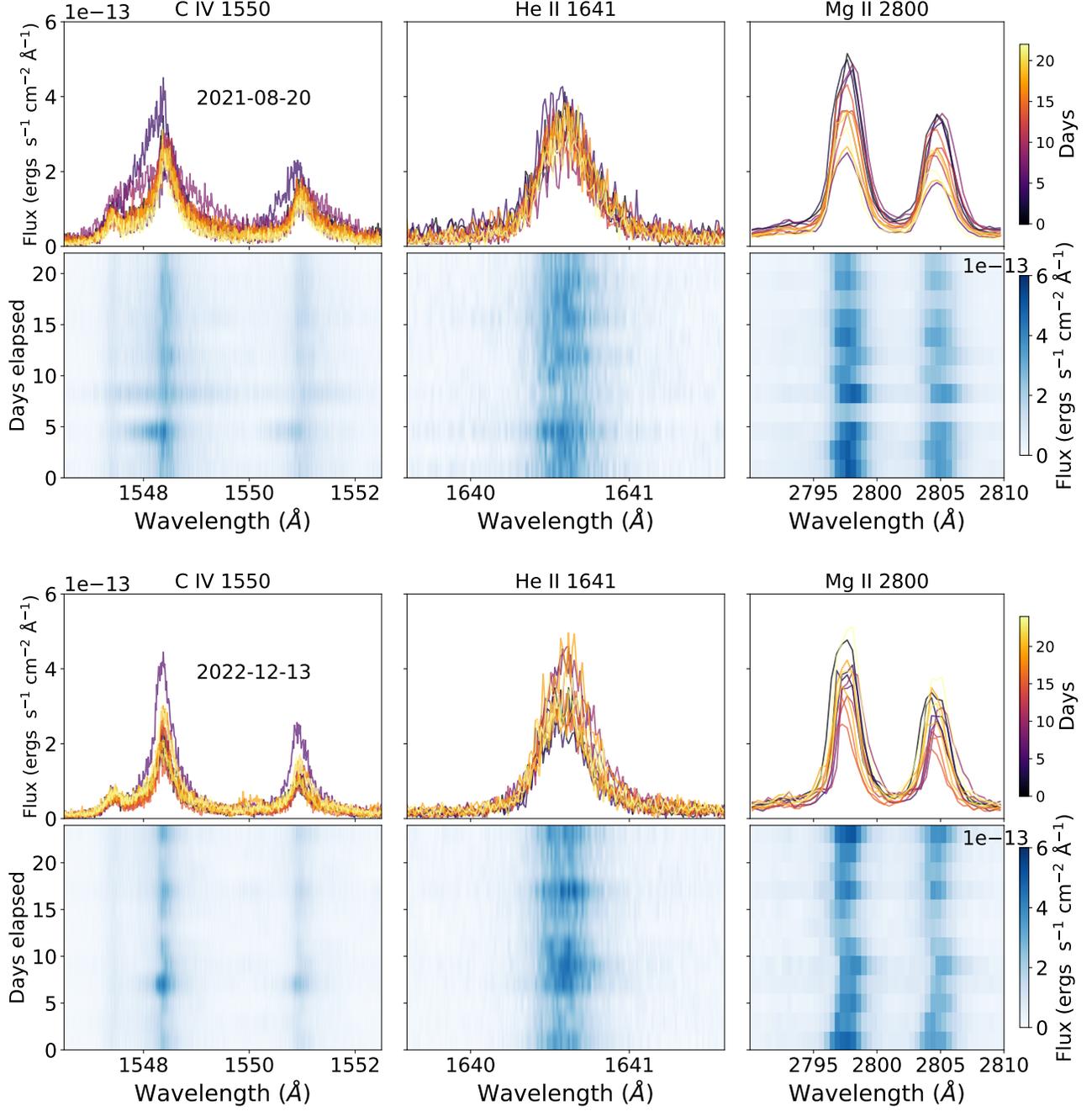

**Figure 5.** Spectral time series of BP Tau around the C IV (left), He II (middle) and Mg II (right) lines. The top two rows correspond to the first epoch of observations (August 2021), while the bottom two rows were obtained in late 2022. The spectral time series are displayed as line plots showing flux variations color coded by days elapsed (top panels), and as images showing the line profiles variations with time (color coded by flux level).

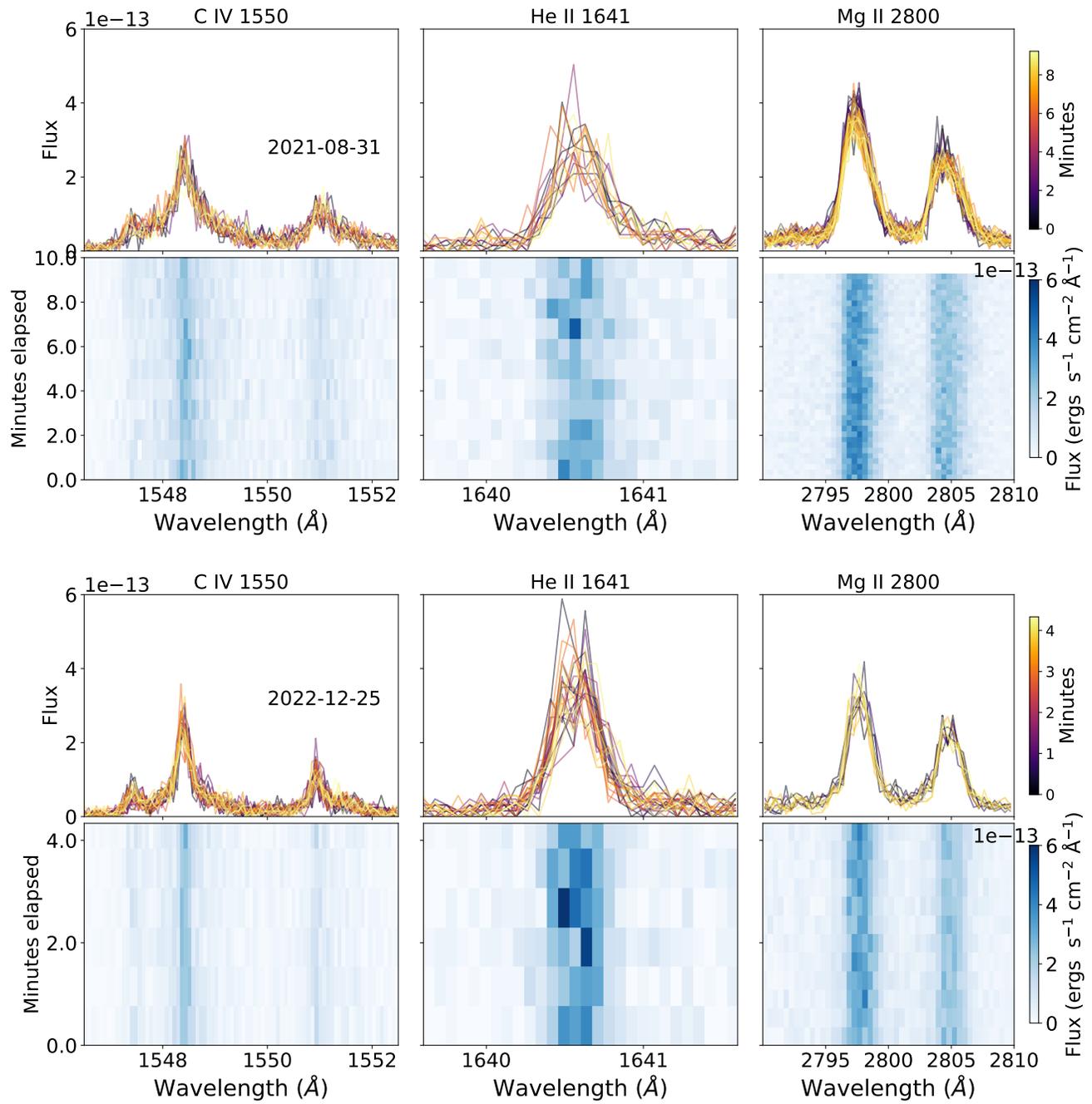

**Figure 6.** Similar to Figure 5, but showing variations on minute timescales during the 5th (out of 12) observations of Epoch 1 (top two rows, August 2021) and Epoch 2 (bottom two rows, December 2022).

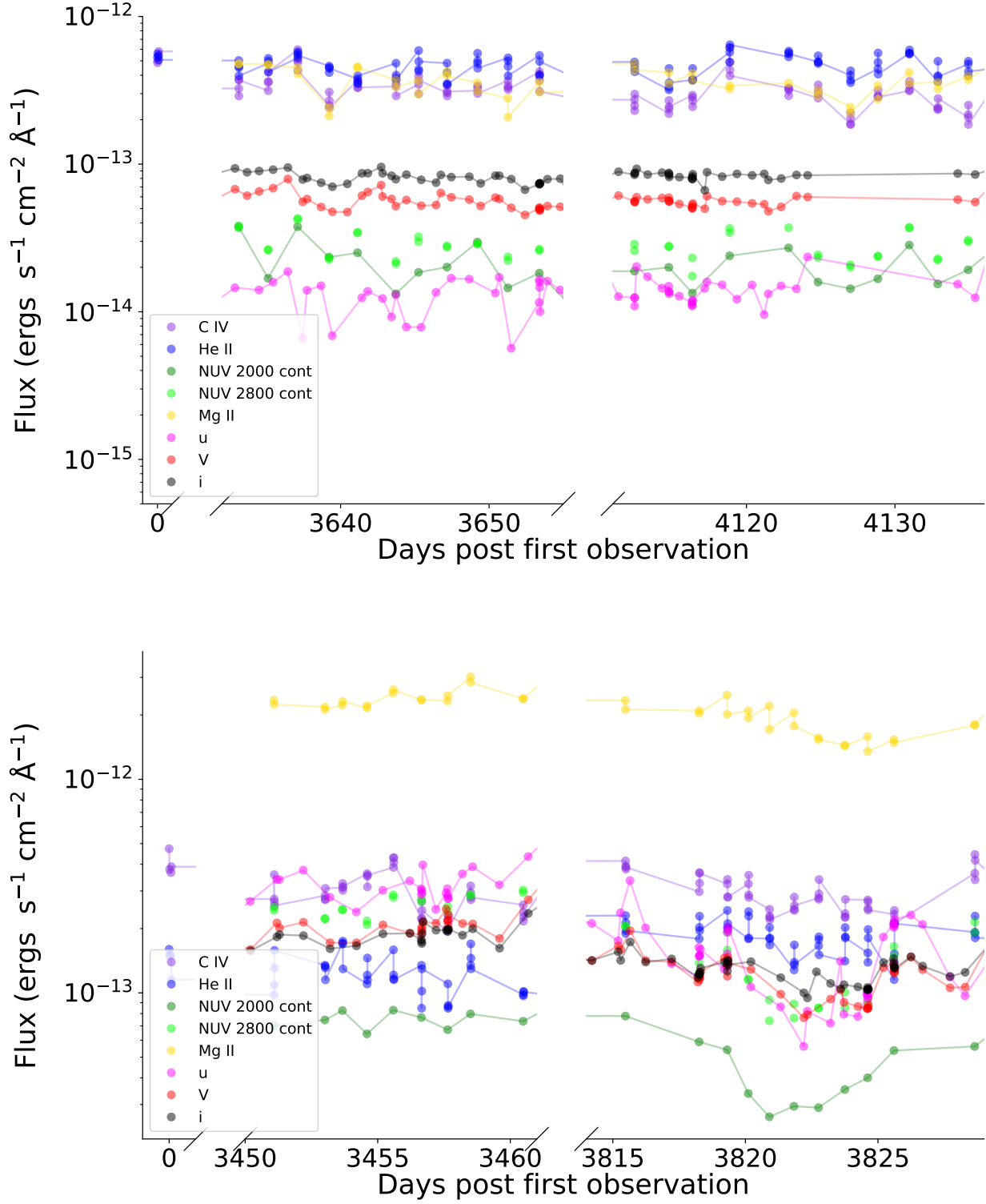

**Figure 7.** Peak fluxes as a function of time for different diagnostics (C IV, He II, NUV continuum at 2000 and 2800 Å, u', V, i') for T Tauri stars BP Tau (top) and RU Lup (bottom).

Surprisingly, the actual, observed peak UV line fluxes of many TTS observed with ULLYSES turned out significantly lower than the predictions based on prior optical spectroscopy, as shown in Figure 8 for the C IV λλ 1548, 1550 doublet. Correspondingly, for many TTS with weaker UV lines than expected from prior optical spectroscopy, the flux level of VLT/X-Shooter spectra taken concurrently with the ULLYSES COS and STIS spectra by the PENELLOPE collaboration (Manara et al. 2021) also often show a significant decrease between the prior epoch of the optical spectroscopy (Alcalá et al. 2017) and the epoch at which the ULLYSES spectra were taken. This effect is illustrated in Figure 9 for TTS Sz 98, for which the first epoch of optical spectroscopy was taken in 2015 and the ULLYSES and PENELLOPE spectra were taken in 2022.

However, on average, the peak UV line fluxes observed by ULLYSES are significantly lower than the predictions from prior optical spectroscopy or photometry, which would not be expected from stochastic variations in accretion rate. While the ratio of actual to predicted C IV peak fluxes does not exhibit any obvious trend with stellar mass, it does show a decreasing trend with increasing accretion rate, as reported in the literature from X-Shooter spectroscopy and optical photometry. This trend is shown in the bottom panel of Figure 8. This suggests a systematic difference between, on the one hand, accretion modeled or derived from STIS NUV-optical spectroscopy in the three templates used to predict UV fluxes (DN Tau, DR Tau, and V836 Tau; Ingleby et al. 2013), and, on the other hand, accretion rates modeled or derived from the X-Shooter optical spectroscopy and optical photometry (Calvet et al. 2005; Alcalá et al. 2014; Manara et al. 2014; Alcalá et al. 2017; Manara et al. 2017b,a). Given the sparsity of the correlation measurements between line emission and accretion rates, which are based on five TTS with accretion rates from multi-column flows (see Figure 17 (Cont.) in Section A.3), it is also entirely possible that those correlations are intrinsically non-linear with accretion rate over the range of accretion rates in the ULLYSES sample, or have secondary dependences on stellar mass or other parameters.

### 8.3. A library of massive stars in the LMC and SMC

Figure 2 confirms that the massive-star component of the ULLYSES program met and often surpassed its goal of observing ∼4 representatives in most of the temperature- and luminosity-class "bins" occupied by O- and B-type stars in the LMC and SMC. Additional archival HST spectroscopy of early-type stars in other nearby, resolved galaxies indicated in Table 4 provides further material to study the systematic effects of reduced metallicity on the atmospheres and winds of early-type stars. When combined with ground-based spectroscopy (e.g., from the X-Shooting ULLYSES program, Vink et al. 2023), these ULLYSES spectra provide the observational foundation required for comprehensive modeling and analysis of the systematic behavior of the photospheres and radiatively-driven winds of massive stars along multiple dimensions, especially temperature, luminosity, and metallicity.

The montages in Figures 10 and 11 illustrate UV spectra for a small subset ULLYSES OB stars in the LMC and SMC, respectively. Although by no means exhaustive, these figures illustrate some of the morphological trends in the strength and shape of wind features that need to be quantified and explained as a function of temperature, luminosity, and metallicity. For example, the upper two panels of Fig. 10 confirm the relative weakness of P Cygni profiles in dwarf stars relative to the denser winds of supergiants, as shown, e.g., by the unsaturated C IV resonance doublet. They also highlight the prominent luminosity effect exhibited by the C III λ1176 multiplet and Si IV resonance doublet, especially for later types. The lower panel shows that except for the C IV doublet, most wind features fade in the B-supergiants, which makes their winds harder to model. These trends have been noted previously by Walborn & Bohlin (1996).

Comparison with the spectra of counterparts in the SMC (Fig. 11) immediately highlights the overall weakness of metallic P Cygni profiles, as expected. This trend is demonstrated even more explicitly in Fig. 12, which shows the increasing strength of the Si IV and C IV resonance doublets in the spectra of B0 supergiants as a function of the increasing ambient metallicity of their host galaxies.

Detailed modeling of this treasure-trove of data is underway, largely (though not exclusively) under the auspices of the X-Shooting ULLYSES collaboration (Vink et al. 2023). These large, multi-dimensional studies will take some time to complete, owing in part to the sheer number of targets, but also to the computational complexity associated with modeling the photospheres and winds of eartly-type stars in a unified manner. Mutiple interations may be required to incorporate new insights that are gained incrementally by the systematic analyses. Nevertheless, the primary goal envisaged by the SAC of improving the accuracy of our knowledge of the fundamental stellar parameters of massive stars as a function of metallicity will ultimately be achieved.

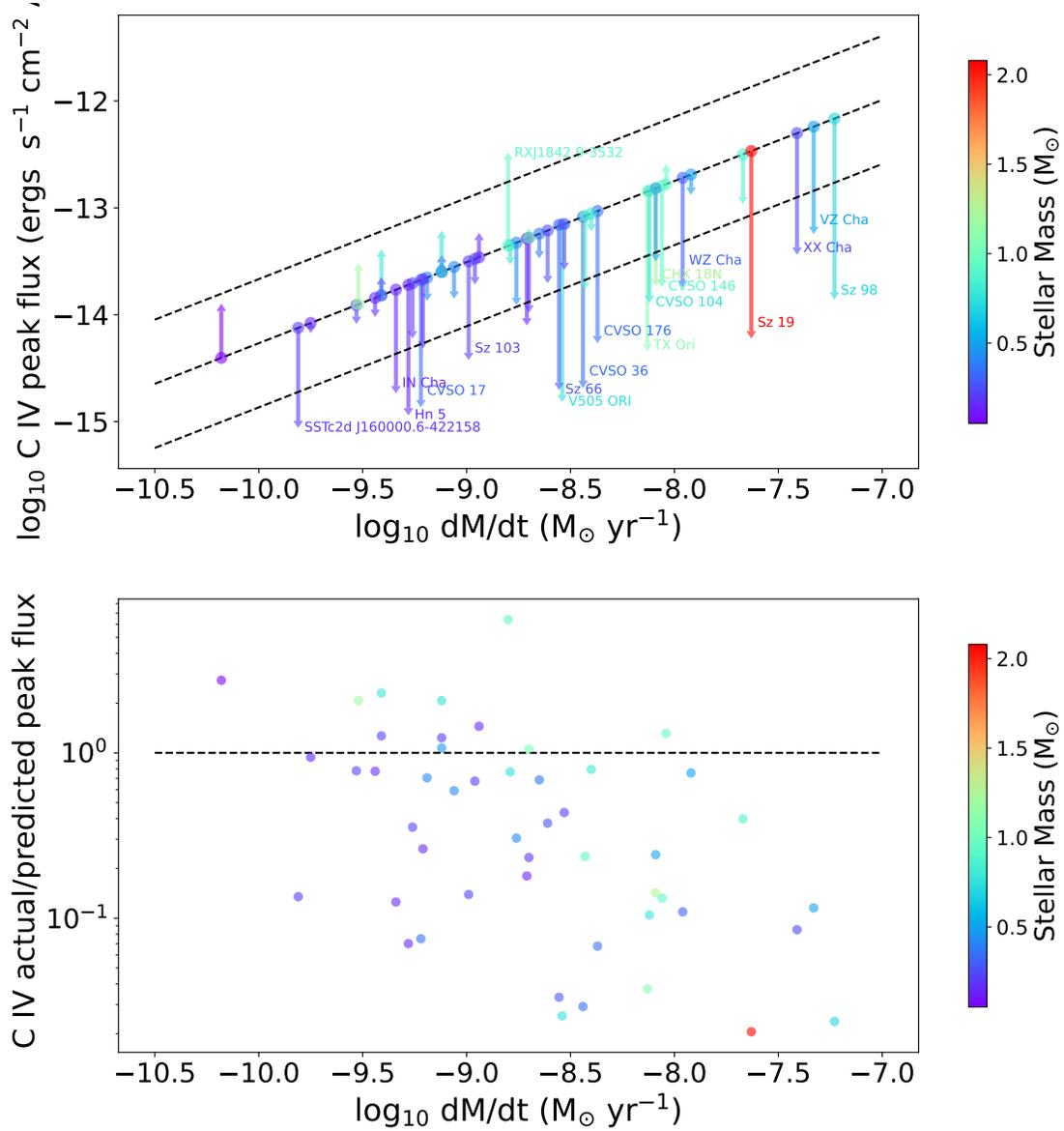

**Figure 8.** (Left) Peak C IV flux as a function of accretion rate for the 49 "survey" TTS in the ULLYSES sample (excluding archival targets) for which a co-added G160M spectrum was created. The central dashed line indicates the peak C IV flux predicted from the accretion rates reported in the literature, using fits in the templates used for exposure time calculations (see Section A.3), while the tip of the arrows indicate the actual measured peak flux. TTS for which the difference between the flux predicted from the previously measured accretion rate and the flux measured in the ULLYSES spectra is over an order of magnitude are indicated in the figure. (Bottom) Ratio of actual to predicted C IV peak flux in the sample sample of TTS as a function of accretion rate reported in the literature.

In the meantime, casual inspection of the data suggests a multitude of other important studies that can also be pursued. Two examples that deal with the detailed structure of the the stellar winds – properties that are not yet captured adequately in the model atmosphere analyses described above – are briefly presented here.

### 8.3.1. *Narrow Absorption Components*

The reduced optical depth in wind lines of stars in the LMC and SMC, particularly the C IV resonance doublet, provides new opportunities to study the phenomenon of "narrow absorption components" (NACs); see, e.g., Lamers et al. (1982) and Prinja & Howarth (1986) for pioneering studies of NACs in Galactic O stars. Figure 13 illustrates the phenomenon for two rep-resentative O4 dwarfs in the SMC, though in fact *all* the SMC O4 dwarfs in the ULLYSES sample exhibit NACs. NACs are thought to represent the end-stage of evolution of recurrent, large-scale perturbations in the wind, see, e.g., Prinja (1987), Prinja & Howarth (1988), and Kaper et al. (1996) for the historical development this notion. However, despite decades of study, some of their properties remain uncertain, in part because at Galactic metallicity the C IV doublet is almost always saturated. Consequently, NACs cannot be observed in this dominant ion in Galactic stars, which relegates assessment of the prevailing ionization state to trace ions, especially Si IV. A systematic study of the many examples of NACs in the ULLYSES spectra of LMC and SMC stars will provide new constraints on the origin of these features and their diagnostic relationship to time-dependent wind structure.

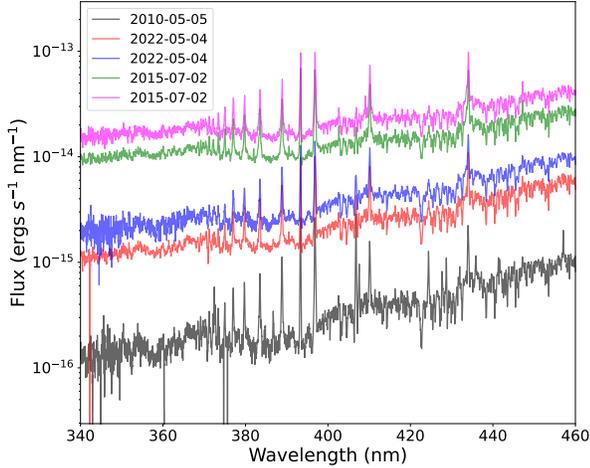

**Figure 9.** VLT/X-Shooter spectra of Sz 98 taken over different epochs: 2010 (Alcalá et al. 2014), 2015 (Alcalá et al. 2017), and 2022 (PENELLOPE collaboration, see Manara et al. 2021)

### 8.3.2. Diversity of Small-Scale Wind Structure

Non-monotonic velocity laws that are the hall-mark of structured stellar winds are typically diagnosed by the presence of "black troughs" and "soft blue edges" in saturated P Cygni profiles, neither of which are predicted to occur in smooth, spherically symmetric, monotonically expanding outflows (Lucy 1982, 1983). The origin of this small-scale velocity dispersion is generally attributed to the action of the powerful line-deshadowing instability (Owocki et al. 1988; Puls et al. 1993). The design of the ULLYSES sample allows morphological diversity to be examined within temperature- and luminosity class "bins," i.e., for stars that might be expected to exhibit similar wind profiles. It is therefore interesting to note that similar stars sometimes exhibit different signatures of small-scale wind structure in saturated C IV profiles, as shown in Figure 14 for a selection of O3-5 dwarfs and giants in the LMC. Detailed investigation of these cases will help to isolate the physical parameters or circumstances that add structure to these outflows.

### 8.4. Constraints on extinction in NGC 3109 and Sextans A

As discussed in Section 5, WFC3 pre-imaging in NGC 3109 and Sextans A was used to estimate the SED of the six ULLYSES targets in those galaxies, and corresponding exposure times for the ULLYSES COS observations. Rough estimates of E(B-V) were thus derived for each of those targets, assuming the "LMC average" extinction curve from Gordon et al. (2003). The LMC average extinction curve was selected for two reasons. First, the WFC3 photometry does not extend far enough into the FUV to robustly differentiate between LMC- or SMC-like extinction. This is demonstrated in Figures 15 and 16, which show that model SEDs with the LMC average or SMC Bar extinction curves are in similar agreements with the WFC3 photometry. Second, it is expected that the extinction toward each target is a mixture of Milky Way foreground and dust located in the target galaxies, likely resulting in an intermediate spectral shape of the extinction between flatter Milky Way-like and steeper low-metallicity-like extinction curve. The assumed spectral types for the initial estimates of the SED and exposure times were within one temperature class of the spectral types reported in the catalogs used for the target selection, which are based on optical/IR spectroscopy. The spectral types and E(B-V) values assumed in the initial estimation of the SED and exposure times based on the WFC3 photometry are reported in Table 7. The corresponding model spectra are shown in Figures 15 and 16, for two stars in each of the Sextans A and NGC 3109 galaxies, respectively.

With the COS spectra of the low metallicity stars in NGC 3109 and Sextans A in hand, we are in a position to further constrain the extinction toward those targets. We performed a $\chi^2$ minimization between the COS spectra combined with the WFC3 photometry, and a grid of Castelli & Kurucz (2004) stellar models spanning the initial $A_V$ estimate (from WFC3 photometry only) $\pm$ 0.2 mag, the LMC average and SMC bar extinction curves, and the original SpT (from catalogs + WFC3 photomtry) $\pm$ two temperature classes (but staying within the same spectral class).

The results of the fitting are shown in Figures 15 and 16, for two stars in each of the Sextans A and NGC 3109 galaxies, respectively. The corresponding best-fit spectral types, E(B-V) values (assuming $R_V = 3.1$), and extinction types are listed in Table 12, and compared to the catalog values and initial estimates based on WFC3 photometry only in the same Table.

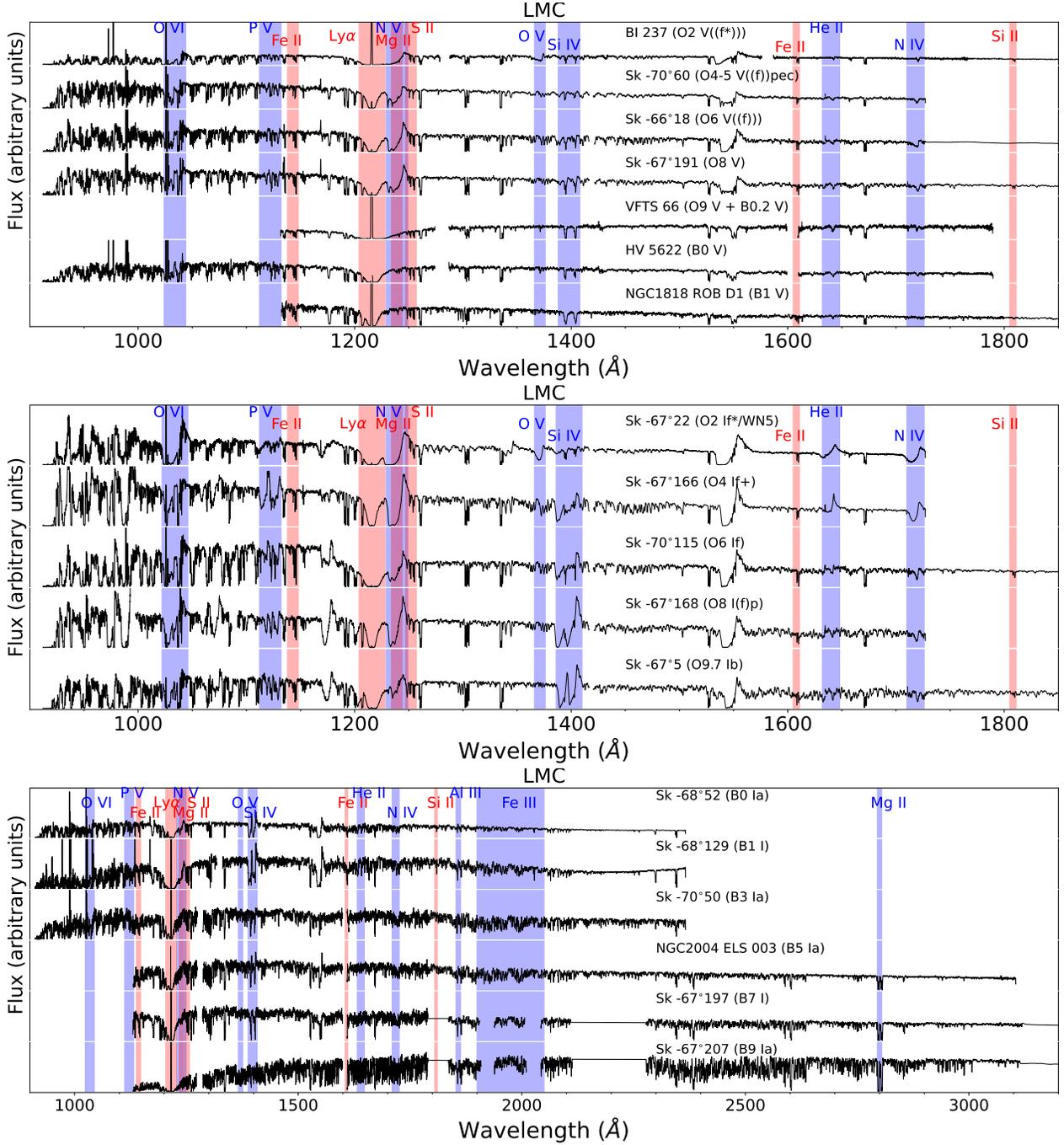

**Figure 10.** Example of UV spectra for O dwarfs (top), O super-giant (middle), and B supergiants (bottom) stars in the LMC, ordered by spectral type. Some important line diagnostics are highlighted in blue (stars) and red (ISM).

In all cases, the fits to the COS spectra yield very small residuals. We do, however, find significant (30-50%) residuals at the longest WFC3 wavelengths (F814W and in some cases F475W) for Sextans A LGN s003 and all three NGC 3109 massive stars. This could be due to deviations of the true extinction curve from the assumed SMC Bar (or LMC average) fiducial curves, or imperfect modeling of the stellar continuum. A more in-depth analysis would be needed to better understand those deviations.

The best-fit temperature classes are generally very close (within one temperature class) to the initial estimate from WFC3 photometry alone and to the literature catalogs. For all but one star (Sextans A LGN s071), the SMC Bar extinction curve is preferred by the data compared to the LMC average extinction curve, resulting in substantially steeper extinction in the UV. This result would be consistent with the dust extinction being dominated by dust located within the low metallicity galaxies. The E(B-V) values derived from the spectral fits do differ from the catalogs and originally derived values, with the maximal difference obtained for NGC 3109 EBU 07 and Sextans A LGN s003, for which the WFC3 photometry initially yielded E(B-V) = 0.065 and 0.13 respectively, but the fit to COS + WFC3 results in E(B-V) = 0.032±0.014 and 0.065±0.014, a factor of two lower in both cases.

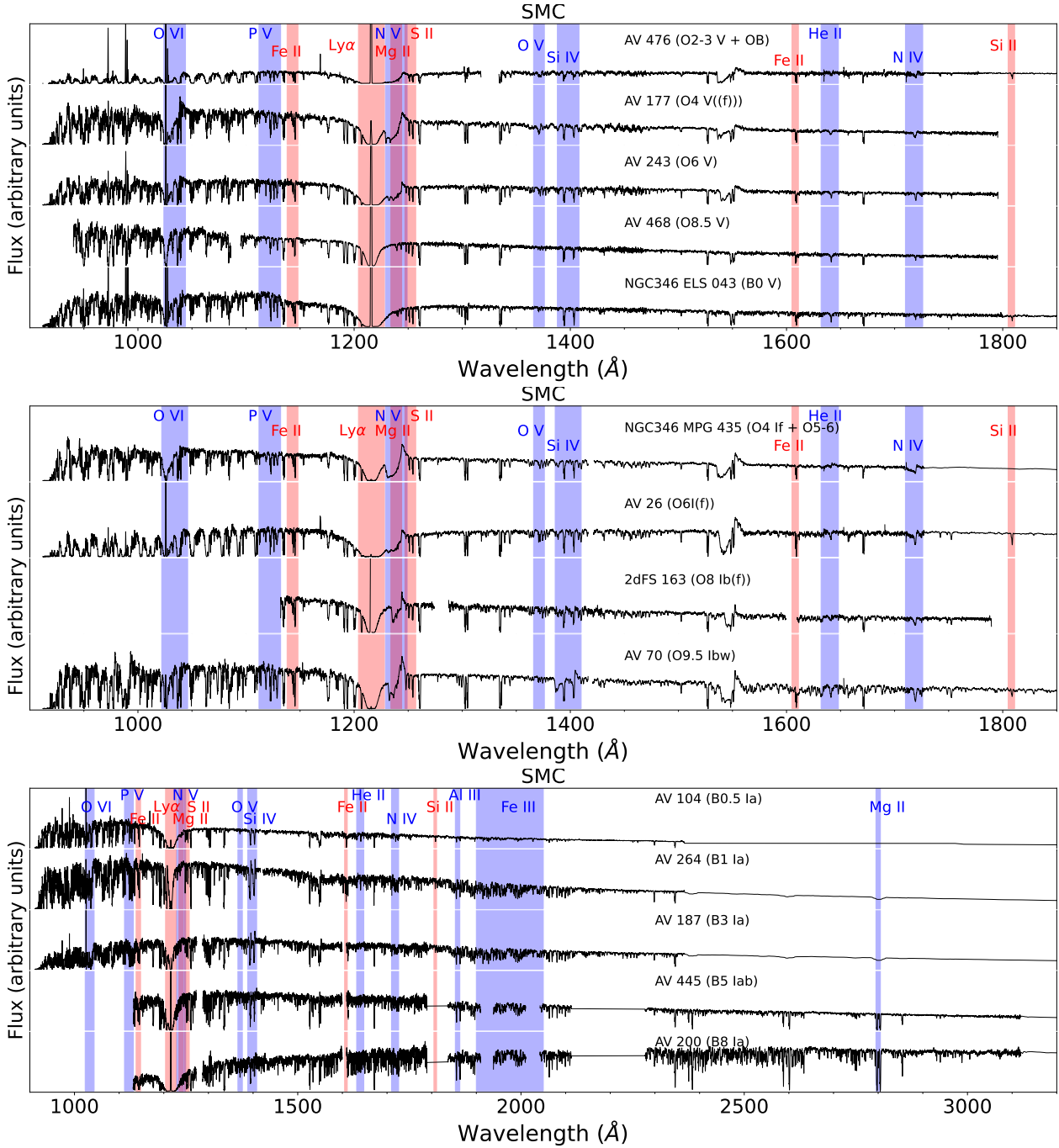

**Figure 11.** Example UV spectra for O dwarfs (top), O super-giant (middle), and B supergiants (bottom) stars in the SMC, ordered by spectral type. Some important line diagnostics are highlighted in blue stars) and red (ISM).

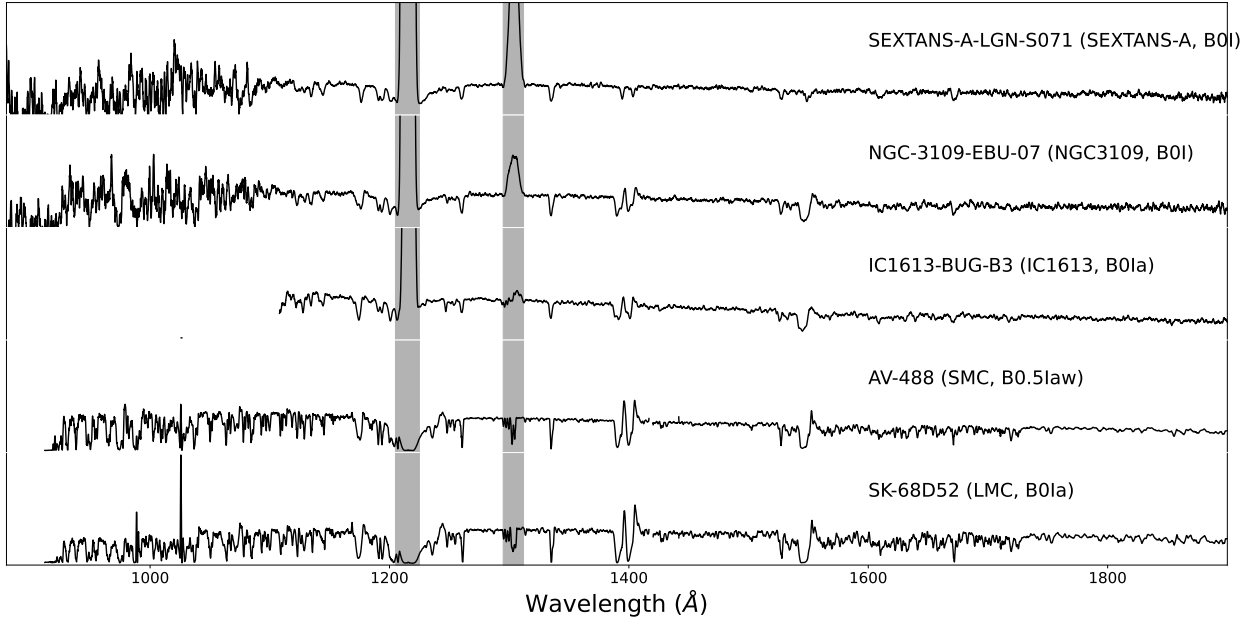

**Figure 12.** Comparison of UV spectra of B0 supergiants as a function of metallicity, going from 10% solar (Sextans A, top) to 10-15% solar (NGC 3109, IC 1613), 20% solar (SMC) and 50% solar (LMC). Spectral regions dominated by airglow in the COS aperture are masked by a vertical gray bar. Note the wind features (Si IV λ1400 and C IV λλ 1548, 1550 disappearing as the metallicity decreases.

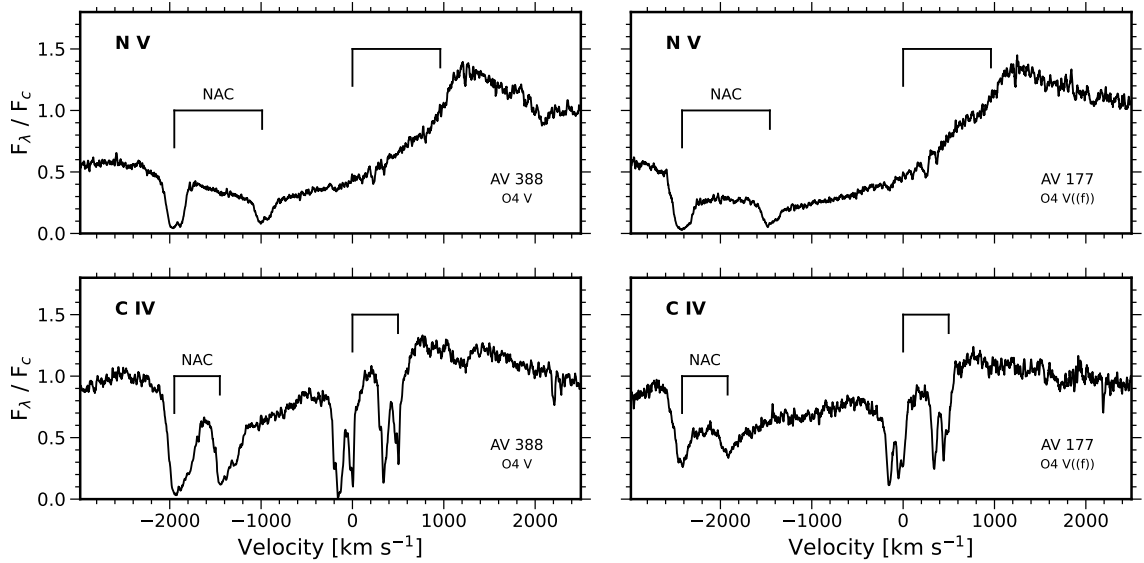

**Figure 13.** Narrow absorption components (NACs) in the N V (left) and C IV doublets of two O4 dwarfs in the SMC. The velocity scale refers to the blue component of the doublet, and has been shifted to remove the systemic velocity of the SMC. The separation of the components of the doublet is indicated by brackets at 0 velocity, and matches the separation of the NACs near the blue edge of the P Cygni absorption trough. Strong interstellar features associated with the Milky Way (near velocity of −170 km s⁻¹ ) and SMC (near 0 velocity) and are visible in the C IV profile.

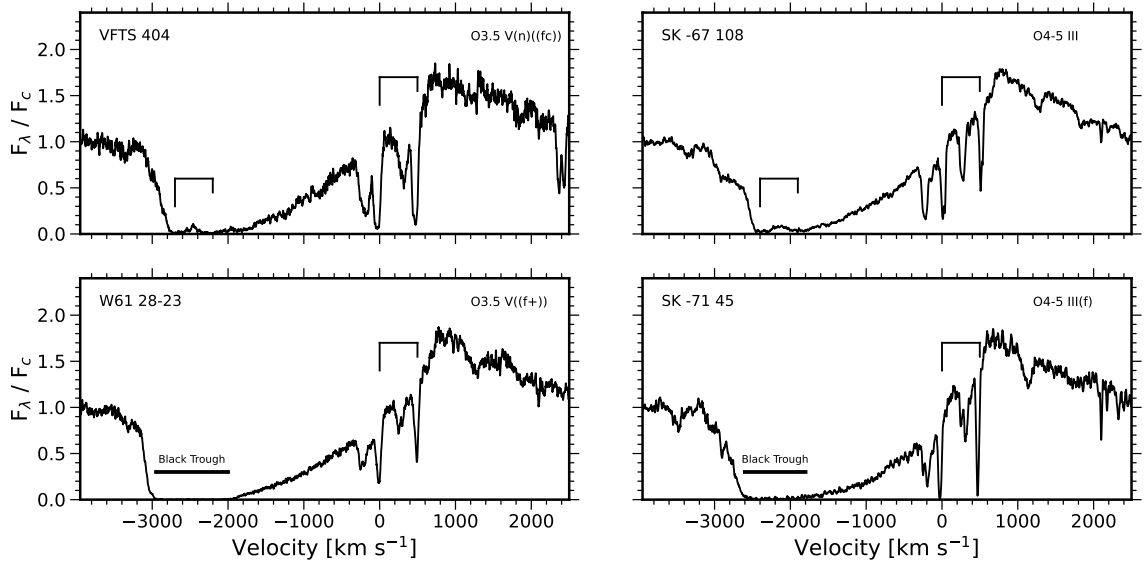

**Figure 14.** P Cygni profiles in the C IV doublet for a pair of O3.5 dwarfs (left) and O4–5 giants (right) in the LMC. The velocity scale refers to the blue component of the doublet, and has been shifted to remove the systemic velocity of the LMC. Although the profiles are broadly similar within each classification "bin," the examples shown in the lower panels exhibit extended regions with no flux. The presence of "black troughs" in some stars but not others indicates that different levels of small-scale velocity dispersion exist within the winds of otherwise similar stars, for reasons that have not yet been explained.

**Table 12.** Comparison of SpT and E(B-V) from literature catalogs based on optical-IR spectroscopy, derived from WFC3 photometry only, and derived from the COS spectra and WFC3 photometry for the ULLYSES targets in Sextans A and NGC 3109

| Star | SpT$_{lit}$ | E(B-V)$_{lit}$ | SpT$_{WFC3}$ | E(B-V)$_{WFC3}$ | SpT$_{COS+WFC3}$ | E(B-V)$_{COS+WFC3}$ | Ext. type (COS+WFC3) |
|---|---|---|---|---|---|---|---|
| | | mag | | mag | | mag | |
| NGC3109 EBU 07 | B0-1 Ia | 0.09 | B0 I | 0.065 | B0 I | 0.032±0.014 | SMC Bar |
| NGC3109 EBU 20 | O8 I | ⋯ | O8 I | 0.13 | O8 I | 0.13±0.014 | SMC Bar |
| NGC3109 EBU 34 | O8 I(f) | ⋯ | O8 I | 0.065 | O8 I | 0.065±0.014 | SMC Bar |
| Sextans A LGN s004 | O5 III | 0.043 | O6 V | 0.045 | O4 V | 0.065±0.014 | SMC Bar |
| Sextans A LGN s003 | O3-5 Vz | 0.23 | O5 V | 0.13 | O6 V | 0.065±0.014 | SMC Bar |
| Sextans A LGN s071 | B1 I | 0.00 | B0 I | 0.077 | B0 I | 0.10±0.014 | LMC average |

Note—Literature catalogs are Evans et al. (2007) for NGC 3109 and Garcia et al. (2019) and Lorenzo et al. (2022) for Sextans A

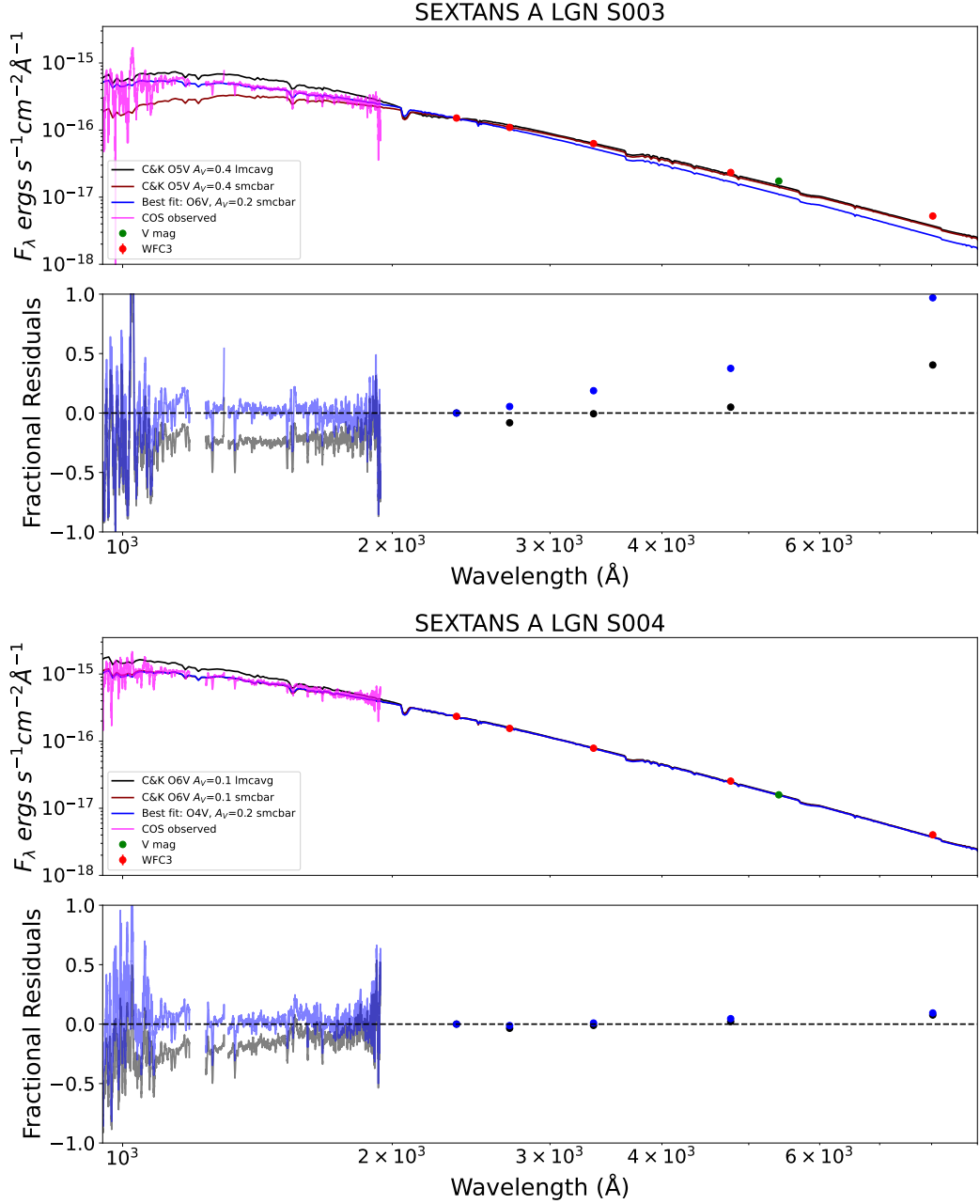

**Figure 15.** The first and third panels show models and observed spectra and photometry of two massive stars in Sextans A (LGN s003 and s004). The observed COS spectra are shown in magenta, while the WFC3 and $V$ band photometry are shown as red and green circles, respectively. The black and dark red curves indicates the original stellar + dust model derived from the WFC3 photometry alone, for the LMC average and SMC Bar extinction curves, respectively. The blue curve corresponds to the best fit to the combined COS spectra and WFC3 photometry. For each model spectrum, the parameters of the model are indicated in the legend. The second and fourth panels show fractional residuals between the observations (COS and WFC3) and the models, with the black and blue points and lines corresponding to the initial model (WFC3 photometry only) and best-fit to COS + WFC3, respectively.

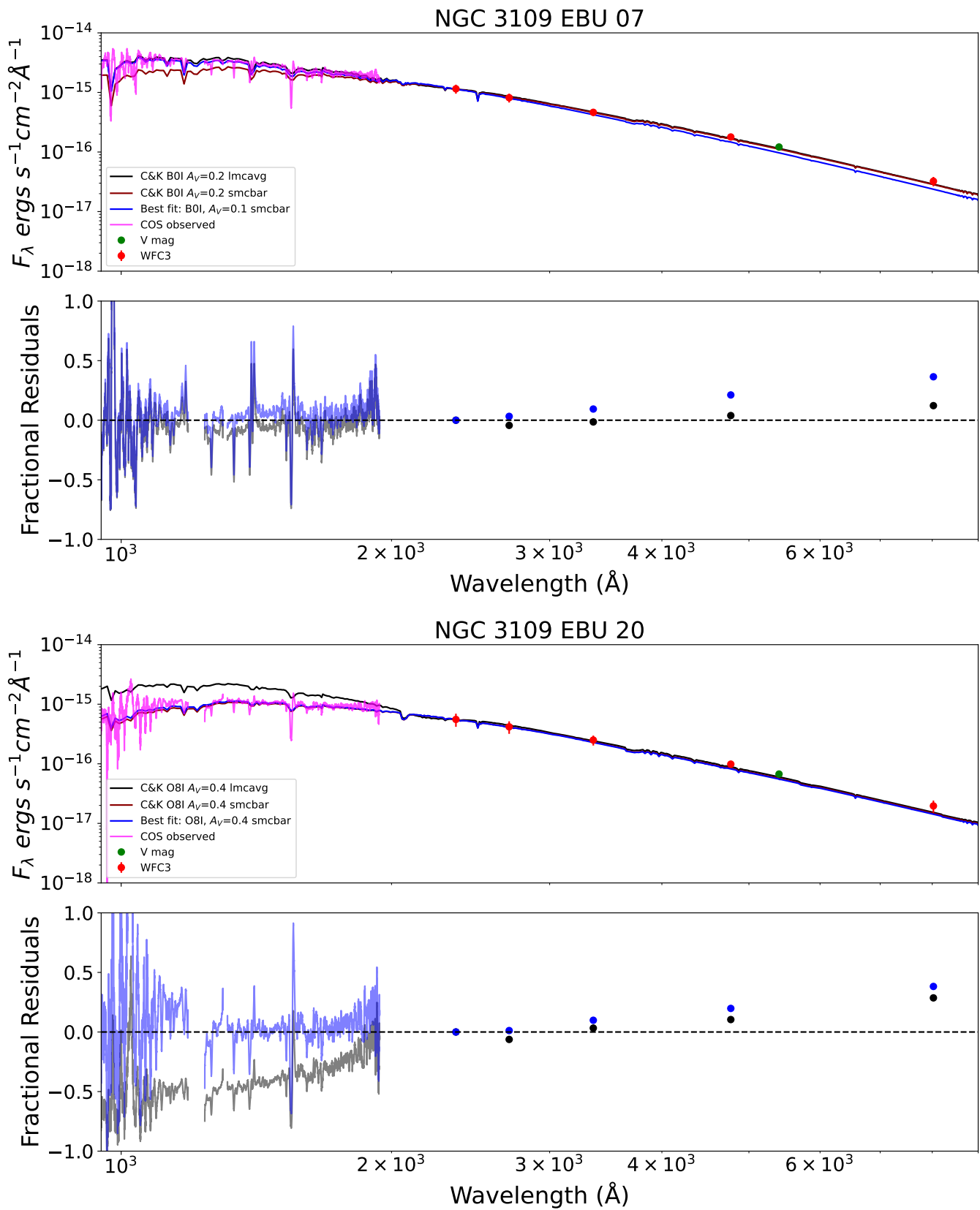

**Figure 16.** Same as Figure 15, but for two stars in NGC 3109 (EBU 07 and 20)

## 9. SUMMARY AND CONCLUSION

In this paper, we present the scientific objectives, observing strategy, target selection, and some initial results for the ULLYSES program. ULLYSES is a 1000-orbit observing program executed with Director's Discretionary time, in the spirit of previous large DD projects (e.g., the Frontier Fields, see Lotz et al. 2017). ULLYSES was designed in close concertation with the community, through a UVLWG. Unlike the previous panchromatic imaging programs, ULLYSES is entirely focused on Hubble's unique UV spectroscopic capabilities. The program was initiated in 2019 and completed in late 2023. The execution of the program, the data reduction and calibration, and the design of high-level science data products will be described in detail in Paper II.

Half of the ULLYSES program obtained medium-resolution COS and STIS UV spectra of 160 massive stars in nearby galaxies. In the LMC and SMC, the targets are strategically selected to uniformly sample spectral type and luminosity class (including WR stars) while leveraging archival observations. The program also explores the few brightest stars in NGC 3109 (10-20% solar metallicity) and Sextans A (10% solar metallicity) to probe even lower metallicities.

The objectives of the massive star component of ULLYSES are to characterize the winds and photospheres of massive stars at low metallicity, and produce a complete spectroscopic library of massive stars that can test stellar evolutionary theory and enable population synthesis in the metallicity range typical of cosmic noon (z ∼ 2; Madau & Dickinson 2014). This effort is particularly timely given the sparsity of massive star UV spectroscopy outside the Milky Way prior to ULLYSES, and the advent of JWST and the ELTs, which will reach unresolved stellar populations in the early Universe. ULLYSES+XShootU is also timely in the context of the Sloan Digital Sky Survey-V Local Volume Mapper (SDSS-V LVM Drory et al. 2024), for which the LMC and SMC are main targets. The study of ionized nebulae around ULLYSES targets has the potential to complete our knowledge about the stars, in particular their ionizing fluxes.

The other half of the program aims at constraining accretion physics in T Tauri stars through single-epoch ("survey") STIS UV-optical-NIR spectroscopy of 58 young low mass stars located in eight star-forming regions of the Milky Way (twelve region including archival data), and COS monitoring of four well-studied T Tauri stars (TW Hya, BP Tau, RU Lup and GM Aur). The spectroscopy of the survey stars utilizes the medium-resolution gratings on COS (FUV) and low-resolution gratings on STIS (NUV, optical, NIR), and the complete wavelength coverage is obtained near-simultaneously to ensure that the accretion and stellar properties can be constrained from the spectra. The 58 survey T Tauri stars were carefully selected to sample accretion rate and stellar mass, particularly below the relatively unexplored regime below 0.5 solar mass. In addition to providing diagnostics of accretion, the UV spectra will provide important constraints in understanding the thermal structure and dispersal of proto-planetary disks, and the evolution and habitability of planets within them. Those constraints are needed to interpret the powerful probes of disk chemistry observed with ALMA and JWST.

The monitoring component relies on the COS G160M and G230L gratings to cover the FUV and NUV in single orbit observations, repeated 12 times through three rotation periods, with the same pattern repeated twice approximately one year apart (for a total of 24 observations per target). The objective of this monitoring component is to constrain the timescales and amplitude of accretion variability, from minute- to year-timescales.

Both the low- and high-mass star components of ULLYSES are complemented by ancillary and coordinated programs, notably with the VLT/X-Shooter instrument (PENELLOPE and X-ShootU programs for the low- and high-mass stars, respectively), which enhance the legacy value of the ULLYSES program considerably.

While the scientific community is leading broad-ranging analyses of the data impacting many fields of astrophysics, we present some immediate results from the ULLYSES data. On the low-mass end, unexpected companions were discovered in the STIS slits of two T Tauri star observations in the Ori OB1 region, one of which turns out to be another T Tauri star. We document the positions of these companions. We examine the variability timescales of the monitoring UV spectra of the four T Tauri star and find that the brightness of the C IV, He II, and Mg II lines can vary by a factor of several over just a few days, and that the relative variations of line and continuum fluxes, even in the optical, generally track each other quite closely. The relative strengths of different lines do change between stars however, with RU Lup exhibiting by far the brightest Mg II line compared to C IV, He II, or N V, contrary to the other three monitored stars, for which those four lines have fluxes with the same order of magnitude.

On the massive star side, we examine how the main wind diagnostics (N V, Si IV, C IV) change with temperature, luminosity class, and metallicity. As expected, the strength of those wind features in early B supergiants decreases considerably from LMC metallic-ity (50% solar) to the metallicity of Sextans A (10% solar).

This paper is dedicated to the memory of Dr. Will Fischer, who passed away unexpectedly on April 16, 2024. Based on observations obtained with the NASA/ESA Hubble Space Telescope, retrieved from the Mikulski Archive for Space Telescopes (MAST) at the Space Telescope Science Institute (STScI). STScI is operated by the Association of Universities for Research in Astronomy, Inc. under NASA contract NAS 5-26555. M. Garcia gratefully acknowledges support by grants PID2019-105552RB-C41, PID2022-137779OB-C41, and PID2022-140483NB-C22, funded by the Spanish Ministry of Science, Innovation and Universities/State Agency of Research MICIU/AEI/10.13039/501100011033 and by "ERDF A way of making Europe". We thank the referee for a very constructive report.

## APPENDIX

### A. APPENDIX A

In this appendix, we describe the methodology used to estimate the UV flux of ULLYSES targets. These estimates were subsequently used in the UBETT to determine the exposure time required to achieve the S/N levels per resel specified in Tables 2 and 3.

#### A.1. *Flux distributions for massive stars in the LMC and SMC*

Several grids of model atmospheres were used to represent the flux distributions of ULLYSES targets, all of which include metallicities appropriate to the LMC ($Z/Z_\odot \sim 0.5$) and SMC ($Z/Z_\odot \sim 0.2$). For observations made in Cycles 27 and 28, the flux distributions came from either the non-LTE grids of WM-basic models (for OB stars; Pauldrach et al. 2001) and CMFGEN models (for WR stars; Hillier & Miller 1998). The models were computed by Smith et al. (2002) or the Castelli & Kurucz (2004) library of LTE models maintained at STScI[11], using the mapping to stellar parameters from Sternberg et al. (2003). Observations obtained in Cycle 29 relied on models computed with the PoWR code (Hainich et al. 2019) to provide non-LTE flux distributions for OB stars with 15 kK $\leq$ T$_{eff}$ $\leq$ 48 kK, while models for the coolest B-type supergiants continued to be drawn from the Castelli & Kurucz (2004) grid. PoWR models for WR stars in either the WC or WN (Todt et al. 2015) sequence were used exclusively in Cycle 29.

Models for specific targets were selected from these grids based on the mapping of their spectral type to (T$_{eff}$, log g) provided by Martins & Plez (2006) for O-type stars and Conti et al.(2008; Table 3.1) for B-type stars. Since OB spectral types are primarily based on line-strength ratios of He lines, the comparatively small effects of metallicity were neglected for the purpose of these exposure-time calculations. However, when previous model atmosphere analyses existed for a particular target, the selection was based on the model in the grid that most-closely matched published values of (T$_{eff}$, log g) rather than the calibration. Published values of stellar parameters were also used to select flux distributions for WR stars.

To account for interstellar extinction, the shape of the flux distribution was reddened by the "law" appropriate to the relevant galaxy as specified by Gordon et al. (2003). For OB stars, the degree of reddening was determined by the estimated color excess, $E(B-V)$, which was determined from published values of Johnson $BV$ photometry and calibrations of intrinsic color as a function of spectral type from Martins & Plez (2006) for O-type stars and Fitzgerald (1970) for B-type stars. For WR stars, estimates of the reddening were generally available from detailed model atmosphere analyses.

Finally, the reddened flux distributions were normalized to observed values in specific wavebands. Measurements from archival UV spectra provided the most reliable normalization of the distribution. For targets without archival spectra, photometric measurements in the Johnson $U$ or $B$ bands (and occasionally the $V$ band) were used to normalize the flux distribution. The UV fluxes were quite uncertain in cases where only Johnson $V$- and $B$-band photometry was available, since these two numbers provide the bare minimum of information required to estimate reddening and normalize the underlying flux distribution.

These normalized and reddened flux distributions were subsequently input to the UBETT, where they were used to calculate the exposure time required to obtain the desired S/N per resel (Table 2) for instrumental configurations selected from Table 1.

#### A.2. *Computation of exposure times for massive stars in Sextans A and NGC 3109*

Exposure times for massive stars in Sextans A and NGC 3109 were estimated similarly to stars in the LMC and SMC, albeit with the additional photometric constraints provided by the WFC3 pre-imaging (see Section 2.2). Accurate extinction values were estimated by reddening WM-BASIC and Castelli & Kurucz (2004) stellar models of the spectral type and luminosity class determined from the original catalogs (Evans et al. 2007; Garcia et al. 2019; Lorenzo et al. 2022) using an "LMCAvg" extinction curve of varying $E(B-V)$, and manually adjusting the $E(B-V)$ to match the WFC3 F225W, F275W, F336W, F475W, and F814W photometry. The resulting extinction values are listed in

**Table 13.** TTS templates used for exposure time estimates

| Template | log(dm/dt) | $A_V$ | $d$ | Range of matched log (dm/dt) |
|---|---|---|---|---|
| | $M_\odot$ yr$^{-1}$ | mag | pc | $M_\odot$ yr$^{-1}$ |
| V836 Tau | -8.96 | 1.5 | 140 | <-8.5 |
| DN Tau | -8.00 | 0.9 | 140 | -8.5 ... -7.5 |
| DR Tau | -7.28 | 1.4 | 140 | >-7.5 |

Section 5, along with the spectral types determined from the original catalogs (Evans et al. 2007; Garcia et al. 2019; Lorenzo et al. 2022) and WFC3 photometry. The resulting "best fit" models were then input into the ETC to estimate exposure times.

### A.3. *Computation of exposure times in T Tauri Stars*

The generation of input TTS models for UBETT was based on the assumption that line and continuum fluxes in these stars scale with accretion rate. This has been demonstrated extensively in the literature (e.g., Gullbring et al. 1998; Muzerolle et al. 1998; Salyk et al. 2013; Rigliaco et al. 2015; Alcalá et al. 2017; Robinson & Espaillat 2019). We chose three well-studied TTS stars as templates: V836 Tau, DN Tau, and DR Tau. The templates were required to have contemporaneous COS FUV and STIS NUV-optical spectra in the MAST archive, with published accretion rates based on those particular spectra (Ingleby et al. 2013). The estimated spectrum of each ULLYSES candidate target is a modified (scaled) version of the template with the closest accretion rate. Table 13 lists the relevant properties of these stars and the range of candidate accretion rates that were matched to it.

The COS spectra for the templates were obtained from the Hubble Spectroscopic Legacy Archive (HSLA; Peeples et al. 2017)[12]. STIS G230L and G430L spectra were obtained from MAST. Simultaneous G750L spectra were not available for the templates, so we instead used a spectrum of GM Aur and scaled it to match each template following the general scaling procedure explained below. The spectra of the templates were de-reddened using the Fitzpatrick & Massa (1990) procedure with parameters appropriate for the line of sight to HD 29647 (Whittet et al. 2004), a B star whose line of sight intercepts the Taurus molecular cloud.

In modifying the templates to simulate particular ULLYSES candidates, we scaled the FUV, NUV, blue optical, and red optical continua independently as well as Ly-$\alpha$, N V $\lambda\lambda1238$, 1242 C II $\lambda1335$, Si IV $\lambda\lambda1393$, 1402, C IV $\lambda\lambda1548$, 1550, and He II $\lambda1640$ lines. The continua and all lines except Ly-$\alpha$ were scaled with relations of the form

$$\log \frac{F_c}{F_t} = a \log \frac{\dot{M}_c}{\dot{M}_t} \qquad (A1)$$

where the subscripts $c$ and $t$ refer to the ULLYSES candidate and the template, respectively, and $F$ refers to the flux in the continuum or line, and $\dot{M}$ corresponds to the accretion rate. The template fluxes ($F_t$) are listed in Table 14. The slope $a$ of the correlation between flux and accretion rate was determined by fitting lines to the relation between log flux versus log accretion rate for 25 HST spectra of five TTS in Robinson & Espaillat (2019). The fits are shown in Figure 17 (Cont.). Table 15 lists the slopes $a$ and intercepts $b$ of the relation between flux and accretion rate for the lines and continua. The modified spectra were then scaled to the appropriate distance and reddened using the appropriate $A_V$ and the extinction law toward HD 29647 reported by Whittet et al. (2004).

Ly-$\alpha$ requires a slightly different procedure, because it is dominated by airglow in the COS templates and is affected by absorption in the ISM and in winds from the star/disk system (e.g. Schindhelm et al. 2012; McJunkin et al. 2014). This makes a simple scaling with accretion rate impossible. Instead, we scale a gaussian profile with a FWHM of 2.5 Å (based on an assessment of profiles in France et al. 2014), and a centroid of 1215.67

---

[12] https://archive.stsci.edu/missions-and-data/hsla

**Table 14.** Continuum and line (peak) fluxes ($F_t$) in the TTS templates

| Feature | $F_t$(V836 Tau) | $F_t$(DN Tau) | $F_t$(DR Tau) |
|---|---|---|---|
| | ergs cm$^{-2}$ s$^{-1}$ Å$^{-1}$ | ergs cm$^{-2}$ s$^{-1}$ Å$^{-1}$ | ergs cm$^{-2}$ s$^{-1}$ Å$^{-1}$ |
| FUV Continuum (1360 Å) | 5.19e-15 | 7.08e-15 | 6.65e-14 |
| NUV Continuum (2200 Å) | 2.06e-15 | 4.51e-13 | 2.06e-13 |
| Blue optical continuum (4225 Å) | 1.57e-14 | 3.39e-14 | 2.30e-13 |
| Red optical continuum (8000 Å) | 5.20e-14 | 9.08e-14 | 2.07e-13 |
| N V λλ1238, 1242 | 2.10e-13 | 3.04e-13 | 3.61e-13 |
| C II λ1335 | 2.28e-13 | 4.96e-13 | 1.14e-12 |
| Si IV λλ1393, 1402 | 1.85e-13 | 2.61e-13 | 9.35e-13 |
| C IV λλ1548, 1550 | 3.22e-13 | 1.46e-12 | 1.18e-12 |
| He II λ1640 | 5.35e-13 | 1.33e-12 | 4.45e-13 |
| Mg II λλ2796, 2803 | 1.08e-13 | 1.53e-13 | 9.79e-13 |

Å. The Gaussian profile is scaled by the accretion rates of the targets using two approaches, to ensure that our bright object protection clearance is robust against different assumptions. In the first approach, we scaled the Gaussian profile using the upper envelope of the relation between Ly-$\alpha$ luminosity and accretion rate measured in France et al. (2017) and shown as a red line in Figure 18. In the second approach, we measured the correlation between accretion rate and peak Ly-$\alpha$ line strength in STIS spectra (see top left panel of Figure 17 (Cont.)), ignoring TW Hya, which is much closer and has very weak interstellar absorption of Ly-$\alpha$. As a worst case scenario, and for the purpose of clearing the brightness of our targets with respect to the instrument's count rate limits, we tied the Ly-$\alpha$ peak strength-accretion rate relation measured in the STIS spectra to the largest Ly-$\alpha$ luminosity in the France et al. (2017) sample, measured for GM Aur ($1.88 \times 10^{32}$ erg s$^{-1}$ for an accretion rate of $9.6 \times 10^{-9}$ $M_\odot$ yr$^{-1}$). The resulting model is shown in Figure 18 as a solid blue line. Figure 19 compares the Ly-$\alpha$ profile observed in RECX-11 ($\eta$ Cha, d = 94 pc) with those modeled using the two approaches described above (approach 1 in red, approach 2 in blue). For this particular object, the first approach overestimates the Ly-$\alpha$ luminosity, while the second approach roughly reproduces the observed profile.

In order to match the rest of the spectrum, the model profile is scaled to the appropriate distance and reddened using the appropriate $A_V$ and the extinction law toward HD 29647 reported by Whittet et al. (2004). Finally, it is subjected to absorption by intervening H I using the method of Roman-Duval et al. (2019). We set

N(H I) = $4.8 \times 10^{21}$ cm$^{-2}$ × $A_V/R_V$, with $R_V = 3.63$, the same as in the extinction law used elsewhere. If $A_V$ = 0, we use log N(H I) = 19.7 cm$^{-2}$, a value typically found for such stars by McJunkin et al. (2014). The model profile then replaces the airglow-dominated profile in the template.

This scaling procedure was tested on a small sample of stars with archival contemporaneous COS or STIS UV spectra. These spectra allowed for the accretion rate and flux in hot gas lines to be measured at the same epoch. Knowing the accretion rate, model templates scaled by accretion rate, distance, and extinction were generated. The line flux of hot gas lines in the scaled templates and observed spectra were then compared. The result of this test is shown in Figure 20. Generally, the estimated line flux is within a factor of four of the actual measured flux.

The modified (scaled) templates were input in the UBETT to calculate the exposure times required to achieve the required S/N (Table 3). To ensure that the targets are observed with sufficient exposure time and S/N to account for the scatter in the relation between accretion rate and line strength (Figure 17 (Cont.)), we padded all COS exposure times by a factor of two.

### A.4. BOP clearance for T Tauri stars (variable targets)

The UV detectors on COS and STIS can be shut down and/or damaged by over-light situations whereby the local or global count rate exceeds the safety limits. In order to prevent such harmful situations, all targets have to be cleared for observations. This is particularly important for variable targets. In those cases, worst case scenarios are assumed to ensure that those objects can be safely observed.

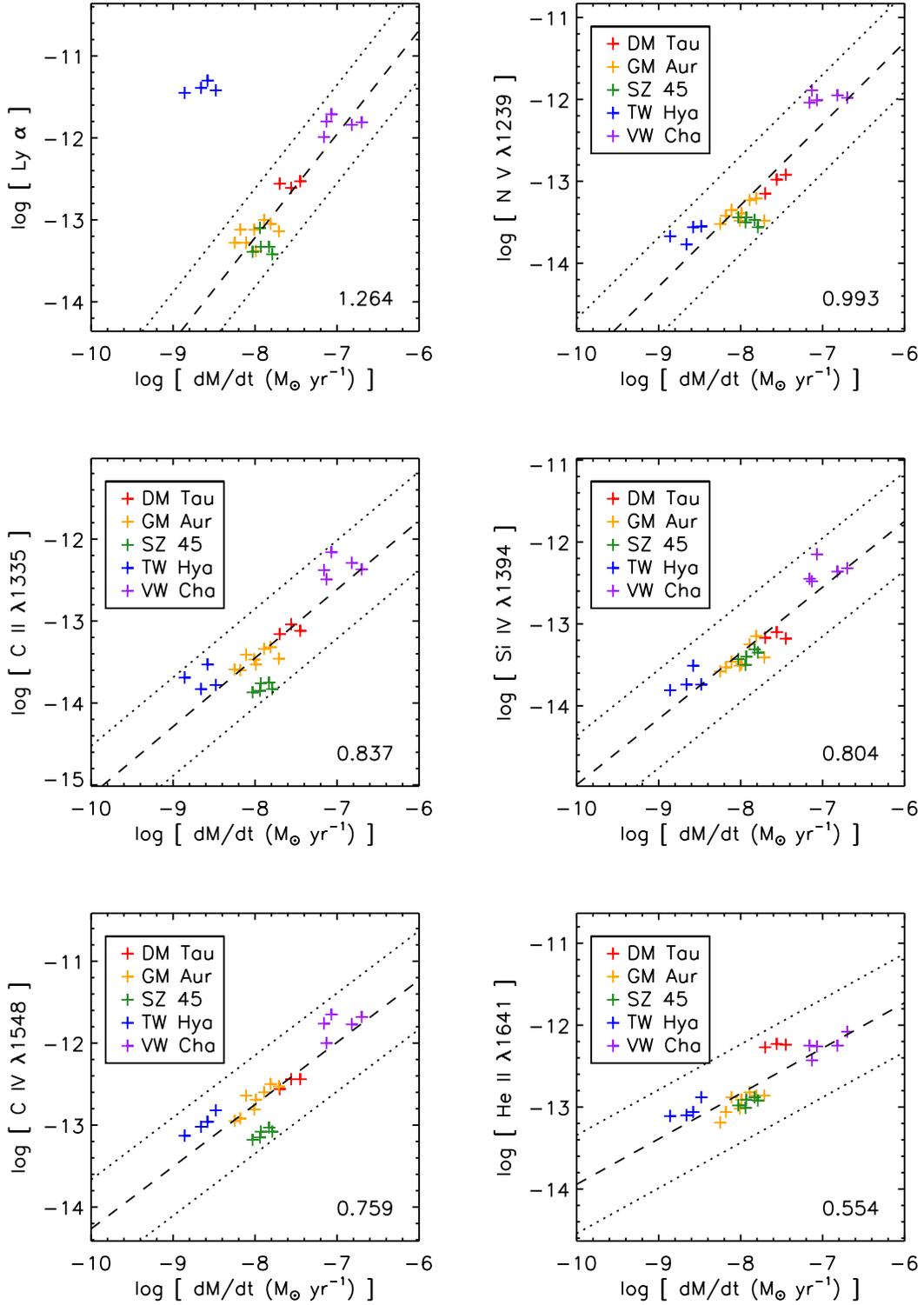

**Figure 17.** Best fit of the peak line strength (in erg cm$^{-2}$ s$^{-1}$ Å$^{-1}$) as a function of accretion rate in the 25 HST spectra of five TTS in Robinson & Espaillat (2019)

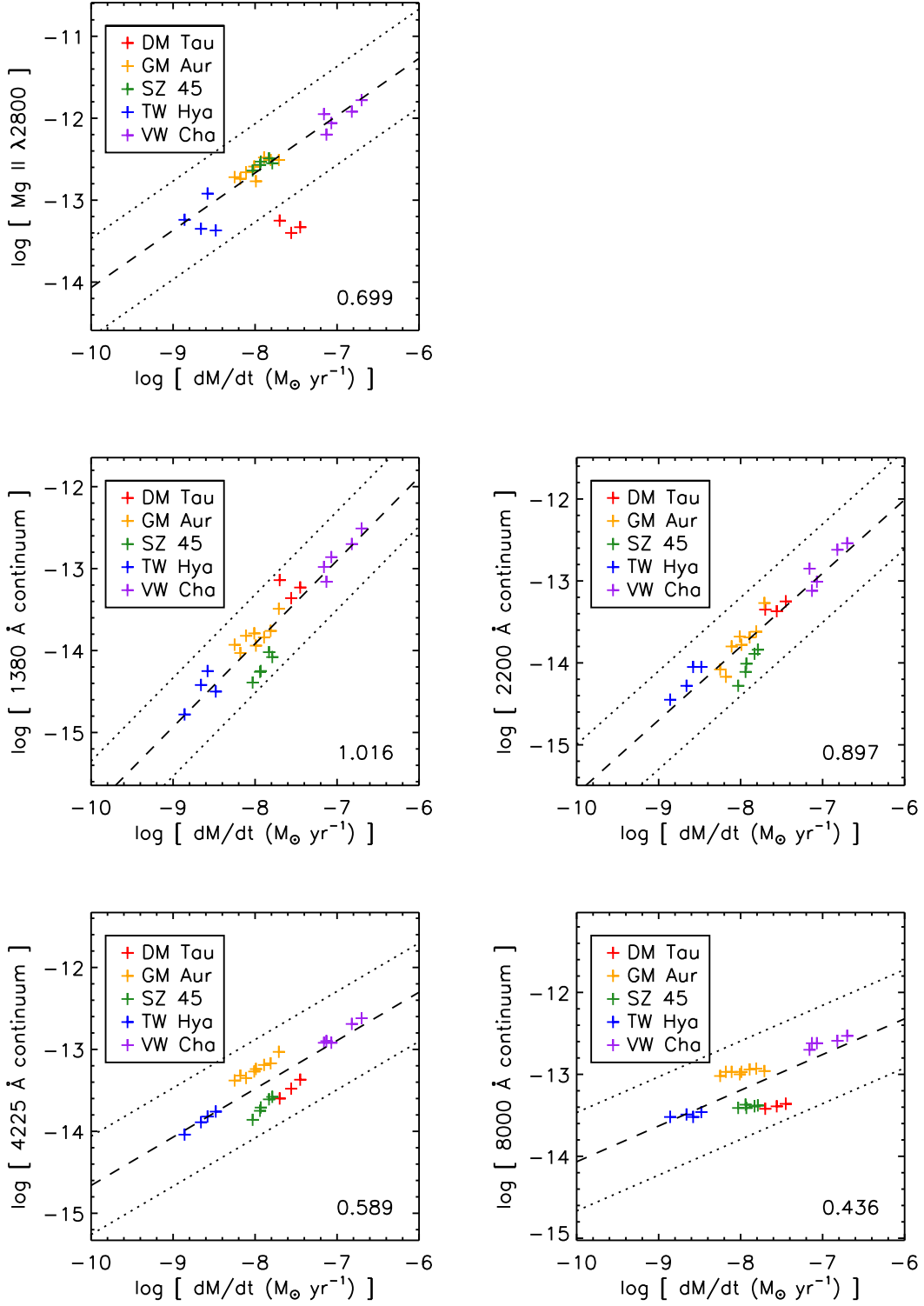

**Figure 17 (Cont.).** Best fit of the peak line strength (in erg cm$^{-2}$ s$^{-1}$ Å$^{-1}$) as a function of accretion rate in the 25 HST spectra of five TTS in Robinson & Espaillat (2019)

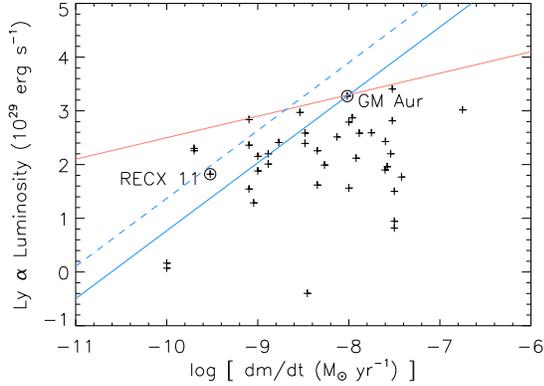

**Figure 18.** Relation between accretion rate and Ly-α luminosity in France et al. (2017) The blue line shows the linear correlation between Ly-α peak line strength and accretion rate measured in STIS spectra from Robinson & Espaillat (2019), tied to the brightest luminosity in France et al. (2017), which is for GM Aur. The blue dashed line includes a factor of 4 above this maximum value. The red line shows the upper enveloped of the accretion rate- Ly-α luminosity relation in France et al. (2017).

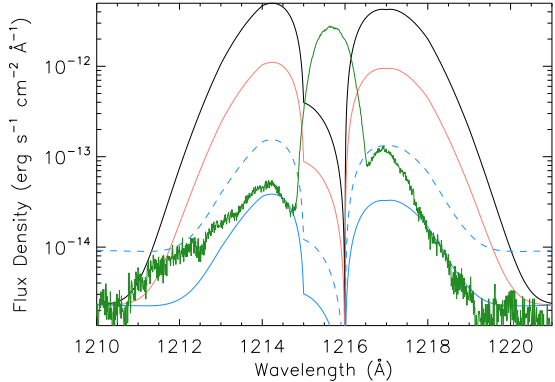

**Figure 19.** Comparison of the observed (green) and modeled Ly-α profiles in RECX-11. The red line corresponds to the profile modeled using the first approach (scaling based on accretion rate and upper envelope of the relation between accretion rate and Ly-α luminosity in France et al. (2017)), while the blue solid line corresponds to the profile modeled using the second approach (scaling based on accretion rate and relation between peak Ly-α flux and accretion rate in Robinson & Espaillat (2019)). The blue dashed line is also generated with the second approach, but the luminosity is multiplied by 4 for BOP checking purposes. The black line is obtained by assuming the same Ly-α luminosity as GM Aur.

**Table 15.** Slope and intercept of the correlation between peak line strength and accretion rate in the 5 TTS from Robinson & Espaillat (2019)

| Feature | $a$ | $b$ |
|---|---|---|
| FUV Continuum (1360 Å) | 1.016 | -5.790 |
| NUV Continuum (2200 Å) | 0.897 | -6.629 |
| Blue optical continuum (4225 Å) | 0.589 | -8.766 |
| Red optical continuum (8000 Å) | 0.436 | -9.704 |
| N V λ1239 | 0.993 | -5.346 |
| C II λ1335 | 0.837 | -6.753 |
| Si IV λ1393 | 0.804 | -6.923 |
| C IV λ1550 | 0.759 | -6.677 |
| He II λ1640 | 0.554 | -8.404 |
| Mg II λλ2796, 2803 | 0.699 | -7.073 |

There are two considerations in the bright object protection (BOP) clearing for ULLYSES targets. First, accretion variability results in UV flux variability. From the Robinson & Espaillat (2019) study and light curves produced from archival COS data of T Tauri stars, the magnitude of this variability is about a factor of four. Therefore, we produced models of the targets assuming an accretion rate four times higher than reported in the literature and cleared those through the ETC. These "×4" templates were also used to compute buffer times for COS and STIS observations.

Second, T Tauri stars of spectral type M have to be cleared against magnetic flaring outbursts (Osten 2017a,b). In main-sequence and possibly pre-main sequence M stars, these magnetic flares are thought to be caused by the rearrangement of magnetic fields in the outer stellar atmosphere through magnetic reconnection processes, and may potentially liberate large amounts of energy. While flares occur on almost all cool stars, M dwarfs have exhibited extreme enhancements, with up to six magnitudes in the $U$ band and even more extreme behaviors observed in the NUV (Robinson et al. 2005). It is still not completely clear whether or not T Tauri stars exhibit flaring properties and frequencies similar to their main-sequence counterparts. Some literature studies based on X-ray emission from T Tauri stars reveal that the magnetic flaring properties of pre-main sequence stars may not be so different from their main-

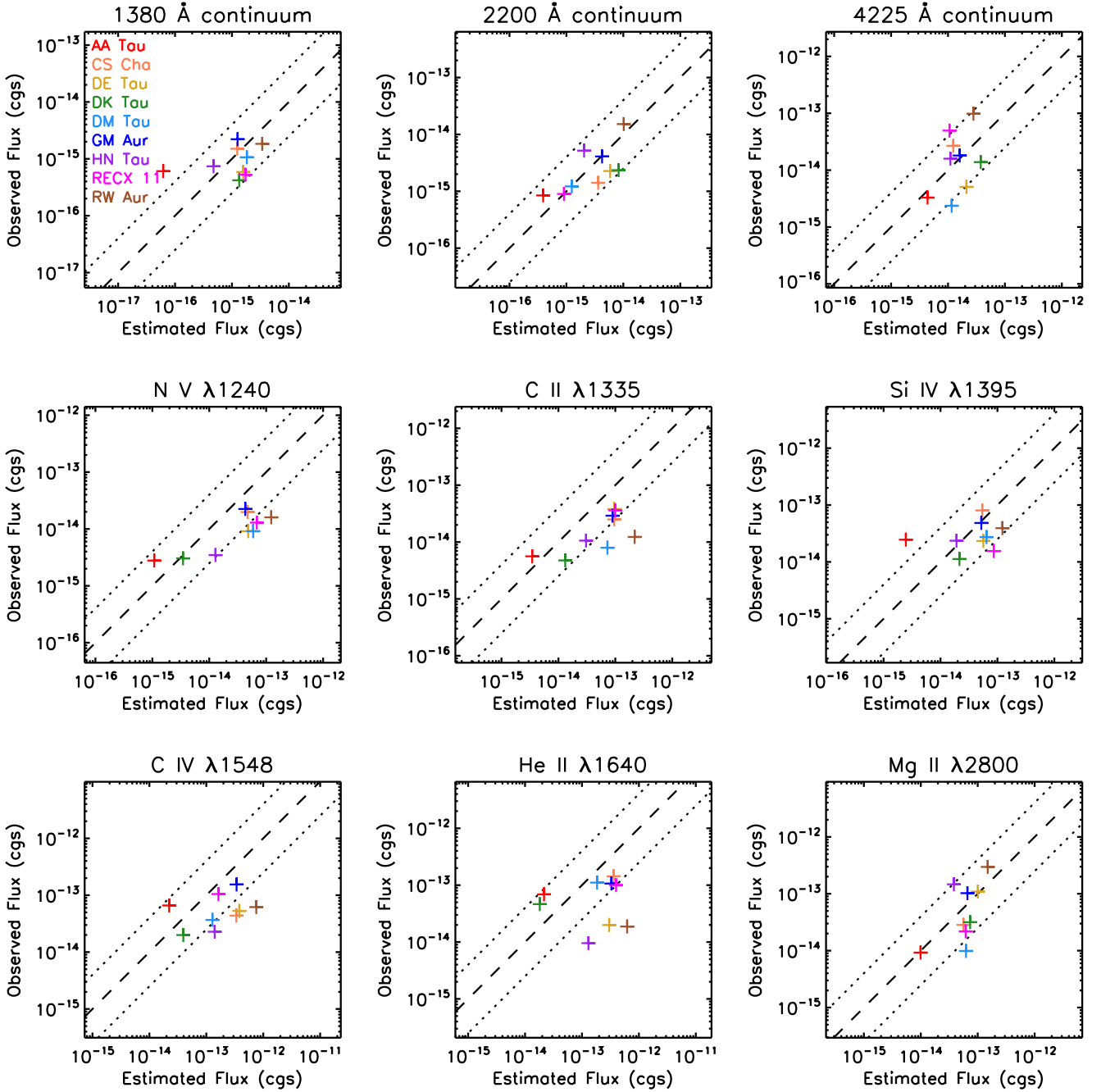

**Figure 20.** Comparison between observed peak line strength and the peak line strength estimated from the accretion rate (see Section A.3). The dashed line indicates a 1:1 relation, while the dotted lines show factors of four deviations.

sequence counterparts (Preibisch et al. 2005; Getman et al. 2008a,b). However, it is worth noting that the UV contribution of flares may be completely dominated by the accretion UV luminosity (Hinton et al. 2022). Given this uncertainty and to preserve the safety of the COS and STIS UV detectors, all T Tauri stars of spectral type M need to be cleared under magnetic flaring conditions following the same approach as for main-sequence

M stars, which is described in Osten (2017a) for COS and Osten (2017b) for STIS.

This approach relies on normalizing a model spectrum to the $U$ band magnitude of a quiescent M star. Because the U-band magnitude of an accreting star is dominated by the accretion flux, we instead estimate the U-band magnitude of a non-accreting star of the same spectral type by normalizing to a longer wavelength measure-ment (normally $V$ band, but $J$ band can be used in cases where the optical is significantly contaminated by accretion flux) and assuming the stellar color $U$-$V$ or $U$-$J$ appropriate for a given spectral type (Ducati et al. 2001). The procedure results in a model spectrum of a flaring M star of the appropriate spectral type.

As for the estimation of exposure times for T Tauri stars described in Section A.3, we apply a dust extinction curve to the magnetic flare model spectrum using the Fitzpatrick & Massa (1990) procedure with the $A_V$ value appropriate of each star and the extinction curve shape parameters appropriate for the line of sight to HD 29647 (Whittet et al. 2004). Additionally, we apply an interstellar H I absorption profile to the magnetic flare model spectrum following the method described in Roman-Duval et al. (2019).

## B. APPENDIX B

This Appendix contains excerpts from the tables listing the target samples for ULLYSES, for the LMC (Table B1), SMC (Table B2), the MC Bridge (Table B3), NGC 3109 (Table B4), Sextans A (Table B5), IC 1613 (Table B6), WLM (Table B7), Leo A (Table B8), Leo P (Table B9), and single-epoch T Tauri stars (Table B10). The full lists of targets for the LMC, SMC, and TTS are available online in machine-readable format. References for astrophysical parameters are only provided in the machine-readable tables in the form of bibcodes.

Table B1. LMC Targets

| Star | RA(J2000) deg | DEC(J2000) deg | SpT | $T_{eff}$ K | log g | $M/M_\odot$ | log $L/L_\odot$ | $R/R_\odot$ | Log $\dot{M}$ $M_\odot$/yr | v sin i km/s | E(B-V) mag | B mag | V mag | AR?[a] |
|---|---|---|---|---|---|---|---|---|---|---|---|---|---|---|
| SK -67 2 | 71.76855431 | -67.1147516 | B1 Ia+ (N wk) | 19910 | 2.32 | | | | | | 0.20 | 11.30 | 11.26 | AR |
| SK -67 5 | 72.57887845 | -67.6605667 | O9.7 Ib | | | | | | | 90 | 0.14 | 11.22 | 11.34 | AR |
| Bl 13 | 73.27710840 | -68.0563326 | O6.5 V | | | | | | | 118 | 0.18 | 13.66 | 13.75 | ULL |
| SK -68 8 | 73.43037118 | -68.7148028 | B5 Ia+ | | | | | | | | 0.16 | 11.09 | 11.02 | ULL |
| SK -70 13 | 73.50485590 | -69.9965344 | O9 V | 22890 | | | | | | 84 | 0.13 | 12.15 | 12.29 | ULL |
| SK -67 14 | 73.63288137 | -67.2568298 | B1.5 Ia | | 2.68 | | 5.84 | | | 88 | 0.08 | 11.42 | 11.52 | AR |
| SK -70 16 | 73.73903106 | -70.0411548 | B4 I | | | | | | | | 0.07 | 13.04 | 13.10 | ULL |
| SK -67 20 | 73.88061005 | -67.5007470 | WN4 b | 158000 | | 25.00 | | 1.10 | -4.48 | | 0.08 | 13.53 | 13.79 | ULL |
| SK -66 19 | 73.97475324 | -66.4164869 | O7 V | | | | | | | | 0.39 | 12.91 | 12.79 | AR |
| SK -66 17 | 73.98124217 | -66.4723902 | OC9.5 II | 29500 | 3.15 | 35.00 | 5.49 | | | 60 | 0.18 | 12.81 | 12.89 | ULL |
| SK -66 18 | 73.99927950 | -65.9749268 | O6 V((f)) | 40200 | 3.76 | 40.70 | 5.55 | 12.20 | -5.97 | 82 | 0.08 | 13.30 | 13.50 | ULL |
| SK -69 42 | 74.01207651 | -69.4559848 | WC4 | 87000 | | | 5.48 | 2.40 | -4.85 | | 0.05 | 13.99 | 14.14 | AR |
| SK -69 43 | 74.04360068 | -69.2606392 | B0.5 Ia | 22845 | 2.62 | | | | | 102 | 0.10 | 11.90 | 11.98 | ULL |
| N11 ELS 033 | 74.04592707 | -66.4734354 | B0 IIIn | 26700 | 3.20 | | 5.05 | 15.70 | -6.60 | 256 | 0.08 | 13.46 | 13.68 | ULL |
| SK -66 21 | 74.04625154 | -66.2925143 | WC4 | 84000 | | | 5.48 | 2.60 | -4.85 | | 0.08 | 14.37 | 14.33 | AR |
| N11 ELS 049 | 74.12318422 | -66.4724952 | O7.5 V | | | | | | | | 0.03 | 13.78 | 14.02 | ULL |
| N11 ELS 051 | 74.12376423 | -66.3607948 | O5 Vn((f)) | 41400 | 3.70 | | 5.42 | 8.60 | -6.39 | 350 | 0.02 | 13.77 | 14.03 | ULL |
| BAT99 10 | 74.14455541 | -66.4742293 | WC4 | | | | | | | | | 13.30 | 13.39 | AR |
| N11 ELS 018 | 74.17105073 | -66.4112613 | O6 II(f+) | | | | | | | 110 | 0.18 | 13.04 | 13.13 | ULL |
| N11 ELS 060 | 74.17557771 | -66.4151979 | O3 V((f*)) | 48000 | 3.97 | | 5.63 | 9.50 | -6.29 | 68 | 0.22 | 14.18 | 14.24 | ULL |
| N11 ELS 031 | 74.17713458 | -66.4217272 | ON2 III(f*) | 56000 | 4.00 | | 6.12 | 12.20 | -5.66 | 71 | 0.27 | 13.67 | 13.68 | ULL |
| PGMW 3070 | 74.18031230 | -66.4173602 | O6 V | | | | | | | 72 | 0.06 | 12.53 | 12.75 | ULL |

[a] "ULL" for targets observed as part of the ULLYSES program; "AR" for purely archival targets

NOTE—The full list of LMC targets is provided online in machine-readable format, which also includes references (not included here). Qualifiers used for spectral classification of massive stars are listed and explained in Sota et al. (2011).

**Table B2.** SMC Targets

| Star | RA(J2000) deg | DEC(J2000) deg | SpT | $T_{eff}$ K | log g | $M/M_\odot$ | $\log L/L_\odot$ | $R/R_\odot$ | Log $\dot{M}$ $M_\odot$/yr | v sin i km/s | E(B-V) mag | B mag | V mag | AR? [a] |
|---|---|---|---|---|---|---|---|---|---|---|---|---|---|---|
| 2dFS 163 | 9.24266705 | -73.3925521 | O8 Ib(f) | | | | | | | | 0.11 | 14.95 | 15.11 | ULL |
| AV 6 | 11.32586390 | -73.2563871 | O9 III | | | | | | | | 0.31 | 13.36 | 13.31 | ULL |
| AV 14 | 11.63597795 | -73.1015510 | O5 V | 42800 | 4.00 | 32.00 | 5.41 | 9.30 | -7.70 | 90 | 0.11 | 13.38 | 13.55 | AR |
| AV 15 | 11.67567080 | -73.4154203 | O6.5 II(f) | 39000 | 3.61 | 47.20 | 5.83 | 18.00 | -5.96 | 120 | 0.10 | 12.93 | 13.12 | AR |
| AV 16 | 11.72930161 | -73.1428217 | sgB0[e] | | | | | | | | 0.07 | 13.10 | 12.97 | AR |
| AV 18 | 11.80091376 | -73.1092036 | B2 Ia | 19000 | 2.30 | 28.00 | 5.44 | 49.00 | -6.64 | 49 | 0.20 | 12.49 | 12.46 | AR |
| AV 22 | 11.91143086 | -73.1302421 | B5 Ia | 14500 | 1.90 | 19.00 | 5.04 | 53.00 | -6.64 | 46 | 0.08 | 12.19 | 12.20 | AR |
| AV 26 | 11.95854773 | -73.1391873 | O6 I(f) | 38000 | 3.52 | 81.00 | 6.14 | 27.50 | -5.60 | 150 | 0.10 | 12.29 | 12.46 | AR |
| AV 39a | 12.12853017 | -73.2627361 | WN5ha | 47000 | | 43.00 | 5.57 | 9.10 | -5.74 | | 0.10 | 14.08 | 14.26 | AR |
| AV 43 | 12.19958504 | -72.7735866 | B0.5 III | 28500 | 3.37 | 22.40 | 5.13 | 15.10 | -7.66 | 200 | 0.08 | 13.76 | 13.88 | AR |
| AV 47 | 12.21454425 | -73.4329340 | O8 III((f)) | 35000 | 3.75 | 42.20 | 5.44 | 14.30 | -7.68 | 60 | 0.05 | 13.26 | 13.44 | AR |
| OGLE J004942.75-731717.7 | 12.42810262 | -73.2883844 | O6 V((f))e | | | | | | | | 0.34 | 14.17 | 14.11 | AR |
| AV 61 | 12.50778471 | -72.1907249 | O6 III((f))e_1 | | | | | | | 228 | 0.09 | 13.36 | 13.54 | AR |
| AV 69 | 12.57170571 | -72.8917444 | OC7.5 III((f)) | 33900 | 3.50 | 39.70 | 5.61 | 18.60 | -6.01 | 70 | 0.11 | 13.09 | 13.27 | AR |
| AV 70 | 12.57552222 | -72.6361269 | O9.5 Ibw | 28500 | 3.10 | | 5.68 | 28.40 | -5.82 | 100 | 0.09 | 12.21 | 12.38 | ULL |
| AV 75 | 12.63503244 | -72.8767948 | O5 III(f+) | 38500 | 3.51 | 51.10 | 5.94 | 21.00 | -5.80 | 120 | 0.14 | 12.55 | 12.70 | AR |
| AV 77 | 12.63965399 | -72.7958413 | O7 III | 37500 | 3.74 | 28.10 | 5.40 | 11.90 | -7.89 | 150 | 0.15 | 13.73 | 13.92 | AR |
| AV 78 | 12.65995443 | -73.4717435 | B1.5 Ia+ | 21500 | 2.40 | 53.00 | 5.92 | 79.00 | -5.64 | 46 | 0.15 | 11.02 | 11.05 | AR |
| AV 81 | 12.68094910 | -73.4516206 | WN6h | 45000 | | 38.00 | 5.78 | 13.00 | -5.18 | | 0.09 | 13.25 | 13.37 | AR |
| AV 80 | 12.68256443 | -72.7948816 | O4-6n(f)p | 38000 | 3.50 | | | | | 325 | 0.15 | 13.19 | 13.32 | AR |
| AV 83 | 12.71668182 | -72.7041557 | O7 Iaf+ | 32800 | 3.26 | 22.10 | 5.54 | 18.30 | -5.64 | 80 | 0.13 | 13.45 | 13.58 | AR |
| AV 83 | 12.71668182 | -72.7041557 | O7 Iaf+ | 32800 | 3.26 | 22.10 | 5.54 | 18.30 | -5.64 | 80 | 0.13 | 13.45 | 13.58 | AR |
| AV 85 | 12.75077817 | -72.8845237 | B1 II-IIIe | | | | | | | | 0.17 | 13.68 | 13.75 | ULL |

[a] "ULL" for targets observed as part of the ULLYSES program; "AR" for purely archival targets

NOTE.—The full list of SMC targets is provided online in machine-readable format, which also includes references (not included here). Qualifiers used for spectral classification of massive stars are listed and explained in Sota et al. (2011).

**Table B3.** MC Bridge Targets

| Star | RA(J2000) | DEC(J2000) | SpT | $T_{\rm eff}$ | log g | $M/M_\odot$ | log $L/L_\odot$ | $R/R_\odot$ | Log $\dot{M}$ | v sin i | E(B-V) | B | V | AR?[a] |
|---|---|---|---|---|---|---|---|---|---|---|---|---|---|---|
| | deg | deg | | K | | | | | $M_\odot$/yr | km/s | mag | mag | mag | |
| D191 719 | 32.97932245 | -74.2103100 | O9 V | 34000 | 4.10 | 19.20 | 4.70 | 6.50 | | 2 | 0.08 | 14.83 | 15.24 | AR |
| Gaia DR3 4637562032451492608 | 33.02010886 | -74.1993675 | O9.5 V | 33000 | 4.00 | 18.00 | 4.72 | 7.00 | | 8 | 0.05 | 15.00 | 15.18 | AR |
| Gaia DR3 4637562376048872448 | 33.04110027 | -74.1740061 | O | | | | | | | | | | | AR |
| D191 739 | 33.07439654 | -74.1642048 | O | | | | | | | | | 13.67 | 14.05 | AR |
| Sk 215 | 33.35593421 | -74.5223119 | B5 Ia | | | | | | | | 0.03 | 12.12 | 12.18 | AR |
| D191 842 | 33.54941711 | -74.0785630 | O8 III | 33000 | 3.50 | 19.30 | 5.25 | 12.90 | | 150 | 0.14 | 13.98 | 14.26 | AR |

[a]"ULL" for targets observed as part of the ULLYSES program; "AR" for purely archival targets

NOTE.—References are not provided here, but are included in the online machine-readable tables. Qualifiers used for spectral classification of massive stars are listed and explained in Sota et al. (2011).

**Table B4.** NGC 3109 Targets

| Star | RA(J2000) | DEC(J2000) | SpT | $T_{eff}$ | log g | $M/M_\odot$ | $\log L/L_\odot$ | $R/R_\odot$ | $\log \dot{M}$ | v sin i | E(B-V) | B | V | AR?[a] |
|------|-----------|------------|-----|-----------|-------|-------------|------------------|-------------|----------------|---------|--------|-----|-----|--------|
| | deg | deg | | K | | | | | $M_\odot/yr$ | km/s | mag | mag | mag | |
| NGC 3109 EBU 07 | 150.72800344 | -26.1499067 | B0-1 Ia | 27000 | 2.90 | 39.70 | 5.82 | 37.00 | | | 0.09 | | 18.69 | ULL |
| NGC 3109 EBU 20 | 150.76373502 | -26.1559282 | O8 I | 31150 | 3.53 | 69.10 | 5.87 | 23.70 | -5.35 | 110 | | | 19.33 | ULL |
| NGC 3109 EBU 34 | 150.80960917 | -26.1549717 | O8 I(f) | 33050 | 3.16 | 25.00 | 5.69 | 21.80 | -5.74 | 82 | | | 19.61 | ULL |

[a]"ULL" for targets observed as part of the ULLYSES program; "AR" for purely archival targets

NOTE—References are not provided here, but are included in the online machine-readable tables. Qualifiers used for spectral classification of massive stars are listed and explained in Sota et al. (2011).

**Table B5.** Sextans A Targets

| Star | RA(J2000) deg | DEC(J2000) deg | SpT | $T_{eff}$ K | log g | $M/M_\odot$ | $\log L/L_\odot$ | $R/R_\odot$ | Log $\dot{M}$ $M_\odot$/yr | v sin i km/s | E(B-V) mag | B mag | V mag | AR? [a] |
|---|---|---|---|---|---|---|---|---|---|---|---|---|---|---|
| Sextans A LGN s014 | 152.72415923 | -4.6869743 | O7.5 III((f)) | 37400 | 3.80 | | | | | | 0.00 | 20.42 | 20.69 | AR |
| Sextans A LGGS J101056.86-044040.8 | 152.73685546 | -4.6780290 | O | | | | | | | | 0.00 | 20.14 | 20.41 | AR |
| Sextans A LGN s004 | 152.74120833 | -4.7195000 | O5 III | | | | | | | | 0.043 | 20.64 | 20.92 | ULL |
| Sextans A LGN s029 | 152.74214670 | -4.7217944 | O8.5 III | 32500 | 3.50 | 21.00 | 5.10 | 12.10 | | 290 | 0.03 | 20.56 | 20.80 | AR |
| Sextans A LGN s003 | 152.74408272 | -4.7247062 | O3-5 Vz | | | | | | | | 0.23 | 20.71 | 20.80 | ULL |
| Sextans A LGN s050 | 152.75271445 | -4.6790009 | O9.7 I | 26300 | 2.90 | | | | | | 0.01 | 19.36 | 19.61 | AR |
| Sextans A LGN s089 | 152.75985926 | -4.6707456 | B2.5 I | | | | | | | | 0.05 | 19.48 | 19.58 | AR |
| Sextans A LGN s037 | 152.76987434 | -4.7067326 | O9 I | 31400 | 3.24 | | | | | | 0.03 | 20.45 | 20.68 | AR |
| Sextans A LGN s021 | 152.76995788 | -4.7058222 | O8 V | | | | | | | | 0.05 | 20.13 | 20.60 | AR |
| Sextans A LGN s022 | 152.77239603 | -4.7111532 | O8 V | 31900 | 3.72 | | | | | | 0.01 | 19.20 | 19.46 | AR |
| Sextans A LGN s071 | 152.77367156 | -4.7037874 | B1 I | | | | | | | | 0.00 | 19.46 | 19.70 | ULL |
| Sextans A LGN s038 | 152.77515852 | -4.7031917 | O9 I((f)) | 29300 | 3.27 | | | | | | 0.03 | 19.26 | 19.49 | AR |

[a] "ULL" for targets observed as part of the ULLYSES program; "AR" for purely archival targets

NOTE—References are not provided here, but are included in the online machine-readable tables. Qualifiers used for spectral classification of massive stars are listed and explained in Sota et al. (2011).

**Table B6.** IC 1613 Targets

| Star | RA(J2000) | DEC(J2000) | SpT | $T_{eff}$ | log g | M/M$_\odot$ | log L/L$_\odot$ | R/R$_\odot$ | Log $\dot{M}$ | v sin i | E(B-V) | B | V | AR?[a] |
|---|---|---|---|---|---|---|---|---|---|---|---|---|---|---|
| | deg | deg | | K | | | | | M$_\odot$/yr | km/s | mag | mag | mag | |
| IC1613 BUG C10 | 16.18075752 | 2.1728840 | B1.5 Ib | | | | | | | | | | 18.82 | AR |
| IC1613 BUG B11 | 16.18263620 | 2.1125449 | O9.5 I | 30000 | 3.25 | 22.30 | 5.45 | 19.70 | -7.82 | | 0.13 | 18.49 | 18.62 | AR |
| IC1613 BUG B4 | 16.24815736 | 2.1545025 | B1.5 Ia | 22500 | 2.60 | 11.00 | 5.20 | 27.00 | | | 0.05 | 18.10 | 18.23 | AR |
| IC1613 GHV 61331 | 16.25067121 | 2.1536638 | O9.7 II | 33000 | 3.80 | | | | | | 0.05 | 18.93 | 19.14 | AR |
| IC1613 GHV 62024 | 16.25272168 | 2.1470408 | O6.5 IIIf | 36500 | 3.60 | 18.30 | 5.29 | 11.10 | -6.37 | | 0.11 | 19.44 | 19.60 | AR |
| IC1613 BUG A10 | 16.25382248 | 2.1782204 | B1 Ia | 25000 | 2.70 | 27.00 | 5.71 | 38.00 | | | 0.08 | 17.28 | 17.42 | AR |
| IC1613 BUG B7 | 16.25824478 | 2.1347658 | O9 II | 35050 | 3.74 | 28.40 | 5.32 | 12.60 | | 270 | 0.06 | 18.76 | 18.96 | AR |
| IC1613 GHV 64066 | 16.25865785 | 2.1578252 | O3 III((f)) | | | | | | | | 0.07 | 18.82 | 19.03 | AR |
| IC1613 BUG B2 | 16.26280211 | 2.1679569 | O7.5 III-V((f)) | | | | | | | | 0.07 | 19.42 | 19.62 | AR |
| IC1613 GHV 67559 | 16.26988241 | 2.1564654 | O8.5 III((f)) | | | | | | | | 0.07 | 19.04 | 19.24 | AR |
| IC1613 GHV 67684 | 16.27035459 | 2.1590737 | O8.5 I | 38500 | 3.80 | | | | | | 0.05 | 18.81 | 19.02 | AR |
| IC1613 BUG A13 | 16.27607586 | 2.1786834 | O3-4 V((f)) | 42500 | 3.75 | 27.60 | 5.62 | 11.90 | -6.60 | | 0.05 | 18.73 | 18.96 | AR |
| IC1613 BUG B3 | 16.27656999 | 2.1587237 | B0 Ia | 24500 | 2.65 | 19.00 | 5.53 | 32.00 | | | 0.09 | 17.54 | 17.69 | AR |

[a]"ULL" for targets observed as part of the ULLYSES program; "AR" for purely archival targets

NOTE—References are not provided here, but are included in the online machine-readable tables. Qualifiers used for spectral classification of massive stars are listed and explained in Sota et al. (2011).

**Table B7.** WLM Targets

| Star | RA(J2000) | DEC(J2000) | SpT | $T_{eff}$ | log g | M/M$_\odot$ | log L/L$_\odot$ | R/R$_\odot$ | Log $\dot{M}$ | v sin i | E(B-V) | B | V | AR?[a] |
|------|-----------|------------|-----|-----------|-------|-------------|-----------------|-------------|---------------|---------|--------|---|---|-----|
| | deg | deg | | K | | M$_\odot$ | | | M$_\odot$/yr | km/s | mag | mag | mag | |
| WLM BPU A 11 | 0.49981162 | -15.4721055 | O9.7 Ia | 29000 | 3.25 | 53.00 | 5.69 | 28.00 | -7.96 | | 0.15 | 18.27 | 18.38 | AR |
| WLM BPU A 15 | 0.50222183 | -15.4978926 | O7 V((f)) | 37500 | 4.00 | 29.00 | 5.30 | 9.50 | | 80 | 0.00 | 19.97 | 20.25 | AR |

[a]"ULL" for targets observed as part of the ULLYSES program; "AR" for purely archival targets

NOTE—References are not provided here, but are included in the online machine-readable tables. Qualifiers used for spectral classification of massive stars are listed and explained in Sota et al. (2011).

**Table B8.** Leo A Targets

| Star | RA(J2000) deg | DEC(J2000) deg | SpT | $T_{eff}$ K | log g | $M/M_\odot$ | $\log L/L_\odot$ | $R/R_\odot$ | $\log \dot{M}$ $M_\odot$/yr | v sin i km/s | E(B-V) mag | B mag | V mag | AR? [a] |
|---|---|---|---|---|---|---|---|---|---|---|---|---|---|---|
| J095921.90+304518.1 | 149.84125000 | 30.7550200 | OB | | | | | | | | | | 19.90 | AR |
| LeoA GWS K1 | 149.86468201 | 30.7493758 | O9 V | 30900 | 3.69 | 19.00 | 4.90 | 9.80 | | 80 | | | 20.10 | AR |
| LeoA GWS K2 | 149.87593492 | 30.7436342 | O9.7 V | 31600 | 4.10 | 17.00 | 4.72 | 7.70 | | 95 | | | 20.20 | AR |

[a] "ULL" for targets observed as part of the ULLYSES program; "AR" for purely archival targets

NOTE—References are not provided here, but are included in the online machine-readable tables. Qualifiers used for spectral classification of massive stars are listed and explained in Sota et al. (2011).

**Table B9.** Leo P Targets

| Star | RA(J2000) | DEC(J2000) | SpT | $T_{\rm eff}$ | log g | $M/M_\odot$ | log $L/L_\odot$ | $R/R_\odot$ | Log $\dot{M}$ | v sin i | E(B-V) | B | V | AR?[a] |
|------|-----------|------------|-----|---------------|-------|-------------|-----------------|-------------|---------------|---------|--------|-----|-----|--------|
| | deg | deg | | K | | | | | $M_\odot$/yr | km/s | mag | mag | mag | |
| Leo-P ECG LP 26 | 155.43800708 | 18.0880361 | O7-8 V | 37500 | 4.00 | 21.00 | 5.10 | 8.20 | | 370 | 0.00 | | 20.62 | AR |

[a]"ULL" for targets observed as part of the ULLYSES program; "AR" for purely archival targets

NOTE—References are not provided here, but are included in the online machine-readable tables.

Table B10. TTS Targets

| Star | RA(J2000) deg | DEC(J2000) deg | SF Region | SpT | $M_*/M_\odot$ | $\mathrm{Log}_{10}\,\dot{M}$ $M_\odot\,\mathrm{yr}^{-1}$ | E(B-V) mag | V mag | AR?[a] |
|---|---|---|---|---|---|---|---|---|---|
| RX J0438.6+1546 | 69.66286169 | 15.7703485 | Taurus | K2 | 1.20 | -9.52 | 0.10 | 10.86 | ULL |
| LkCa 15 | 69.82418065 | 22.3508638 | Taurus | K5 | 1.10 | -8.51 | 0.35 | 12.03 | AR |
| CVSO 109 | 83.13610687 | -1.2294375 | Ori OB1b | M0 | 0.56 | -7.60 | 0.00 | 13.97 | ULL |
| CVSO-109A | 83.13610687 | -1.2294375 | Ori OB1b | | | | | | COMP |
| CVSO-109B | 83.13599552 | -1.2295746 | Ori OB1b | | | | | | COMP |
| TX Ori | 84.64037132 | -2.7372705 | Sigma Ori | K4.5 | 1.09 | -8.13 | 0.00 | 12.06 | ULL |
| RECX 1 | 129.23353980 | -78.9458900 | eta Cha | K6.0 | 0.75 | -9.12 | 0.00 | 10.46 | AR |
| RECX 5 | 130.61220090 | -78.9631891 | eta Cha | M4.5 | 0.15 | -9.89 | 0.00 | 15.20 | ULL |
| RECX 6 | 130.66092400 | -78.9117512 | eta Cha | M3 | 0.30 | -10.60 | 0.00 | 14.10 | ULL |
| ECHA J0843.3-7915 | 130.82678310 | -79.0882684 | eta Cha | M4.0 | 0.20 | -9.12 | 0.00 | 13.97 | AR |
| Hn 5 | 166.67372190 | -76.5969654 | Cha I | M5 | 0.16 | -9.28 | 0.00 | | ULL |
| TWA 3A | 167.61565310 | -37.5310572 | TWA | M3 | 0.30 | -10.01 | 0.00 | 12.57 | AR |
| Sz 45 | 169.40362790 | -77.0772622 | Cha I | M0.5 | 0.51 | -8.09 | 0.23 | 13.50 | ULL |
| HD 104237E | 180.03787310 | -78.1951076 | eps Cha | K5.5 | 0.90 | -9.07 | 0.32 | 12.08 | ULL |
| TWA 27 | 181.88907900 | -39.5484432 | TWA | M9 | 0.02 | -10.68 | 0.00 | 19.95 | AR |
| PDS 70 | 212.04213470 | -41.3980439 | Upper Scorpius | K7 | 0.76 | -9.88 | 0.02 | 12.18 | AR |
| Sz 66 | 234.86777440 | -34.7717866 | Lupus I | M3 | 0.29 | -8.54 | 0.32 | 15.00 | ULL |
| RX J1556.1-3655 | 239.00868170 | -36.9246207 | Lupus II | M1 | 0.50 | -7.92 | 0.32 | 13.85 | ULL |
| Sz 84 | 239.51043100 | -37.6008607 | Lupus I | M5 | 0.16 | -9.21 | 0.00 | 16.16 | ULL |
| Sz 130 | 240.12925820 | -41.7270450 | Lupus IV | M2 | 0.41 | -9.19 | 0.00 | 14.71 | ULL |
| Sz 98 | 242.09366790 | -39.0796721 | Lupus III | K7 | 0.70 | -7.23 | 0.32 | 13.66 | ULL |
| RXJ1842.9-3532 | 280.74159610 | -35.5453496 | CrA | K2 | 0.93 | -8.80 | 0.13 | 11.77 | ULL |

[a] "ULL" for targets observed as part of the ULLYSES program; "AR" for purely archival targets; "COMP" for companions in the STIS slit

NOTE—The full list of TTS targets is provided online in machine-readable format, which also includes references (not included here)